\documentclass{amsart}

\usepackage[T1]{fontenc}
\usepackage{amsmath, amsthm, amssymb, bm, bbm}
\usepackage{amsfonts}
\usepackage{geometry}
\usepackage{url,color}
\usepackage{graphicx}

\usepackage{algpseudocode}
\usepackage{algorithm}
\usepackage{comment}
\RequirePackage[colorlinks,citecolor=blue,linkcolor=blue,urlcolor=blue,pagebackref]{hyperref}
\usepackage{tikz}

\allowdisplaybreaks
\usepackage{setspace}
\usepackage{booktabs}

\usepackage{subcaption}
\tikzset{
glow/.style={preaction={#1, draw, line join=round, line width=3pt, opacity=1}}
}
\usepackage{multirow}
\usepackage{babel}

\usepackage[authoryear,round]{natbib}

\usepackage{mathtools}
\usepackage{threeparttable}

\numberwithin{equation}{section}

\theoremstyle{plain}
\newtheorem{theorem}{Theorem}

\newtheorem{lemma}{Lemma}
\newtheorem{ass}{Assumption}  
\newtheorem{thm}{Theorem} 
\newtheorem{cor}{Corollary} 

\newtheorem{prop}[thm]{Proposition}

\theoremstyle{remark}
\newtheorem{defn}{Definition}
\newtheorem{exmp}{Example} 
\newtheorem{remark}{Remark}

\newcommand{\EE}{{\mathbb E}}

\newcommand{\PP}{{\mathbb P}}
\renewcommand{\Pr}{{\mathbb P}}

\newcommand{\RR}{{\mathbb R}}

\newcommand{\ZZ}{{\mathbb Z}}

\newcommand{\calA}{{\mathcal A}}

\newcommand{\calC}{{\mathcal C}}

\newcommand{\calE}{{\mathcal E}}

\newcommand{\calG}{{\mathcal G}}

\newcommand{\calI}{{\mathcal I}}
\newcommand{\calJ}{{\mathcal J}}
\newcommand{\calK}{{\mathcal K}}

\newcommand{\calP}{{\mathcal P}}

\newcommand{\calR}{{\mathcal R}}
\newcommand{\calS}{{\mathcal S}}
\newcommand{\calT}{{\mathcal T}}

\newcommand{\calV}{{\mathcal V}}

\newcommand{\calX}{{\mathcal X}}

\begin{document}

\title[Network Measurement Error and Non-robustness of Diffusion Estimates]{Non-robustness of diffusion estimates on networks with measurement error}

\author[Chandrasekhar]{Arun G. Chandrasekhar$^{\ddagger,\dagger,\star}$}
\author[Goldsmith-Pinkham]{Paul Goldsmith-Pinkham$^{\circ,\star}$}
\author[McCormick]{Tyler H. McCormick$^{\S}$}
\author[Thau]{Samuel Thau$^{\ddagger}$}
\author[Wei]{Zeyu Wei$^{\S}$}

\thanks{We gratefully acknowledge Isaiah Andrews, Abhijit Banerjee, Haoge Chang, Jishnu Das, Krishna Dasaratha, Matt Jackson, Matt Gentzkow, Ben Golub, Ed Kaplan, Julianne Meisner, Jim Moody, Karl Rohe, Adam Szeidl, Alex Volfovsky, and Juan Pablo Xandri.}
\thanks{$^{\ddagger}$Department of Economics, Stanford University}
\thanks{$^{\dagger}$NBER}
\thanks{$^{\star}$J-PAL}
\thanks{$^{\circ}$Yale School of Management}
\thanks{$^{\S}$Department of Statistics, University of Washington}

\begin{abstract}
Network diffusion models are used to study disease transmission, information spread, technology adoption, and other socio-economic processes. We show that estimates of these diffusions are highly non-robust to mismeasurement. First, even when the \emph{network} is measured perfectly, small and local mismeasurement in the initial seed generates a large shift in the locations of the expected diffusion. Second, if instead the initial seed is known, even a vanishingly small share of missed links causes diffusion forecasts to be significant under-estimates. Forecast failure depends critically on the \emph{geometry} of measurement error: we provide sufficient conditions for catastrophic failure when missing links bridge distant network regions (acting as shortcuts), and sufficient conditions for robustness when missing links are a uniformly, randomly thinned subset of the full network (preserving network structure). Such failures exist even when the basic reproductive number is consistently estimable. We explore difficulties implementing possible solutions and conduct simulations on synthetic and real networks.
\end{abstract}

\maketitle

Researchers and policymakers use network data to estimate models of diffusion---quantifying the extent of illness or technology adoption, summarizing dynamics (e.g., $\calR_{0}$), and targeting interventions. See \cite{anderson1991infectious}, \cite{jackson2009}, \cite{jackson2011diffusion}, and \cite{sadlere2023} for applications, including settings with strategic behavior.

This paper shows that even tiny mismeasurement of the network creates large forecast errors in diffusion. We identify a general mechanism: on networks with polynomial expansion---capturing structured interactions from geography, social groups, or institutions---even a vanishingly small number of missing links that are geometrically unaligned with the observed network change the expansion properties, generating catastrophic forecast errors. This holds even when the econometrician has perfect knowledge of either the network or the initial seed, and over policy-relevant intermediate horizons: long enough for diffusion to matter, short enough that it has not saturated the network. The numerical observation of \cite{watts1998collective}---that random shortcuts on a ring lattice dramatically compress path lengths---is a special case of this principle; our framework explains \emph{why} it occurs and characterizes precisely \emph{when} it does and does not arise.

Crucially, not all missing links cause these problems. When missing links mirror the structure of observed links---what we term \emph{aligned} error---forecasts remain asymptotically correct. Forecast failure occurs precisely when missing links are more dispersed than observed links: they act as shortcuts bridging distinct parts of the observed graph, accelerating expansion beyond what the observed network would predict. We call this \emph{unaligned} error. This characterization provides actionable guidance for applied researchers: the key question is not \emph{how many} links are missing, but whether data collection is more likely to miss links that resemble observed links (aligned---safe for forecasting) or links that bridge different network regions (unaligned---dangerous).

We show five key results: (i) predictions of \emph{where} diffusion goes are sensitive to local uncertainty of the initial seeding (Theorem~\ref{thm:sensitive-dep}); (ii) predictions of diffusion counts will be catastrophically under-estimated with even vanishingly small unaligned measurement error of the network (Theorem~\ref{thm:main-polynomial}); (iii) when measurement error is instead aligned, forecast failure vanishes (Theorem~\ref{thm:no-failure-mar}); (iv) while aggregated quantities such as the basic reproductive number $\calR_0$ can be estimated consistently, they provide limited information for disaggregated forecasts; (v) because the measurement error is so small, most data augmentation approaches are ineffectual.

For intuition, consider a network where connections depend on observable proximity (geography, school, work) or latent factors \citep{hoffrh2002}. A ball around the initial seed enumerates possibly affected nodes, expanding polynomially. Even with a perfectly known network, nearby seeds produce substantially diverging balls---misleading conclusions about where diffusion goes. The mismeasured seed can be within a neighborhood proportional to the forecast horizon, so the error vanishes in relative terms. 

Now suppose the seed is known but some unaligned links are missed. If any missed link reaches beyond the seed's ball on the observed graph, the diffusion escapes to an unexposed region. This jump need not span a large physical distance---it must only bridge parts far apart in \emph{graph distance} on the observed network. Missing links function as escape hatches that alter the effective geometry of diffusion---a mechanism that underlies the \cite{watts1998collective} observation, but which operates on any polynomial-expansion network, not just ring lattices. Our general theorems allow each node to link to a vanishing fraction of the population with arbitrary structure, nesting cases of only local mismeasurement that is geometrically distinct from $L_n$. 

The interaction between both issues, seed sensitivity and jump links, is multiplicative. Since different local seeds influence disjoint local regions, they encounter different error links and jump to different distant parts of the network. This causes the predictions to then diverge further as each secondary wavefront generates its own independent jumps. A local seed perturbation
cascades into forecast errors spanning the entire network.

This mechanism also clarifies why aligned error is comparatively benign. When missing links are aligned, they do not bridge distant regions of the graph, so they furnish no escape hatches. Provided the number of such omissions remains small, the diffusion process stays confined to the neighborhood of the seed's ball, and forecasts remain accurate because no secondary wavefront can leap to an unexposed part of the network.

Missing links are a common concern \citep{wang2012measurement,sojourner2013identification,chandrasekharl2010,advani2018credibly,griffith2022name}, but our paper highlights the impact of even the smallest errors on diffusion forecasts. Mismeasurement arises in three ways. First, aggregation into compartments (e.g., location-by-age-by-occupation) may match average interaction patterns but miss heterogeneity and cross-compartment connections \citep{acemoglu2021optimal,farboodi2021internal,fajgelbaum2021optimal,candogan2025network}. Second, surveys may focus on local connections (within a school or village) while ignoring others. Third, a network snapshot may not capture links relevant to diffusion by the time the process reaches an individual -- in many cases, the network itself is forecasted from mobility or other interaction data. All three sources tend to produce \emph{unaligned} error: they disproportionately miss links that bridge distant parts of the observed network while retaining local ones. 

Formally, we study $n$ agents in an undirected, unweighted network $G_n$ with $n \to \infty$. A SIR diffusion proceeds for $T_n$ periods: each activated node transmits i.i.d.\ with probability $p_n$ to neighbors and is then removed. The true network $G_n = L_n \cup E_n$, where $L_n$ is the observed base network (with polynomial expansion) and $E_n$ is an unobserved error graph.\footnote{We focus on missing links, the primary concern in practice \citep{griffith2022name}. When the econometrician both misses and incorrectly adds links, the problem is more complex.} Polynomial expansion is natural for latent-space models \citep{hoffrh2002}; we verify our conditions for random geometric graphs in Example~\ref{ex:rgg} and Appendix~\ref{sec:rgg-proof}. We use \emph{activated} to nest ``infected,'' ``informed,'' and ``adopted'' \citep{jackson2007diffusion}.

Sections~\ref{sec:sensitive-dependence} and~\ref{sec:forecasting-errors} establish the negative results: seed sensitivity (Theorem~\ref{thm:sensitive-dep} and Corollary~\ref{cor:sensitive-iid}) and catastrophic under-estimation from vanishing measurement error (Theorem~\ref{thm:main-polynomial}). Section~\ref{sec:mar} establishes the positive result: under aligned error, forecast failure vanishes (Theorem~\ref{thm:no-failure-mar}), contrasting sharply with Theorem~\ref{thm:main-polynomial}. Section~\ref{sec:estimation} shows that $p_n$ and $\calR_0$ remain consistently estimable despite forecast failures, while Section~\ref{sec:possible-solutions} shows that potential remedies are ineffectual.\footnote{\cite{alimohammadi2023epidemic} make a similar point, showing that a local estimation algorithm on sampled network data asymptotically identifies the correct SIR compartment proportions.} Section~\ref{sec:empirical-applications} examines these results through three empirical applications: a CA/NV mobility exercise showing that \emph{how} links are missed matters more than \emph{how many}; NYC's micro-cluster zoning, where a degree-matched mobility network identifies ${\sim}30\%$ different buffer zone neighborhoods and better predicts case growth; and the targeting methodology of \cite{beaman2021can}, which is most fragile in the villages where it appears most valuable. Simulations on synthetic networks confirming finite-sample behavior appear in Appendix~\ref{sec:sims}.

\section{Model}
\paragraph*{Environment} 
Consider a set of $n$ observed nodes $V_n$. We model the network as a random undirected and unweighted graph $G_n \coloneqq (V_n, L_n \cup E_n)$, where $L_n$ represents the ``base'' links and $E_n$ represents the missing links. The base network $L_n$ is fixed and perfectly observed by the econometrician. Each link in $E_n$ is formed independently according to $\mathrm{Ber}(\beta_{ij,n})$, where the link probability $\beta_{ij,n}$ may vary across pairs. For expositional simplicity, we focus in the main text on the homogeneous case where $\beta_{ij,n} = \beta_n$ for all pairs $ij$. Appendix~\ref{sec:proofs} expands to a heterogeneous case for $\beta_{ij,n}$, which requires additional assumptions on the ``expansiveness'' of the support of $E_n$.\footnote{Formally, the assumption requires that the support of links in $E_n$ provides sufficiently broad coverage across the node set. Nonetheless, realized networks $E_n$ remain sparse with high probability. The i.i.d.\ case trivially satisfies this expansive support condition. See Remark~\ref{rem:ws} for the connection to the small-worlds framework of \cite{watts1998collective}.} 

The links in $E_n$ are unobserved, so randomness in $G_n$ stems from the random realization of $E_n$. Our model focuses exclusively on mismeasurement due to link omission; we rule out false positives (spurious links). By a slight abuse of notation, we use $L_n$ and $E_n$ to denote both edge sets and the spanning subgraphs on $V_n$.

The diffusion process spreads over the network $G_n$ following a standard Susceptible-Infected-Removed (SIR) model with i.i.d.\ passing probability $p_n$. Each node, once activated, remains active for a single period during which it independently transmits the infection to each neighbor with probability $p_n$. After this period, the node is removed and cannot be reactivated. 

To formalize the stochastic dynamics, we define $P_n(G_n)$ as a random percolation on $G_n$: a directed graph in which each undirected edge of $G_n$ generates two directed edges, each activated independently with probability $p_n$. Under this representation, the diffusion process from an initial seed is equivalent to a deterministic traversal through the realized percolation $P_n(G_n)$. Consequently, all randomness in the diffusion is fully captured by the random realization of $P_n(G_n)$. Similarly, we define $P_n(L_n)$ as the percolation restricted to the base network $L_n$, obtained by restricting $P_n(G_n)$ to edges in $L_n$.

Our analysis is asymptotic in both $n$, the number of nodes, and $T$, the forecast horizon. We consider a sequence of graphs $\{G_n\}_{n=1}^\infty = \{(L_n, E_n)\}_{n=1}^\infty$ where each $E_n$ is drawn randomly and the base network $L_n$ grows with $n$. The forecast horizon is a function of network size, $T = T(n)$, increasing in $n$. The precise growth rate of $T(n)$ is specified below; for notational simplicity, we suppress the dependence of $T$ on $n$ throughout.

The upper bound on the forecast horizon ensures diffusion has not saturated the network; the lower bound ensures a nontrivial forward-looking period. Together they isolate the policy-relevant intermediate regime.\footnote{The upper bound also ensures that diffusion wavefronts remain localized enough to analyze independently across network regions.}

\begin{ass}[Forecast Horizon]\label{ass:forecast-time}
For each \(n\), the forecast horizon \(T_n\) lies in a nonempty interval \([\underline{T}_n,\overline{T}_n]\) where, for the same \(q>0\) as in Assumption~\ref{ass:disease},
\[
\overline{T}_n = o\!\left(n^{\frac{1}{2q+3}}\right)
\quad\text{and}\quad
\underline{T}_n = \omega(1).
\]
Here \(a_n=o(b_n)\) means \(a_n/b_n\to 0\) as \(n\to\infty\), and \(a_n=\omega(1)\) means \(a_n\to\infty\).
\end{ass}

We impose structural assumptions on $L_n$. First, define the activated set:

\begin{defn}\label{defn:infected-set}
Let $I(i_0, T, P_n)$ denote the set of nodes activated by a diffusion process starting at seed $i_0$ and running for $T$ periods through percolation $P_n$. When clear from context, we suppress dependence on $n$ and $P_n$ for notational simplicity.
\end{defn}

Next, we make assumptions on $L_{n}$.
\begin{ass}[Base Network Structure]\label{ass:disease}
Fix \(q>0\). Assume:

\begin{enumerate}
\item \textbf{Polynomial volume growth.} For every node \(j\) and integer \(t\ge 1\), let \(B_j(t)\) denote the ball of graph-distance radius \(t\) centered at \(j\) in \(L_n\). Then for all $t \leq C\overline T_n$ for any $0<C<\infty$:
\[
a_{\min}\, t^{q+1} \le |B_j(t)| \le a_{\max}\, t^{q+1},
\]
where $1<a_{\min}\leq a_{\max} < \infty$. 

\item \textbf{Supercriticality and uniform spread.} The diffusion process is assumed to be supercritical on $L_n$. Furthermore, there exist constants $\alpha < 1/3, \theta \in (0,1)$ such that for any seed $j$ and all $z\in B_j((1-\alpha)t)$ and for all $t \leq \overline T_n$
\begin{align*}
    \mathbb{P}\bigl(|I(j, t)\cap B_z(\alpha t)| > \theta|B_z(\alpha t)|\bigr) > \varepsilon > 0.
\end{align*}
\end{enumerate}
\end{ass}

Part~1 imposes polynomial expansion of order $q+1$: at $t=1$ it yields degree bounds $d_{\min} = a_{\min} - 1$ and $d_{\max} = a_{\max} -1$; for $t > 1$ it bounds reachable nodes within graph distance $t$. The condition permits substantial heterogeneity within polynomial envelopes.

Part~2 requires supercriticality ($p_n > p_c > 0$, where $p_c$ is the percolation threshold for the graph) and uniform spread: for any ball $B_z(\alpha t)$ within the $(1-\alpha)t$-neighborhood of the seed, at least fraction $\theta$ of the ball is reached with positive probability. This ensures diffusion does not systematically avoid local neighborhoods. Regular lattices and polynomial-expansion graphs satisfy this property.\footnote{We conjecture Part~2 can be derived from Part~1, supercriticality, and isoperimetric inequalities using \citet{contreras2024supercritical}'s renormalization techniques.} See \citet{Grimmett2002} and \citet{contreras2024supercritical} for treatments of general graph classes satisfying Part~1.

\begin{remark}[Role of $\alpha < 1/3$]\label{rem:alpha-restriction}
The parameter $\alpha$ controls the \emph{resolution} at which the diffusion fills space. It partitions the reachable radius $t$ into two zones: nodes within $(1-\alpha)t$ of the seed serve as candidate centers $z$, and the ball $B_z(\alpha t)$ around each must be filled with positive probability. Small $\alpha$ imposes a \emph{local} filling condition---the diffusion must reach a positive fraction of moderate-sized neighborhoods throughout most of the reachable region. Large $\alpha$ instead requires filling of a single large ball from near the seed, a coarser, more global condition.

The bound $\alpha < 1/3$ is used in the proof of Theorem~\ref{thm:sensitive-dep} to ensure the geometric construction admits a valid choice of parameters (see Appendix~\ref{sec:proofs}). This restriction is not binding for the graph classes of primary interest. For the random geometric graph (Example~\ref{ex:rgg}), the block-renormalization proof of Theorem~\ref{thm:rgg-spread} establishes Part~2 for \emph{some} $\alpha \in (0,1)$, but the underlying shape-theorem machinery---chemical-distance control and positive cluster density in supercritical percolation---yields uniform spread at \emph{every} scale: for any $\alpha' \in (0,1)$, the conclusion holds with constants $\theta(\alpha')$ and $\varepsilon(\alpha')$ that remain positive. The same is true for integer lattices and, more broadly, any polynomial-expansion graph where supercritical percolation satisfies a shape theorem. A graph that satisfies Part~2 only for large $\alpha$ would exhibit ``lumpy'' diffusion---reliable spread at coarse scales but systematic gaps at finer resolution---a pathology that does not arise in spatially grounded networks.
\end{remark}

By Assumption~\ref{ass:disease}, the radius-\(t\) ball around any node contains \(\Theta(t^{q+1})\) nodes. Consequently, at the maximal forecast horizon,
\[
|B(\overline{T}_n)| = \Theta\!\left(\overline{T}_n^{q+1}\right)
= o\!\left(n^{\frac{q+1}{2q+3}}\right)
= o(n^{1/2})
= o(n),
\]
implying the diffusion explores a vanishing fraction of the network—an ``intermediate'' asymptotic regime. The lower bound \(\underline{T}_n=\omega(1)\) rules out trivially short horizons while permitting arbitrarily slow growth.

\begin{remark}[Connection to the S-Curve]\label{rem:s-curve}
The intermediate regime corresponds to the steep, accelerating portion of the S-shaped epidemic curve, where diffusion grows as $\Theta(t^{q+1})$. Jump links supercharge expansion, creating a wedge between true and predicted S-curves that widens as $T$ grows. Once saturation dominates, forecasts reconverge. Measurement error is most damaging during the policy-critical window when interventions have the greatest potential impact.\footnote{In the simulations and empirical examples we consider in Section \ref{sec:empirical-applications} and Appendix \ref{sec:sims}, the diffusion processes can be well-approximated by standard SIR curves. One can fit SIR curves to the data via moment conditions. In sample, standard SIR difference equation models fit the data well; however, forward projections using the estimated SIR model have error due to mismatch between the exponential SIR model and the underlying polynomial diffusion process.}
\end{remark}

We now introduce notation for the expected spread of diffusion on the base network \(L_n\). Let \(\mathcal{E}_t\) denote the expected number of nodes activated by time \(t\) under percolation \(P_n\):
\[
\mathcal{E}_t
:=
\mathbb{E}_{P_n}\Bigl[\bigl|\{\, j \in V(L_n): j \text{ is activated by time } t \,\}\bigr|\Bigr].
\]
We define \(\mathcal{S}_t := \mathcal{E}_t - \mathcal{E}_{t-1}\) as the expected number of new activations at integer time \(t\ge1\).

\begin{lemma}[Polynomial Diffusion Growth]\label{lem:diffusion}
Suppose Assumptions~\ref{ass:forecast-time} and \ref{ass:disease} hold. Then, uniformly over \(t\in[\underline{T}_n,\overline{T}_n]\),
\[
\mathcal{E}_t = \Theta\bigl(t^{q+1}\bigr).
\]
\end{lemma}

All proofs appear in Appendix~\ref{sec:proofs}. Supercritical percolation fills balls proportional to their volume: cumulative activations grow as $\Theta(t^{q+1})$, with the diffusion frontier expanding into the boundary of the growing ball. Uniform spread ensures diffusion does not stall in bottlenecks. This polynomial baseline is precisely what missing links disrupt: because diffusion on $L_n$ grows \emph{only} polynomially, even a tiny number of jump links produces a proportionally enormous effect.

We provide two examples illustrating the scope of  Assumptions~\ref{ass:forecast-time} and \ref{ass:disease}, and how they can fail to hold.

\begin{exmp}[Line Graph]
The line graph (and ring graphs, as in the  base \citealt{watts1998collective} case) violates both parts of Assumption~\ref{ass:disease}. Starting from seed $i_0$, volumes grow linearly, $|B_{i_0}(t)| = \Theta(t)$, which corresponds to $q=0$, which is ruled out by assumption. However, consider the set of nodes at distance $(1-2\alpha)T$ to $T$ from $i_0$. Since there exists a unique path from $i_0$ to this set, the probability the diffusion reaches it is at most $p_n^{(1-2\alpha)t}$. If $p_n < 1$, this probability vanishes as $t \to \infty$, violating the uniform spread condition. This is consistent with \citealt{watts1998collective}, who show that increasing ``randomness'' is necessary to achieve non-trivial global infection.
\end{exmp}

\begin{exmp}[Complete $k$-ary Tree]
Consider a complete $k$-ary tree with $k>1$ and seed $i_0$ at the root. This graph violates Assumption~\ref{ass:disease}.1. The volume of a radius-$t$ ball expands as
\[
|B_{i_0}(t)| = \sum_{s=0}^t k^s = \frac{k^{t+1} - 1}{k-1} = \Theta(k^t),
\]
exhibiting exponential rather than polynomial growth in $t$.
\end{exmp}

Alternatively, latent-space models---where nodes link based on proximity in an underlying space \citep{hoffrh2002}---satisfy our assumptions and encompass geographic contact networks, spatial interaction models, and latent-position models. We verify Assumption~\ref{ass:disease} for the canonical case: the random geometric graph (RGG), which transparently connects latent-space geometry to expansion properties. The RGG is the most favorable case for our assumptions; if measurement error causes problems here, it will likely be worse in more complex networks.

The result extends to boundedly inhomogeneous Poisson intensities (Remark~\ref{rem:rgg-inhomogeneous} in Appendix~\ref{sec:rgg-proof}): polynomial volume growth and uniform spread hold as long as no region is completely empty or infinitely dense. Assumption~\ref{ass:disease} does \emph{not} cover scale-free networks or networks with exponential volume growth. We consider graphs with exponential volume graph in Appendix \ref{sec:extensions}. Moreover, the data-collection procedures that produce polynomially-expanding measured networks---geographic proximity proxies, interaction thresholds, survey boundaries---are precisely those most likely to create unaligned measurement error. 

\begin{exmp}[Random Geometric Graph]\label{ex:rgg}
Fix dimension $d \geq 2$ and let $G_n$ be a random geometric graph on $n$ points from a homogeneous Poisson process of intensity $\lambda$ on a $d$-dimensional torus of volume $n/\lambda$, with connection radius $r > 0$ in the supercritical regime of continuum percolation. Taking $L_n$ to be the giant component $\mathcal{C}_n$ and $q = d - 1$ so that $t^{q+1} = t^d$, both parts of Assumption~\ref{ass:disease} hold for polylogarithmic horizons and at least a $(1-\varepsilon)$ fraction of vertices for any $\varepsilon$. The proof, a block-renormalization argument using the Liggett--Schonmann--Stacey domination theorem, appears in Appendix~\ref{sec:rgg-proof}.
\end{exmp}

\begin{remark}[Typical vs.\ uniform seeds]\label{rem:typical-seeds}
Assumption~\ref{ass:disease} requires polynomial volume growth and uniform spread for \emph{every} node and seed. Example~\ref{ex:rgg} verifies these properties for $(1-\varepsilon)$-typical vertices in the giant component; the remaining $\varepsilon$-fraction near irregular regions (e.g., the boundary of the giant component) may violate the uniform bounds. Accordingly, the random geometric graph example should be read as validating the theory for typical seeds rather than literally for every seed. All subsequent theorems hold verbatim when Assumption~\ref{ass:disease} is satisfied; for graph classes where the assumption holds only for typical vertices, the results apply with at most an $\varepsilon$-probability exception set that can be made arbitrarily small.
\end{remark}

Finally, we specify the distribution of missing links $E_n$. Our baseline analysis focuses on i.i.d.\ link formation with probability $\beta_n$; Appendix~\ref{sec:proofs} extends the results to settings where each node's linking capacity is bounded. The key requirement is that the support of $E_n$ provides sufficient coverage at medium-range distances relative to $L_n$—a condition trivially satisfied under i.i.d.\ formation. Despite this expansive \textit{support}, \textit{realized} networks $E_n$ remain sparse with high probability.\footnote{We thank an anonymous referee whose comments on an earlier draft motivated the treatment of bounded linking capacity in Appendix~\ref{sec:proofs}.}

\begin{ass}[Missing Link Distribution]\label{ass:beta}
For all $n$ and all pairs $i,j \in V_n$, $E_{ij} \stackrel{\mathrm{iid}}{\sim} \mathrm{Ber}(\beta_n)$ with
\[
\beta_n = \omega\left(\frac{1}{p_n \underline{T}_n^{q+1} n}\right),\quad  \beta_n = o\left(\frac{1}{n}\right)
\]
\end{ass}

Both bounds on $\beta_n$ vanish as $n\to\infty$. Combined with Assumptions~\ref{ass:disease} and~\ref{ass:forecast-time}, the lower bound ensures that $p_n \underline{T}_n^{q+1} n \beta_n \to \infty$: the expected number of successful transmissions through missing links originating from the initial diffusion ball diverges, guaranteeing that the diffusion escapes its local polynomial neighborhood with probability approaching one. The upper bound $\beta_n \ll n^{-1}$ ensures sparsity: the expected degree from $E_n$ is $o(1)$, implying $E_n$ is asymptotically disconnected and contains no giant component with high probability.

To calibrate, consider a geographic contact network with $q = 2$ (balls grow as $t^3$), $n = 10^6$ nodes, and passing probability $p_n = 0.4$. The key finite-sample diagnostic is the ratio $T_n^{2q+3}/n$, which must be small for the asymptotic approximations to hold.\footnote{Asymptotic bounds involve unspecified constants, so finite-sample calibrations are illustrative. We choose parameters where $T_n^{2q+3}/n$ is well below 1.} Taking $T_n = 4$: the ratio evaluates to $4^7/10^6 = 16{,}384/10^6 \approx 0.016 \ll 1$, comfortably within the valid regime. Assumption~\ref{ass:beta} then requires $\beta_n \gg 1/(0.4 \cdot 64 \cdot 10^6) \approx 3.9 \times 10^{-8}$ and $\beta_n \ll 10^{-6}$. Setting $\beta_n = 10^{-7}$, the expected number of missing links per node is $n\beta_n = 0.1$---roughly one missing link per ten nodes---yet our theorems show this suffices for catastrophic forecast failure. As $n$ grows the requirements become less stringent: at $n = 10^8$ the same assumptions permit $T_n$ up to $6$ (with $6^7/10^8 \approx 0.003$) and $\beta_n$ as small as $10^{-10}$.

The forecast errors we characterize arise not from dense unobserved networks or hidden giant components---settings where errors would be unsurprising---but from sparse, fragmented, idiosyncratic \emph{jump links}, each connecting two otherwise distant parts of $L_n$. Substantial forecast error emerges even when mismeasurement takes its mildest form. These links are dangerous because their placement ignores the geometry of $L_n$: they connect nodes far apart in graph distance as readily as nodes nearby, functioning as escape hatches that vault diffusion past the predicted frontier.

\begin{remark}[Relationship to Small-Worlds Models]\label{rem:ws}
The \cite{watts1998collective} model is a low-dimensional analogue of our framework: $L_n$ is a ring lattice and $E_n$ is generated by uniform rewiring. The ring lattice does not satisfy Assumption~\ref{ass:disease}---volume growth is linear ($q = 0$, excluded) and uniform spread fails (see Example~1)---so the \cite{watts1998collective} model is not a literal special case of our assumptions. Nonetheless, the qualitative mechanism is the same: random shortcuts dramatically compress path lengths and accelerate diffusion. Our framework generalizes this observation to the class of polynomial-expansion networks that \emph{do} satisfy Assumption~\ref{ass:disease}, and additionally characterizes when shortcuts are dangerous (geometrically unaligned with the expansion structure of $L_n$) and when they are not (when missing links are aligned with $L_n$, i.e., resemble a random thinning of the observed network)---a distinction absent from the small-worlds literature. ``Long-range'' here refers to graph distance on $L_n$, not necessarily physical distance---a missing link between geographically nearby nodes functions as a long-range shortcut if $L_n$ provides no short path between them.
\end{remark}

\paragraph*{Econometrician's Goals}
The econometrician has two objectives: predicting \textit{which} nodes are reached by diffusion and predicting \textit{how many}, both by time $T$.

Let $y_{jt}$ indicate whether node $j$ has been activated by time $t$ for a diffusion starting at seed $i_0$ under percolation $P_n(G_n)$; we suppress dependence on $G_n$, $P_n(G_n)$, and $i_0$ when clear from context. The set of activated nodes by period $T$ is
\[
I_{P_n(G_n)}(i_0,T) = \{j \in V_n : y_{jT} = 1\}.
\]
The econometrician observes $L_n$ but not $E_n$ or $P_n$, so we study the distribution over $P_n(G_n)$ induced by both sources of randomness. We further assume the econometrician knows $T$, $q$, and $L_n$ perfectly---heroic assumptions favorable to prediction, so our impossibility results characterize a best-case scenario.

\begin{remark}[Mismeasured Forecast Horizon]\label{rem:mismeasured-T}
In practice $T$ may itself be uncertain, with effects analogous to seed mismeasurement. Under polynomial expansion, a small error $\Delta T$ yields relative forecast error of order $\Delta T / T$---potentially large when $T$ is moderate. Moreover, a longer-than-expected $T$ amplifies the impact of jump links by allowing more time for cascading secondary wavefronts. Our impossibility results, which assume $T$ is known, therefore represent a lower bound on realistic forecast errors.
\end{remark}


\section{Sensitive Dependence on the Seed}\label{sec:sensitive-dependence}

Small errors in identifying the initial seed generate substantial prediction errors in both the spatial extent and composition of activated nodes.

Fix a percolation realization $P := P_n(G_n)$ and consider two seeds: the true seed $i_0$ and a nearby counterfactual seed $j_0$. Let $I_P(i_0,T)$ and $I_P(j_0,T)$ denote the sets of nodes activated by period $T$ from each seed under the same transmission realization $P$.

We measure forecast sensitivity using a modified Jaccard index \citep{jaccard1901etude}:
\[
\Delta_n(i_0,j_0) := \frac{|I_P(i_0,T) \cap I_P(j_0,T)|}{|I_P(i_0,T) \cup I_P(j_0,T)|}.
\]
Values bounded away from one indicate that diffusions from nearby seeds $i_0$ and $j_0$ reach substantially different sets of nodes, despite their proximity.

\begin{theorem}[Seed Sensitivity]\label{thm:sensitive-dep} 
Suppose Assumptions~\ref{ass:forecast-time} and \ref{ass:disease} hold. Fix a graph $L_n$. Then for every seed $i_0$, there exists a forecast horizon $T_n \in [\underline{T}_n, \overline{T}_n]$ and constants $C, c' \in (0,1)$ such that:
\begin{enumerate}
    \item There exists a nontrivial set of alternative seeds near $i_0$: let $U_{n,i_0} := B_{i_0}(a_n)$ for some $a_n = \Theta(T_n)$, and let $J_{n,i_0} \subseteq U_{n,i_0}$ satisfy $|J_{n,i_0}|/|U_{n,i_0}| \geq C$.
    \item For all $j_0 \in J_{n,i_0}$, with probability bounded away from zero over percolation realizations $P_n(L_n)$:
    \[
    \Delta_n(i_0,j_0) \leq c'.
    \]
\end{enumerate}
\end{theorem}

All proofs appear in Appendix~\ref{sec:proofs} unless otherwise noted.

\begin{figure}[ht]
 \centering
 \scalebox{0.4}{
\begin{tikzpicture}[font=\Large]
  \begin{scope}
    \fill (0,0) circle (2pt) node[below left] {$i_0$};
    
    \draw[dashed,red] (0,0) circle (3cm);
    \draw[dashed,red] (0,0) -- node[midway, above] {$a_n$} ++(150:3cm);
    
    \fill (3.18,3.18) circle (2pt) node[above right] {$k$};
    \draw[dotted] (0,0) -- node[midway, below right] {$>a_n$} (3.18,3.18);
    
    \draw[dashed,red] (3.18,3.18) circle (3cm);
    \draw[dashed,red] (3.18,3.18) -- node[midway, above right] {$a_n$} ++(80:3cm);
    
    \node at (2.25, -4) {(a)};
  \end{scope}
  
  \begin{scope}[xshift=12cm]
    \fill (0,0) circle (2pt) node[below left] {$i_0$};
    \draw[dashed,red] (0,0) circle (3cm);
    \draw[dashed,red] (0,0) -- node[midway, above] {$a_n$} ++(150:3cm);
    
    \fill (3.18,3.18) circle (2pt) node[above right] {$k$};
    \draw[dashed,red] (3.18,3.18) circle (3cm);
    \draw[dashed,red] (3.18,3.18) -- node[midway, above right] {$a_n$} ++(80:3cm);
    
    \begin{scope}
      \clip (0,0) circle (3cm);
      \fill[red!20] (3.18,3.18) circle (3cm);
    \end{scope}
    \node[black] at (1.55, 1.55) {$J_{i_0}$};
    
    \node at (2.25, -4) {(b)};
  \end{scope}
  
  \begin{scope}[yshift=-13cm]
    \fill (0,0) circle (2pt) node[below left] {$i_0$};
    \draw[dashed,red] (0,0) circle (3cm);
    
    \fill (3.18,3.18) circle (2pt) node[above right] {$k$};
    \draw[dashed,red] (3.18,3.18) circle (3cm);
    
    \begin{scope}
      \clip (0,0) circle (3cm);
      \fill[red!20] (3.18,3.18) circle (3cm);
    \end{scope}
    
    \fill (2.1, 1) circle (2pt) node[above right] {$j_0$};
    
    \draw (0,0) circle (3.75cm);
    \draw (0,0) -- node[midway, above left] {$T_n$} ++(180:3.75cm);
    
    \draw (2.1, 1) circle (3.75cm);
    \draw (2.1, 1) -- node[midway, above] {$T_n$} ++(0:3.75cm);
    
    \node at (2.25, -4) {(c)};
  \end{scope}
  
  \begin{scope}[xshift=12cm, yshift=-13cm]
    \begin{scope}
      \begin{scope}
        \clip (2.1, 1) circle (3.75cm);
        \fill[blue!20] (-5,-5) rectangle (10,10);
      \end{scope}
      \fill[white] (0,0) circle (3.75cm);
    \end{scope}
    
    \fill (0,0) circle (2pt) node[below left] {$i_0$};
    \draw[dashed,red] (0,0) circle (3cm);
    
    \fill (3.18,3.18) circle (2pt) node[above right] {$k$};
    \draw[dashed,red] (3.18,3.18) circle (3cm);
    
    \begin{scope}
      \clip (0,0) circle (3cm);
      \fill[red!20] (3.18,3.18) circle (3cm);
    \end{scope}
    
    \fill (2.1, 1) circle (2pt) node[above right] {$j_0$};
    
    \draw (0,0) circle (3.75cm);
    \draw (0,0) -- node[midway, above left] {$T_n$} ++(180:3.75cm);
    
    \draw (2.1, 1) circle (3.75cm);
    \draw (2.1, 1) -- node[midway, above right] {$T_n$} ++(0:3.75cm);
    
    \node at (2.25, -4) {(d)};
  \end{scope}
  
\end{tikzpicture}
}
\caption{Construction of the sensitivity argument. (a) Balls $B(i_0, a_n)$ and $B(k, a_n)$ for some node $k$ at distance $>a_n$ from $i_0$. (b) The intersection $J_{i_0}$ (shaded red). (c) For $j_0 \in J_{i_0}$, draw balls of radius $T_n$ around $i_0$ and $j_0$. (d) The blue region is reachable from $j_0$ in $T_n$ steps but not from $i_0$.}
\label{fig:sensitive-dep}
\end{figure}

This sensitivity is intrinsic to polynomial-expansion networks---it does not require shortcuts or missing links. The intuition is spatial: on a polynomially expanding network, seeds $i_0$ and $j_0$ each command a ``cone'' of influence that expands outward. Because $j_0$ is displaced from $i_0$, part of $j_0$'s cone extends into regions that $i_0$'s diffusion cannot reach within $T$ steps (see Figure~\ref{fig:sensitive-dep}, panel~(d)). The displaced seed does not merely add noise to the activation set---it redirects a substantial portion of the diffusion into an entirely different part of the network. This divergence occurs for a \emph{constant fraction} of alternative seeds within a local neighborhood---not just for adversarially chosen ones.

Network mismeasurement amplifies this sensitivity. Without missing links, the two diverging wavefronts from $i_0$ and $j_0$ remain within their respective local neighborhoods on $L_n$---they diverge, but only locally. Jump links change this picture. As each wavefront expands, it encounters unobserved connections to other parts of the network. When a jump link fires, it seeds a new wavefront in a remote region, outside the econometrician's predicted set. Because the two original wavefronts already diverged locally (Theorem~\ref{thm:sensitive-dep}), their respective jump links reach \emph{different} remote regions, amplifying a local discrepancy into a global one. To formalize this notion of disjoint regions, we divide the graph into disjoint \textit{tiles} (the construction of which is detailed in Lemma \ref{lem:tiles}). Collectively, the tiles cover a constant fraction of the graph without overlap. 

\begin{cor}[Global Seed Sensitivity]\label{cor:sensitive-iid}
Suppose Assumptions~\ref{ass:forecast-time}, \ref{ass:disease},  and \ref{ass:beta} hold. Consider a local perturbation from seed $i_0$ to alternative seed $j_0$ as in Theorem~\ref{thm:sensitive-dep}. Then with strictly positive probability, the following events occur jointly:
\begin{enumerate}
    \item \textbf{Shortcut activation.} Both diffusions utilize missing links in $E_n$: there exist nodes $e_1, e_2 \in V_n$ such that the diffusion from $i_0$ reaches $e_1$ via $L_n$ and then transmits through a missing link to some node outside $B_{i_0}(T_n)$, and similarly the diffusion from $j_0$ reaches $e_2$ and transmits outside $B_{j_0}(T_n)$.
    
    \item \textbf{Macroscopic divergence.} Let $s := \max\{d_{L_n}(i_0, e_1), d_{L_n}(j_0, e_2)\}$ denote the maximum graph distance to the shortcut nodes. Conditional on the shortcut activations in part~(1), the expected cumulative number of disjoint tiles reached by the two diffusions over the remaining $T_n - s - 1$ steps is at least
    \[
    2C\, n\beta_n p_n (T_n - s - 1)^{q+1},
    \]
    where $C > 0$ is the constant from Lemma~\ref{lem:regions-corr}.
\end{enumerate}
\end{cor}

This is stated under the i.i.d.\ case (Assumption~\ref{ass:beta}); the appendix establishes the same result under the more general bounded-capacity assumption (Corollary~\ref{cor:sensitive}).

The interaction between seed sensitivity and jump links is multiplicative. Since seeds $i_0$ and $j_0$ influence disjoint local regions (Theorem~\ref{thm:sensitive-dep}), they encounter different escape hatches into $E_n$, jump to different distant parts of the network, and then diverge further as each secondary wavefront generates its own independent jumps. A local seed perturbation cascades into forecast errors spanning the entire network.

\section{Sensitivity to Network Mismeasurement}\label{sec:forecasting-errors}

Forecasts based on the observed network $L_n$ systematically underestimate diffusion on the true network $G_n$, even when the measurement error is vanishingly small.

Consider an econometrician who observes $i_0$ and $L_n$ perfectly but assumes $E_n \equiv \emptyset$. The forecast for expected activations is
\[
\hat{Y}_T(L_n) := \mathbb{E}_{P_n(L_n)}\left[\sum_{j=1}^n y_{jT} \,\bigg\vert\, L_n, i_0\right],
\]
where the expectation integrates over diffusion realizations on the base network $L_n$ alone.

This estimator reflects standard practice: survey constraints truncate reported connections, mobility data impose distance or frequency thresholds, and social media studies omit offline interactions. Each effectively assumes $E_n \equiv \emptyset$, making $\hat{Y}_T(L_n)$ the natural benchmark.

Accounting for $E_n$ is infeasible even when researchers suspect its presence. Assumption~\ref{ass:beta} requires $\beta_n = o(n^{-1})$, so the expected number of missing links is $o(n)$---as we show in Proposition \ref{prop:sampling_beta}, there is so little error that it is hard to estimate. The misspecified forecast $\hat{Y}_T(L_n)$ is therefore not merely a stylized benchmark but the practical reality.\footnote{Note that computing the probability of activation for any given node is NP-complete \citep{shapiro2012finding}.}

As a benchmark, consider an oracle econometrician who knows the distribution of $E_n$ and computes
\[
\hat{Y}_T(G_n) := \mathbb{E}_{E_n, P_n(G_n)}\left[\sum_{j=1}^n y_{jT} \,\bigg\vert\, L_n, i_0\right],
\]
integrating over both missing link realizations and diffusion outcomes. Comparing $\hat{Y}_T(L_n)$ against this (computationally intractable) oracle isolates the effect of ignoring sparse measurement error. Despite knowing $L_n$, $i_0$, $T$, and $q$ perfectly, the misspecified forecast exhibits catastrophic error:

\begin{theorem}[Asymptotic Forecast Failure]\label{thm:main-polynomial}
Suppose Assumptions~\ref{ass:forecast-time}, \ref{ass:disease}, and \ref{ass:beta} hold. Then as $n\to\infty$,
\[
\frac{\hat{Y}_T(L_n)}{\hat{Y}_T(G_n)} \to 0.
\]
\end{theorem}

On $L_n$, diffusion expands polynomially, never escaping its local neighborhood. Each missing link in $E_n$ is an escape hatch that sends diffusion to a new region, where it spreads polynomially. As the wavefront grows, it encounters more escape hatches, accelerating the cascade. Even though any given node has vanishing probability of generating an escape, the cumulative effect overwhelms the locally-predicted spread. The small-worlds phenomenon of \cite{watts1998collective}---demonstrated numerically on ring lattices---is one instance of this cascade mechanism. Our framework identifies the general principle: forecast failure occurs whenever missing links are geometrically unaligned with the observed network's expansion structure.

The wavefront grows by $\Theta(t^{q+1})$ nodes at each step, each independently probing for jump links. A jump link firing at time $s$ seeds a new wavefront contributing $\Theta((T-s)^{q+1})$ additional activations. The econometrician's forecast captures only $\Theta(T^{q+1})$ from the original source, while the oracle aggregates the original plus all secondary outbreaks. The ratio diverges because cumulative satellite contributions grow faster than the single-source prediction. The proof formalizes this cascade by partitioning $L_n$ into regions and tracking how jump links seed separate wavefronts (Appendix~\ref{sec:proofs}).

\section{When Does Mismeasurement Matter?}\label{sec:mar}

Not all mismeasurement is so problematic. The geometry of $E_n$---whether missing links preserve or disrupt the expansion structure of $L_n$---is critical in determining the impact of missing links. When measurement error is \emph{aligned}---missing links resemble a random thinning of the observed network---forecast failure vanishes and seed sensitivity is not amplified. 


\begin{ass}[Aligned Edge Censoring]\label{ass:mar}
For each $n$, let the \emph{true} network be $G_n$, where $G_n$ satisfies Assumption~\ref{ass:disease} and the forecast horizon $T_n$ satisfies Assumption~\ref{ass:forecast-time}.

The econometrician observes a thinned network $L_n$ obtained by independent edge censoring: for each undirected edge $e\in G_n$, include $e$ in $L_n$ with probability $1-\varepsilon_n$ and delete it with probability $\varepsilon_n$, independently across edges. Define the unobserved edges as $E_n := G_n \setminus L_n$, so the true diffusion network is $G_n = L_n \cup E_n$ augmented by the missing edges $E_n$.

Assume the censoring rate satisfies
\[
\varepsilon_n T_n^{q+1} \;\longrightarrow\; 0 \quad\text{as } n\to\infty.
\]
\end{ass}

Assumption \ref{ass:mar} combines two pieces. The first component assumes correct model specification -- the researcher only misses links within the context of the correct network geometry. While there may be some long range links missed, they are long range links within the correctly specified structure. The second component ensures that there cannot be too much censoring -- this ensures we do not have dramatic changes in the geometry due to censoring. 

\begin{theorem}[No Asymptotic Forecast Failure under Aligned Error]\label{thm:no-failure-mar} 
Suppose Assumptions~\ref{ass:forecast-time}, \ref{ass:disease}, and \ref{ass:mar} hold. Then for any fixed seed $i_0$,
\[
\mathbb{P}\bigl( I(i_0,T_n,G_n) = I(i_0,T_n, L_n)\mid P_n \bigr)
\;\longrightarrow\; 1.
\]
In particular, the econometrician's forecast based on $L_n$ exhibits no asymptotic forecast failure:
\[
\frac{\bigl| I(i_0,T_n, G_n) \triangle I(i_0,T_n, L_n) \bigr|}
     {\bigl| I(i_0,T_n, G_n) \bigr|}
\;\xrightarrow{\mathbb P}\; 0,
\quad
\frac{\mathbb{E}\bigl[|I(i_0,T_n, L_n)|\bigr]}
     {\mathbb{E}\bigl[|I(i_0,T_n, G_n)|\bigr]}
\;\longrightarrow\; 1.
\]
Where we consider the same percolation $P_n$ for the symmetric difference. The expectation ratio is a deterministic consequence.
\end{theorem}

The contrast with Theorem~\ref{thm:main-polynomial} is striking: forecast failure depends on the \emph{geometry} of missing links, not their quantity. Under aligned error, missing links are not escape hatches---they are redundant copies of existing paths. The true diffusion uses $O(T_n^{q+1})$ edges, each independently missing with probability $\varepsilon_n$, so the expected number of ``missing but needed'' edges is $O(\varepsilon_n T_n^{q+1}) \to 0$. With high probability, the two diffusions produce identical activation sets. Under Assumption~\ref{ass:beta}, by contrast, missing links bridge distant regions. The practical question is whether a data-collection process disproportionately misses links resembling observed links (aligned---safe) or links bridging different network regions (unaligned---problematic). Since activation sets coincide with high probability, any functional is preserved. In particular, aligned error cannot amplify seed sensitivity beyond what is intrinsic to the true network.

\begin{remark}[Scope of the aligned/unaligned contrast]\label{rem:scope-contrast}
Theorems~\ref{thm:main-polynomial} and~\ref{thm:no-failure-mar} provide sufficient conditions for forecast failure and robustness, respectively, under two distinct error-generating processes. They are not exhaustive: intermediate cases---such as missing links that are partially correlated with $L_n$, or error processes that combine aligned and unaligned components---are not covered by either theorem and remain an open question. The key qualitative insight is that the geometric relationship between missing and observed links, rather than the volume of missing links alone, governs forecast reliability.
\end{remark}

The condition $\varepsilon_n T_n^{q+1} \to 0$ deserves interpretation. For $q = 2$ and $T_n = 10$, this requires $\varepsilon_n \ll 1/1000$. This is sufficient but not necessary: the proofs use it to ensure the diffusion path on $G_n^\star$ lies entirely in $L_n$ with high probability. In many survey settings, random nonresponse rates satisfy the condition---a $5\%$ link censoring rate works when $T_n^{q+1} < 20$. The condition becomes restrictive only when the forecast horizon is long relative to the censoring rate.

\section{Estimation and Potential Solutions}

Forecast failure occurs under unaligned error (Theorem~\ref{thm:main-polynomial}) but not under aligned error (Theorem~\ref{thm:no-failure-mar}). Can standard econometric approaches help in the unaligned case? We consider three strategies---estimating aggregate parameters ($p_n$, $\calR_0$), estimating the missing link probability $\beta_n$, and widespread testing---and show that each is insufficient.

\subsection{Consistent Estimation of Aggregate Parameters}\label{sec:estimation}

Despite the forecast failures above, certain aggregate parameters remain consistently estimable---though this provides no protection against the spatial and volumetric errors documented in Theorems~\ref{thm:sensitive-dep} and~\ref{thm:main-polynomial}.

Assume the econometrician observes all activations perfectly.\footnote{We do not characterize the optimal estimator, as computing maximum likelihood estimates for network diffusion is NP-complete \citep{shapiro2012finding}. Instead, we demonstrate consistency using a simple moment-based approach.} Given $L_n$ and the activation history $\{y_{j,s}\}_{j \in V_n, s \leq t-1}$, a consistent estimator $\hat{p}_n$ can be constructed. Consistency of $\hat{p}_n$ immediately yields consistent estimation of $\calR_0$, the expected number of secondary activations from a single seed in a susceptible population.\footnote{In a fully mixed population, $\calR_0 = p_n \bar{d}$ where $\bar{d}$ is the average degree.} The natural estimator is $\hat{\calR}_0 = \hat{p}_n d_L$, where $d_L$ is the mean degree in $L_n$. The true reproduction number $\calR_0(G_n) = p_n(d_L + \beta_n n)$ exceeds this by the contribution of missing links.

\begin{remark}\label{rem:r0_consistency}
Suppose Assumptions~\ref{ass:forecast-time}, \ref{ass:disease}, and~\ref{ass:beta} hold, and that $\calR_0(L_n) = p_n d_L$ remains bounded as $n \to \infty$. We construct a consistent estimator $\hat{p}_n \to_p p_n$ below (Eq.~\ref{eq:pn_estimator}). Given $\hat{p}_n$ and the observed mean degree $d_L$, the natural estimator $\hat{\calR}_0 = \hat{p}_n d_L$ satisfies
\begin{equation*}
    \frac{\hat{\calR}_0}{\calR_0(G_n)} \to_p 1.
\end{equation*}
\end{remark}

\begin{proof}
Decompose the true reproduction number as
\begin{equation*}
    \calR_0(G_n) = p_n d_L + p_n \beta_n  n = p_n d_L \left(1 + \frac{\beta_n n}{d_L}\right).
\end{equation*}
By Assumption~\ref{ass:beta}, $\beta_n = o(1/n)$, which implies $\beta_n n = o(1)$ and hence $\beta_n n/d_L = o(1)$ under the maintained assumption that $d_L$ is bounded away from zero and infinity. Therefore,
\begin{equation*}
    \calR_0(G_n) = \calR_0(L_n) \cdot (1 + o(1)).
\end{equation*}
Since $\hat{p}_n \to_p p_n$ by assumption and $d_L$ is observed, we have $\hat{\calR}_0 \to_p \calR_0(L_n)$ by the continuous mapping theorem. The result follows by combining these convergences.
\end{proof}

$\calR_0$ governs whether diffusion spreads or dies out, but conditional on spread, it reveals nothing about \emph{where} or \emph{how far} diffusion reaches. The reason is simple: $\calR_0$ measures the average local transmission rate, which is virtually identical on $L_n$ and $G_n$ since missing links contribute $o(1)$ to average degree. The forecast errors arise from the \emph{spatial reach} of jump links, which redistribute diffusion without materially changing $\calR_0$.

A simple consistent estimator for $p_n$ illustrates the construction. Let $\calI(i,t) \subseteq V_n$ denote the neighbors of node $i$ \emph{in $L_n$} activated at time $t$. Consider
\begin{equation}\label{eq:pn_estimator}
    \hat{p}_n := \frac{\sum_{t=1}^T \sum_{i=1}^n y_{it} \mathbbm{1}\{y_{i,t-1} = 0, |\calI(i,t-1)| = 1\}}{\sum_{t=1}^T \sum_{i=1}^n \mathbbm{1}\{y_{i,t-1} = 0, |\calI(i,t-1)| = 1\}}.
\end{equation}
The estimator restricts attention to susceptible nodes with exactly one activated $L_n$-neighbor, which---absent hidden exposures---isolates independent Bernoulli trials with success probability $p_n$. However, the econometrician does not observe $E_n$: a node classified as having exactly one activated $L_n$-neighbor may have additional activated neighbors through unobserved missing links, creating hidden exposures that could bias the estimator. We now show this contamination is asymptotically negligible.

At any time $t \leq \overline{T}_n$, the number of activated nodes is at most $\calE_t = O(t^{q+1})$ by Lemma~\ref{lem:diffusion}. For a susceptible node $i$, the probability that $i$ has \emph{any} activated $E_n$-neighbor is bounded by the union bound:
\[
\PP(i \text{ has an activated } E_n\text{-neighbor at time } t) \;\leq\; \calE_t \cdot \beta_n \;=\; O\!\left(\frac{t^{q+1}}{n}\right) \;=\; o(1),
\]
where we used $\beta_n = o(1/n)$ from Assumption~\ref{ass:beta}. Let $N_t^{\text{clean}}$ denote the number of susceptible nodes at time $t$ with exactly one activated $L_n$-neighbor and \emph{no} activated $E_n$-neighbor (``clean'' observations), and let $N_t^{\text{contam}}$ denote those with exactly one activated $L_n$-neighbor \emph{and} at least one activated $E_n$-neighbor (``contaminated'' observations). By the bound above and Markov's inequality, $\EE[N_t^{\text{contam}}] / \EE[N_t^{\text{clean}}] = o(1)$: the contaminated fraction vanishes. Each clean observation is an independent $\mathrm{Ber}(p_n)$ trial, so $\hat{p}_n \to_p p_n$ by the law of large numbers applied to the clean subsample, which dominates the denominator. This estimator exploits knowledge of $L_n$ through $\calI(i,t)$ and is inefficient, discarding nodes with multiple activated neighbors.\footnote{The contamination bound $O(t^{q+1}/n) = o(1)$ holds uniformly over $t \leq \overline{T}_n$ since $\overline{T}_n^{q+1} = o(n^{(q+1)/(2q+3)}) = o(n^{1/2}) = o(n)$ by Assumption~\ref{ass:forecast-time}.}

\subsection{Potential Remedies and Their Limitations}\label{sec:possible-solutions}

A policymaker might try to (i) estimate $\beta_n$ via supplementary network surveys or (ii) detect activated regions through widespread testing. Each strategy fails.

\subsubsection{Estimating $\beta_n$ via Network Surveys}\label{sec:survey_beta}

Suppose the econometrician samples $m_n$ nodes uniformly at random and checks whether each of the $\binom{m_n}{2}$ pairs has a link in $G_n$. Since $\beta_n$ is vanishingly small, this is a best-case scenario (i.i.d.\ links are easiest to find). Two regimes emerge: with a slowly growing sample, no missing links are found; with a larger sample, some may be found but consistent estimation remains impossible.

\begin{prop}\label{prop:sampling_beta}
    Under Assumption \ref{ass:beta}, if:
    \begin{enumerate}
        \item  $m_n=o(\sqrt{n})$,
 $
        \PP\left(\text{{\rm No links amongst }}\binom{m_n}{2} \text{ {\rm found}}\right) \rightarrow 1.
  $
    \item $m_n=  O(1/\sqrt{\beta_n})$, 
    there exists $\epsilon >  0$ and $c\in(0,1)$ such that $\Pr(|\hat \beta_n / \beta_n - 1| <\epsilon ) < c$.
    \end{enumerate}
\end{prop}

For a population of $n = 10^6$: with $\beta_n = 1/(n(\log n)^2)$ (admissible under our assumptions with $T_n = \log n$, $q = 2$), Part~(1) implies that a survey of $m_n = o(\sqrt{n}) = o(1{,}000)$ nodes yields no missing links with probability approaching one. For larger surveys up to $m_n = O(1/\sqrt{\beta_n}) \approx O(13{,}816)$ nodes, Part~(2) shows that consistent estimation of $\beta_n$ remains impossible. With $\beta_n = \log(n)/(p_n n^{(2q+1)/(q+1)})$ and $q = 4$, Part~(2) implies that even surveying 68{,}000 individuals---6.8\% of the population---cannot produce a consistent estimator. Directly estimating $\beta_n$ through supplementary data collection is infeasible in most settings.

\subsubsection{Widespread Testing}\label{sec:testing}

Can widespread testing identify which regions contain activated nodes? Using the tiling construction from Lemma~\ref{lem:tiles}, we model regions as disjoint tiles and consider an idealized regime: instantaneous random tests across all $n$ nodes, each activated node detected independently with probability $\gamma_n$. This abstracts from resource constraints, compliance, and delays, providing an upper bound on feasible protocols.

\begin{theorem}\label{thm:regionalDetection}  
Suppose Assumptions~\ref{ass:forecast-time}, \ref{ass:disease}, and~\ref{ass:beta} hold. Consider a time period $T$ and testing regime with detection probability $\gamma_n \to 0$ satisfying $\gamma_n = O(T^{-(q+1)})$ with implied constant satisfying $\gamma_n a_{\max} T^{q+1} < 1$. 
Let $K_T^\star$ denote the expected number of tiles containing at least one activated node at time $T$, and let $\hat{K}_T$ denote the expected number of tiles in which at least one activation is detected under the testing regime. If each activated individual is observed independently with probability $\gamma_n$, then as $n \to \infty$,
\begin{equation*}
    \frac{\hat{K}_T}{K_T^\star} \leq \gamma_n \, a_{\max} \, T^{q+1} < 1.
\end{equation*}
\end{theorem}

The reason is tied to the escape-hatch mechanism: jump links seed distant regions that initially contain few activated nodes. Detection requires observing at least one activation per region, but recently-seeded regions have had little time for local spread. Since jump links continue seeding new regions throughout the forecast horizon, a substantial fraction are always in this ``recently seeded, hard to detect'' state.

Since universal, instantaneous testing is already an upper bound on feasible protocols, and real-world testing faces resource constraints, incomplete compliance, and reporting delays, spatial detection provides limited value for medium-run forecasting. Testing is most needed precisely in the unaligned setting where shortcuts create unexpected distant activations; under aligned error, forecasts based on $L_n$ are already correct. 

\section{Empirical Applications}\label{sec:empirical-applications}

The following applications are designed to illustrate the \emph{mechanisms} identified by our theory---sensitivity to error geometry, the relative harmlessness of aligned error, and the fragility of structured networks---in empirically relevant settings. They do not constitute formal tests of the theorems, which require asymptotic regimes ($n \to \infty$, intermediate horizons) not available in finite data. We note below where specific design choices depart from the formal assumptions.

Appendix~\ref{sec:sims} confirms the finite-sample relevance of our asymptotic results through Monte Carlo simulations on synthetic lattice networks: with $n = 4{,}000$ and $q = 2$, the policymaker underestimates diffusion by 83\% at the worst point, and nearby seeds produce nearly disjoint activation sets ($\calJ = 0.055$ for $q = 4$). 

Here we turn to real-world data. Three applications illustrate these results, moving from controlled simulation to policy failure: a cell-phone mobility exercise showing that \emph{how} links are missed matters more than \emph{how many}; New York City's micro-cluster zoning, where geography-based policy used the wrong network; and the optimal targeting methodology of \cite{beaman2021can}, where measurement error is most damaging precisely where targeting is most valuable.

\subsection{Network Truncation on Real Mobility Topology}\label{sec:canv-application}
First, we consider the spread of COVID-19 in the American West and Southwest. Using SafeGraph cell-phone mobility flows \citep{kang2020multiscale} across California, Nevada, and part of Arizona (${\sim}11{,}000$ Census tracts), we construct $L_n$ by pruning flows below the 92nd percentile of tract to tract travel. We simulate SIR diffusion ($\calR_0 = 2.5$) on both and measure sensitive dependence and forecast underestimation (details in Appendix~\ref{sec:covid}). The first exercise compares \emph{two error structures} with the same number of missing links. The \emph{pruned} $E_n = G_n \setminus L_n$ consists of geographically structured moderate-flow connections between nearby tracts. We construct $G_n$ by pruning travel flows at the 91st percentile. For the \emph{i.i.d.} case, we generate $E_n$ as an Erd\H{o}s--R\'{e}nyi graph calibrated to match the edge count of the pruned $E_n$. 

\begin{figure}[ht]
\centering
\begin{subfigure}[t]{0.48\textwidth}
\centering
\includegraphics[width=\textwidth]{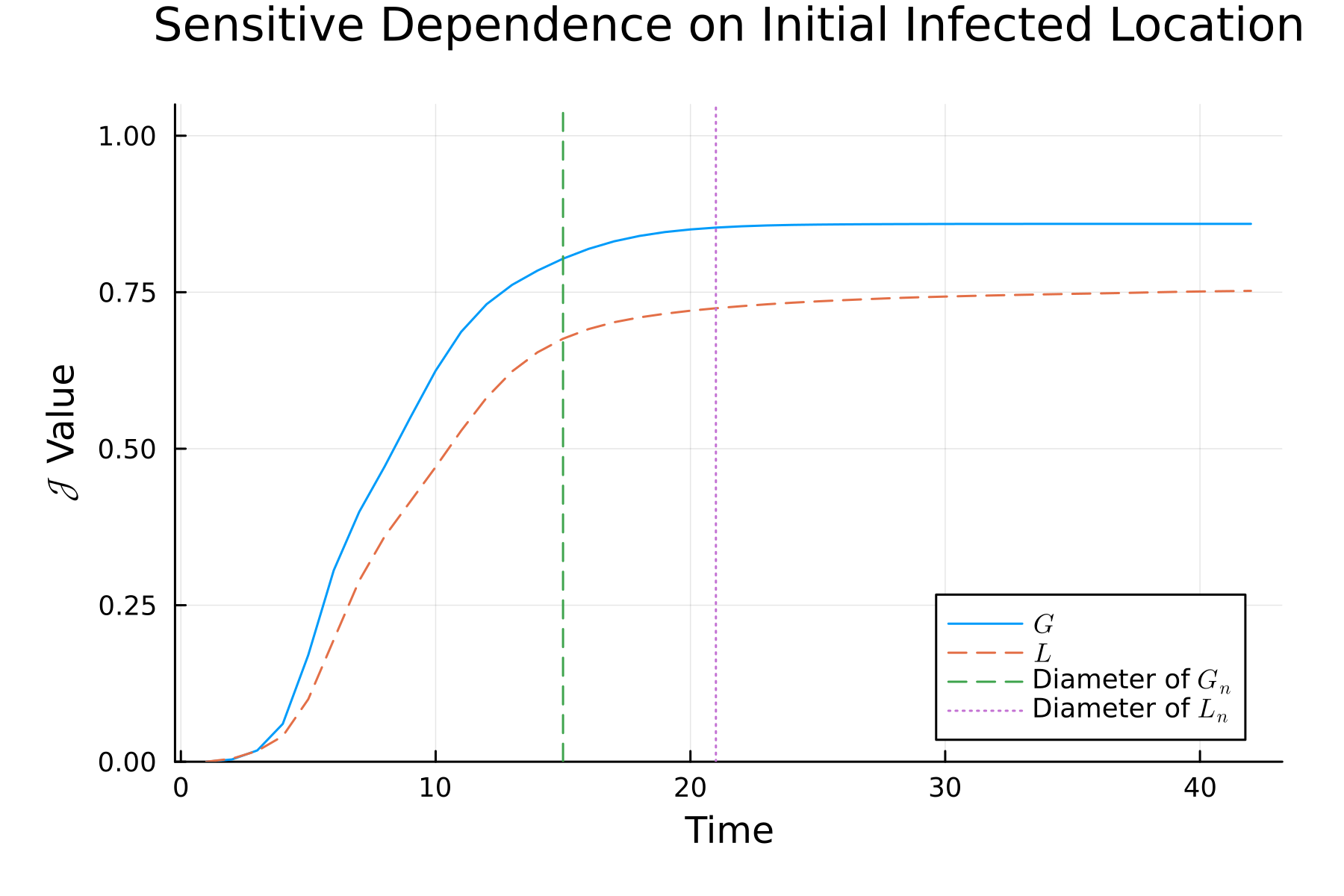}
\caption{Pruned (geographically correlated) error}
\label{fig:canv-sens-pruned}
\end{subfigure}\hfill
\begin{subfigure}[t]{0.48\textwidth}
\centering
\includegraphics[width=\textwidth]{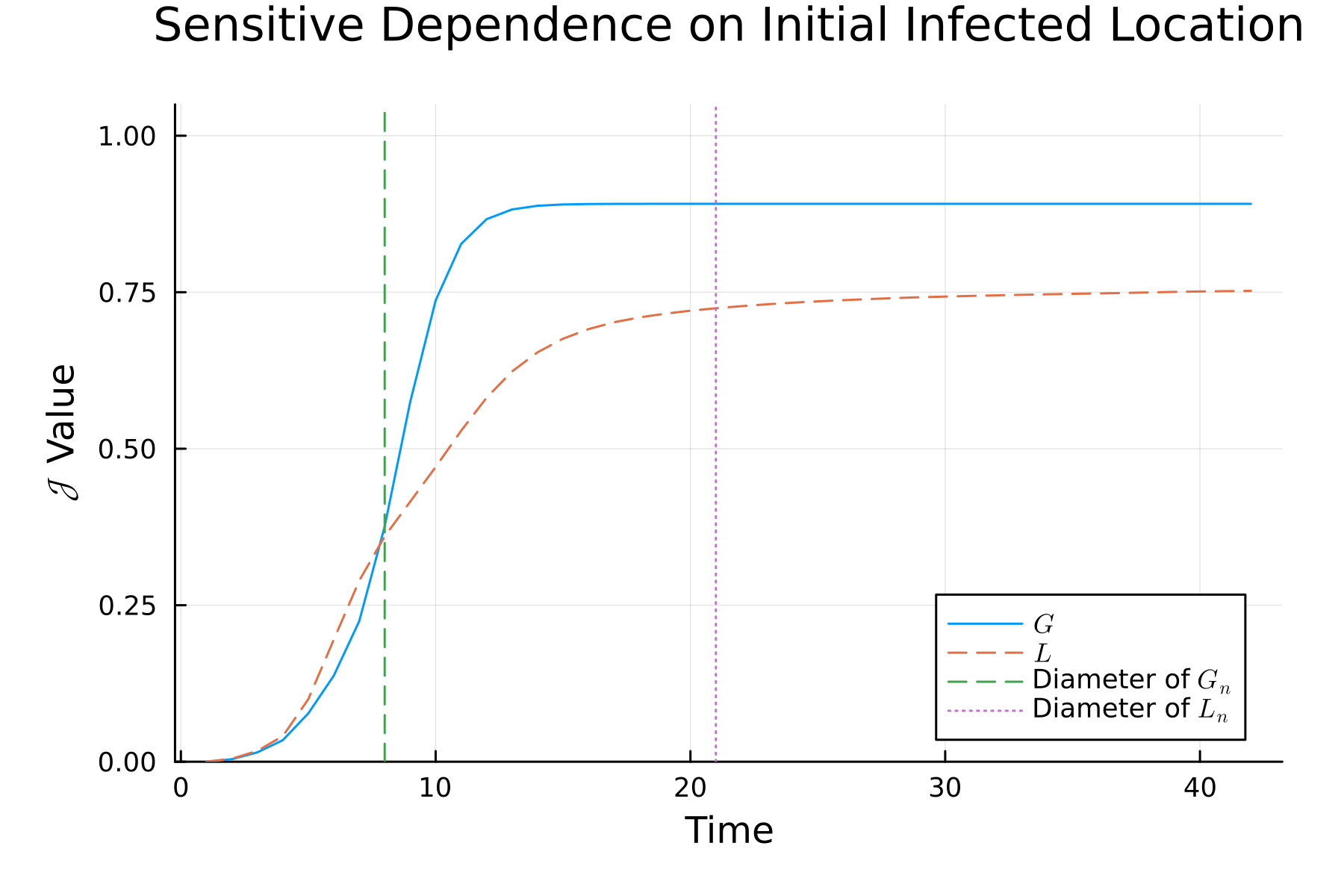}
\caption{I.I.D.\ (dispersed) error, same expected edge count}
\label{fig:canv-sens-iid}
\end{subfigure}
\caption{Figures show the average Jaccard index for the diffusion processes running from $i_0$ and $j_0$. Dashed lines mark the diameters of $G_n$ and $L_n$. I.I.D.\ errors (right panel) are far more damaging than pruned errors (left panel) despite producing the same number of missing links, because i.i.d.\ errors create long-range shortcuts in graph distance while pruning misses only local, redundant connections.}
\label{fig:canv-sens}
\end{figure}

We begin by analyzing sensitive dependence. We compare diffusion from $i_0$ and a given $j_0$. We select $j_0$ uniformly at random from a neighborhood of radius two around $i_0$ in $L_n$, which makes up 2.5\% of the graph. We then compute the expected Jaccard index of diffusion from each seed. Figure~\ref{fig:canv-sens} shows the sensitive dependence of the diffusion on the seed. In all cases, the average Jaccard index is well below 1. We consider the time  halfway to the diameter of $L_n$ as a benchmark: for $L_n$, the average Jaccard index is 0.47 at time step ten. For $G_n$ generated via pruning, the average Jaccard index is 0.62 -- the local links cause additional overlap (corresponding to how the local linking dampens the global spread in Corollary \ref{cor:sensitive}). For $G_n$ generated via i.i.d. links, the average Jaccard index is 0.73. However, unlike in the pruned case, the Jaccard index for $L_n$ is higher than for the i.i.d. $G_n$, up to the diameter of $G_n$ -- this is consistent with Corollary \ref{cor:sensitive-iid}, where missing links increase separation. 

\begin{figure}[ht]
\centering
\begin{subfigure}[t]{0.48\textwidth}
\centering
\includegraphics[width=\textwidth]{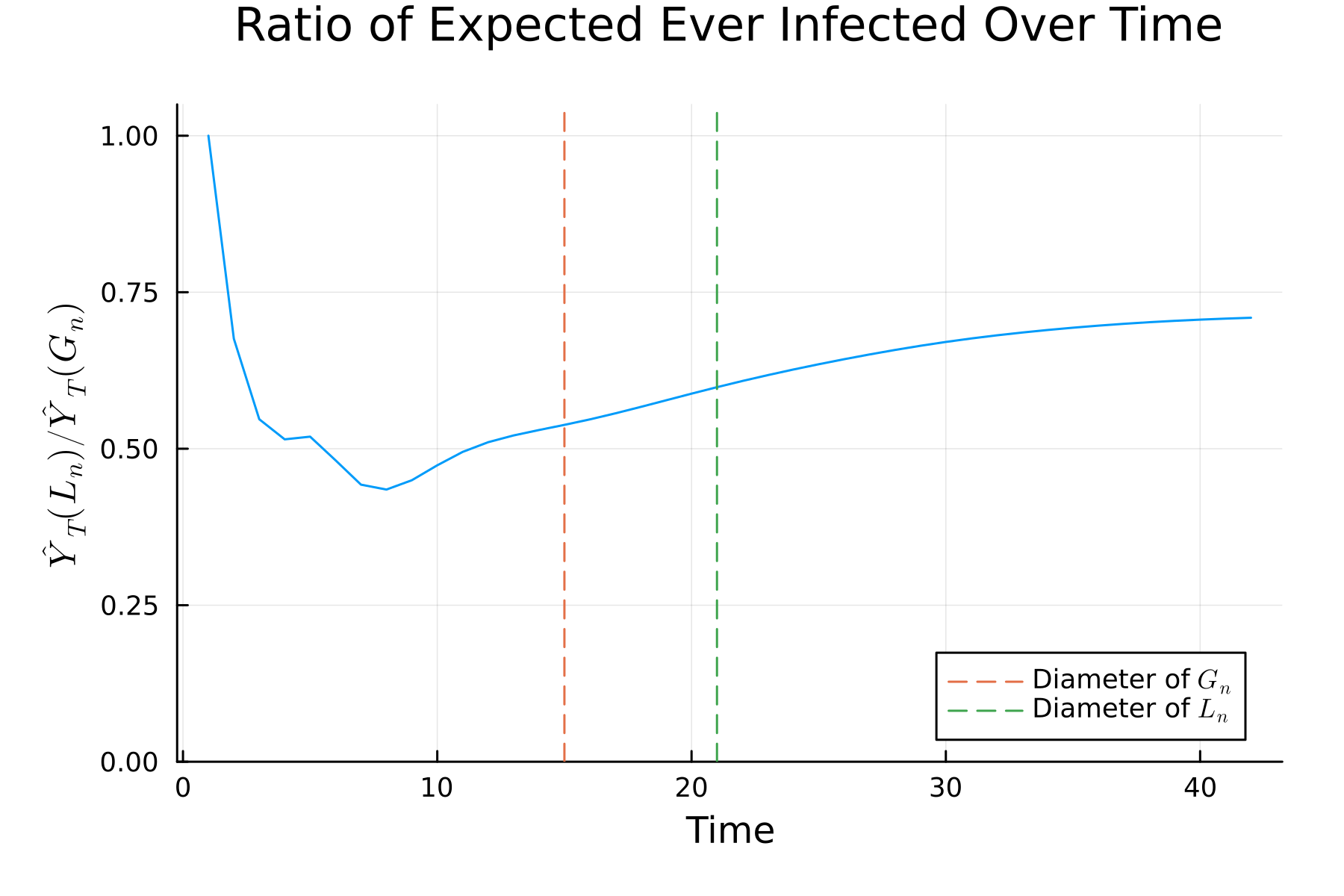}
\caption{Pruned (geographically correlated) error}
\label{fig:canv-ratio-pruned}
\end{subfigure}\hfill
\begin{subfigure}[t]{0.48\textwidth}
\centering
\includegraphics[width=\textwidth]{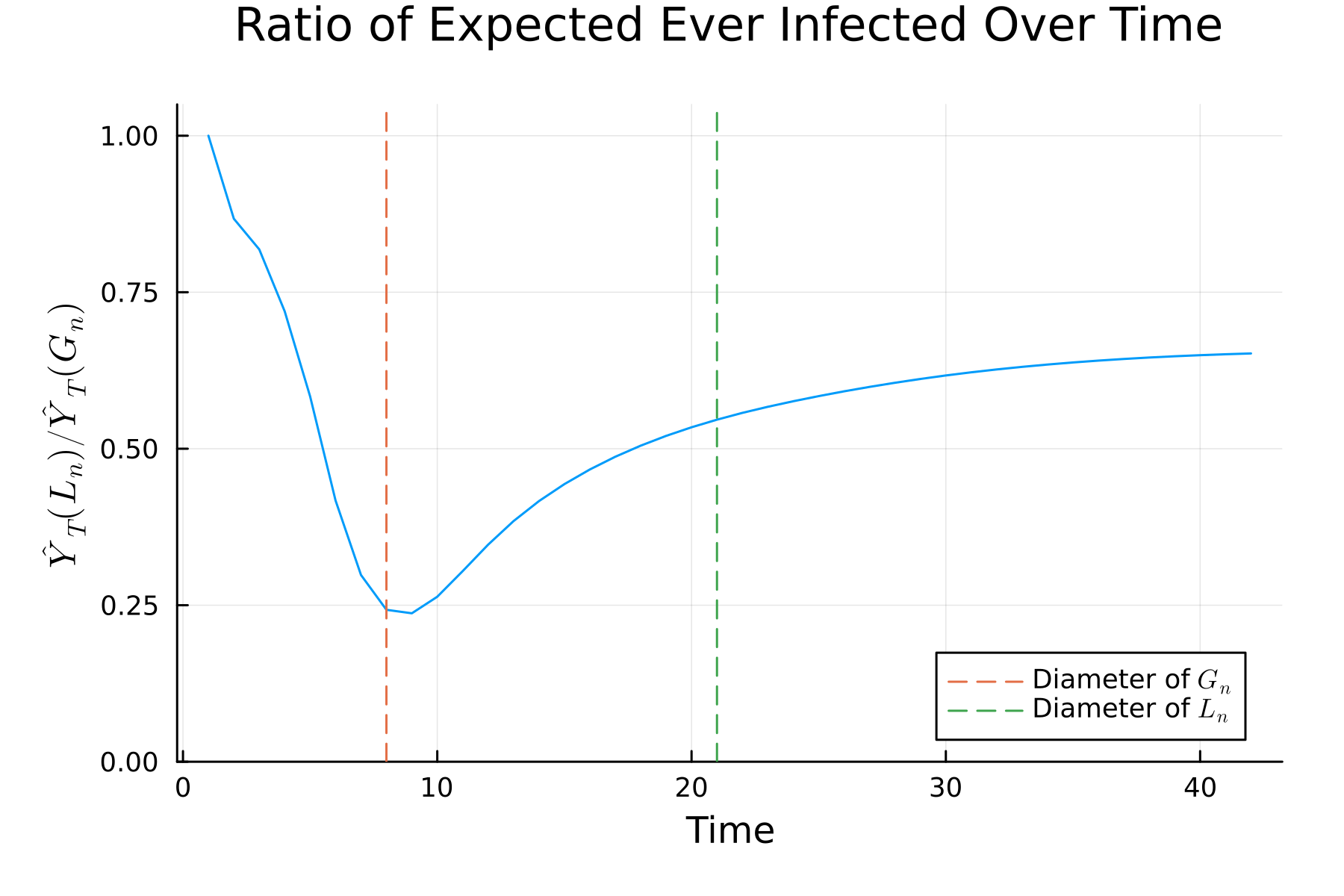}
\caption{I.I.D.\ (dispersed) error, same expected edge count}
\label{fig:canv-ratio-iid}
\end{subfigure}
\caption{Forecast underestimation on the CA/NV mobility network under two error structures. The ratio $\hat{Y}_T(L_n)/\hat{Y}_T(G_n)$ measures how much the econometrician underestimates cumulative infections. Dashed lines mark the diameters of $G_n$ and $L_n$. I.I.D.\ errors (right panel) are far more damaging than pruned errors (left panel) despite producing the same number of missing links, because i.i.d.\ errors create long-range shortcuts in graph distance while pruning misses only local, redundant connections.}
\label{fig:canv-comparison}
\end{figure}

Figure~\ref{fig:canv-comparison} shows $\hat{Y}_T(L_n)/\hat{Y}_T(G_n)$ under both error structures (Table~\ref{tab:graph_stats_sw_pruned}). Under pruning, the minimum ratio is ${\sim}0.45$ (55\% underestimate). Under i.i.d.\ errors with the same edge count, it drops to ${\sim}0.24$ (76\% underestimate). The i.i.d.\ errors create shortcuts between distant tracts; pruned errors add redundant connections between nearby ones. Both add the same volume of edges, but i.i.d.\ errors compress average path length to 4.03 vs.\ 5.87 for pruned (and 7.25 for $L_n$).

\begin{figure}[ht]
\centering
\begin{subfigure}[t]{0.48\textwidth}
\centering
\includegraphics[width=\textwidth]{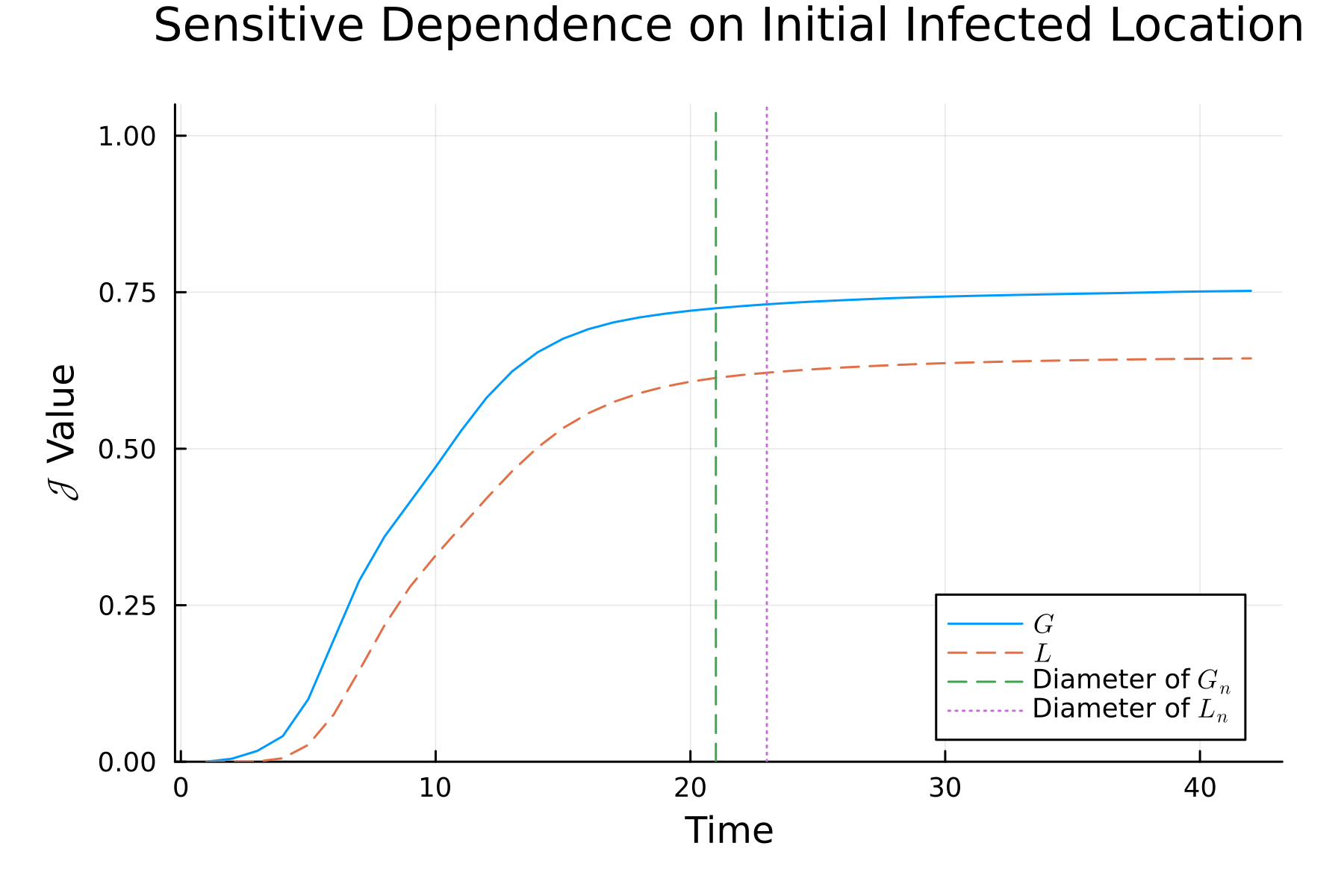}
\caption{Sensitive dependence under aligned error} 
\label{fig:mar-canv-sens}
\end{subfigure}\hfill
\begin{subfigure}[t]{0.48\textwidth}
\centering
\includegraphics[width=\textwidth]{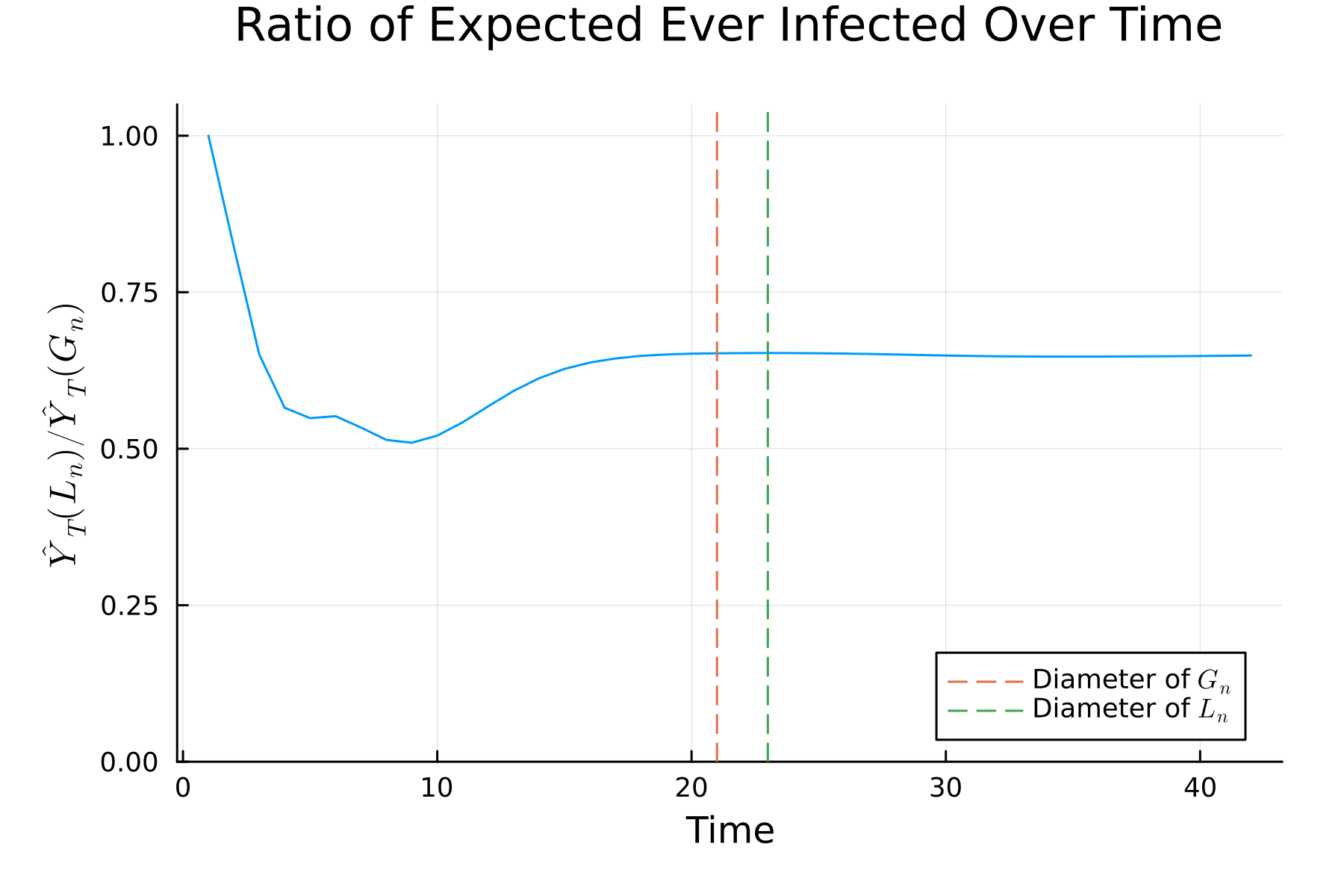}
\caption{Forecast error under aligned error} %
\label{fig:mar-canv-ratio}
\end{subfigure}
\caption{Figures show the average Jaccard index for diffusion starting at $i_0$ and a specific $j_0$ and diffusion ratio $\hat{Y}_T(L_n)/\hat{Y}_T(G_n)$ under aligned errors.} 
\label{fig:mar-canv}
\end{figure}

A natural complement is an \emph{aligned random-deletion} simulation: randomly deleting edges from $G_n$ to form $L_n$. Theorem~\ref{thm:no-failure-mar} predicts negligible forecast errors, completing the comparison: i.i.d.\ (catastrophic) $\gg$ pruned (substantial) $>$ aligned (less harm). The \emph{aligned} case constructs $L_n$ from $G_n$ dropping links i.i.d. at random---the missing links are a uniform thinning of the true network, preserving its geometric structure. We calibrate the probability of dropping links so that the implied $E_n$ (the gap between $L_n$ and $G_n$) has the same number of links as in the pruned and i.i.d. cases.\footnote{In a second case, we specifically set $\varepsilon_{n} = \frac{1}{\text{diam}(G_n)^3}$, to match Theorem \ref{thm:no-failure-mar}. Here, the gap between the Jaccard indices at time step 10 is 0.017, at 0.47 and 0.45, and the minimal ratio of total infections is $\sim0.85$, a much higher minimum.} Figure \ref{fig:mar-canv} shows the results. Despite the same volume of links being missed, aligned errors are less damaging than the pruned case. For sensitive dependence, the gap between $L_n$ and $G_n$ is smaller at the tenth step (0.47 vs 0.33, compared to 0.47 vs 0.62)\footnote{Note that $G_n$ in the aligned case is $L_n$ in pruned and i.i.d. cases.}. Furthermore,  the minimum ratio of total infections is $\sim 0.52$ ($48\%$ underestimate), a smaller underestimate. 

\subsection{NYC Micro-Cluster Initiative: When Policy Used the Wrong Network}\label{sec:nyc-application}
\begin{figure}[ht]
\centering
\includegraphics[width=\textwidth]{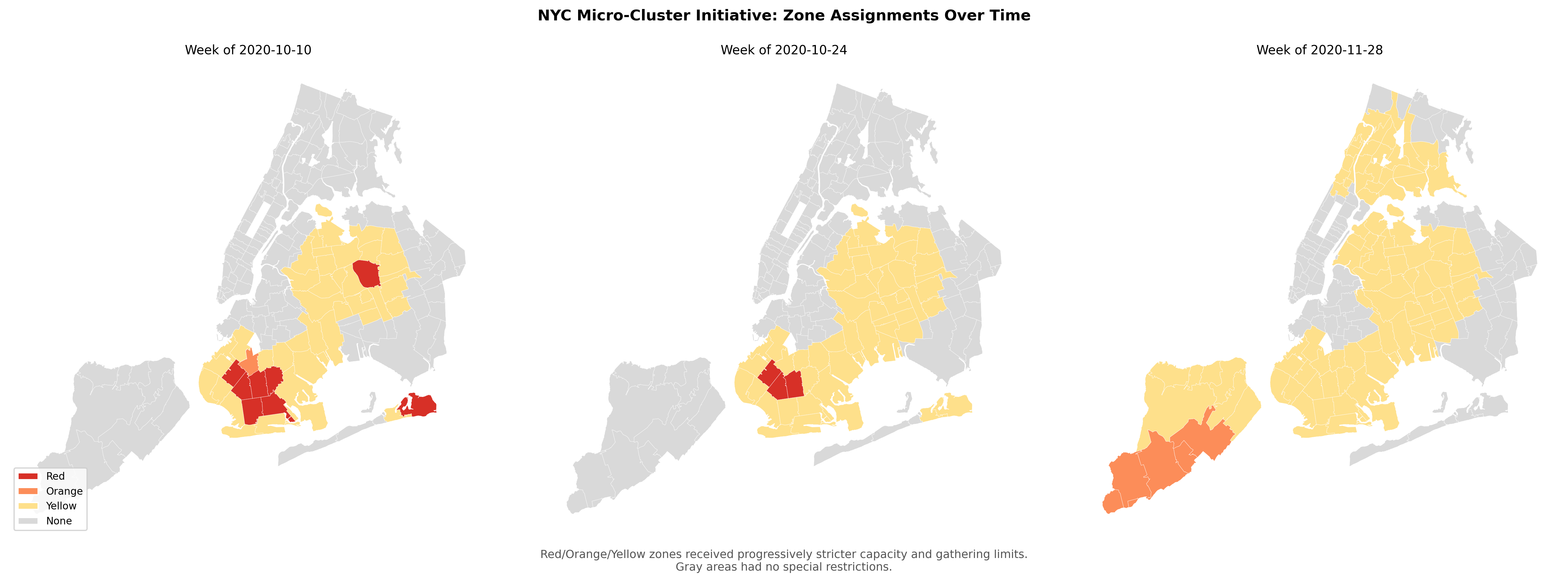}
\caption{NYC micro-cluster zone assignments over time. Each map shows MODZCTA-level tier designations for one week. Red/Orange/Yellow zones received progressively stricter capacity and gathering limits; gray areas had no special restrictions. The initial outbreak cluster in southern Brooklyn expands across the city by late November.}
\label{fig:nyc-policy-tiers}
\end{figure}

In October 2020, New York State implemented a ``micro-cluster'' strategy to contain COVID-19 outbreaks \citep{cuomo2020microcluster}, assigning each Modified ZIP Code Tabulation Area (MODZCTA) to a tier---Red, Orange, Yellow, or None---based on local case rates and those of \emph{geographically adjacent} areas. Red and Orange zones faced school and business closures; Yellow zones received moderate restrictions. The program covered 177 MODZCTAs over 24 weeks (October 2020--March 2021), producing 4{,}175 MODZCTA-weeks (Figure~\ref{fig:nyc-policy-tiers}; Appendix~\ref{sec:nyc-appendix} provides the full timeline).

The implicit network was geographic adjacency---a planar-like graph with polynomial expansion (${\sim}t^2$). Our theory predicts this could be misleading: if actual transmission follows a mobility network, the absent mobility links function as jump links, connecting areas far apart in adjacency-graph distance. 

We construct two alternative mobility networks from GeoDS cell-phone mobility data \citep{kang2020multiscale}, which records weekly population-scaled visitor flows between Census tracts, aggregated to the MODZCTA level.\footnote{The geographic adjacency network $L$ connects every pair of MODZCTAs that share a physical boundary (${\sim}420$ edges, ${\sim}4.7$ neighbors per node, row-normalized). The mobility flow structure is stable across weeks (average pairwise Spearman $\rho = 0.93$; see Appendix~\ref{sec:nyc-appendix}).} The \emph{degree-matched mobility} network $G_{\text{match}}$ retains, for each MODZCTA $i$, the top-$d_i$ mobility connections by flow weight, where $d_i$ is $i$'s geographic degree in $L$. This produces a network with the same per-node out-degree as $L$ but different edges (${\sim}70\%$ overlap), so that any difference in predictive power reflects \emph{which} connections matter rather than how many. The \emph{geo-superset mobility} network $G_{\text{super}}$ keeps every geographic neighbor and adds mobility connections above a per-node threshold, so $L \subset G_{\text{super}}$ by construction. This second comparison asks whether mobility information helps even when geography is fully retained.

For each network $X \in \{L, G_{\text{match}}, G_{\text{super}}\}$, we define exposure as the row-normalized weighted-average neighbor case rate:
\begin{equation}\label{eq:nyc-exposure}
\text{Exposure}_{X,i,w} = \sum_{j} X_{ij} \cdot \text{CaseRate}_{j,w},
\end{equation}
where $X_{ij}$ are the row-normalized adjacency weights and $\text{CaseRate}_{j,w}$ is the weekly case rate per 100{,}000 population. Exposures are $z$-scored within week.

\begin{figure}[ht]
\centering
\includegraphics[width=0.65\textwidth]{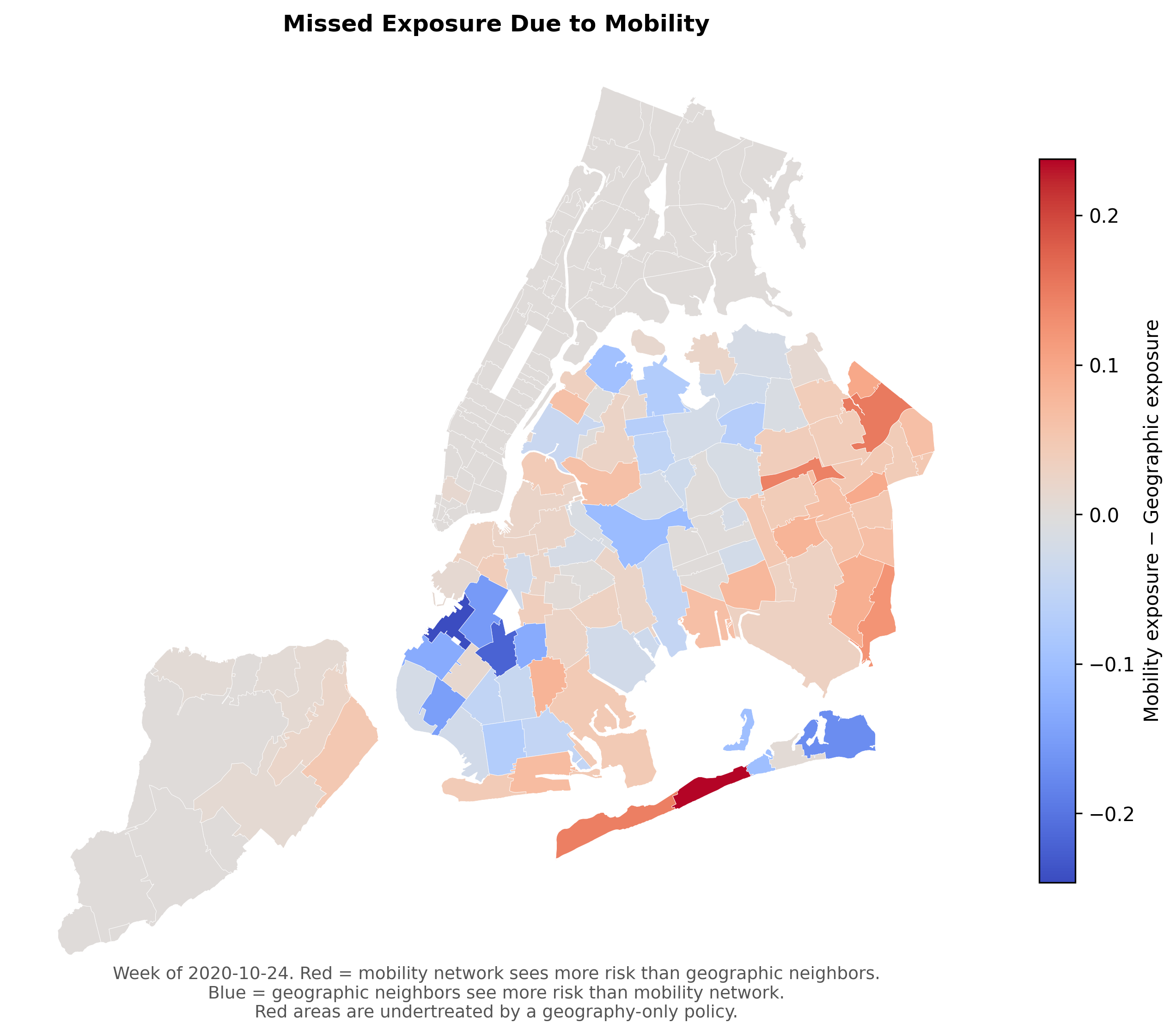}
\caption{Hidden threat: risk invisible to geography-only policy (week of 2020-10-24). Colors show the difference between mobility-based and geographic exposure for each MODZCTA. Warm colors indicate neighborhoods where mobility-based risk exceeds what geography alone suggests (undertreated by geography-only policy); cool colors indicate the reverse (potentially overtreated).}
\label{fig:nyc-hidden-threat}
\end{figure}

\paragraph*{Where the networks disagree.} Figure~\ref{fig:nyc-hidden-threat} maps the difference between mobility-based and geographic exposure (week of October 24, 2020). Red areas face more risk through mobility than geography suggests (undertreated); blue areas face more from geography (potentially overtreated). Disagreement is greatest at the outbreak periphery---precisely where getting the network right matters most. The rank correlation between exposures averages 0.74 across weeks (Appendix Figures~\ref{fig:nyc-scatter} and~\ref{fig:nyc-bridges} provide additional detail).

We now ask which network actually predicts subsequent case growth. For each horizon $h \in \{1, 2, 3, 4\}$ weeks, we estimate:
\begin{equation}\label{eq:nyc-horserace}
\Delta_{h,i,w} = \beta_L \, \text{Exp}_{L,i,w}^z + \beta_G \, \text{Exp}_{G,i,w}^z + \bm{\theta}' \bm{Z}_{i,w} + \gamma \, \text{CaseRate}_{i,w} + \alpha_i + \delta_w + \varepsilon_{i,w},
\end{equation}
where $\Delta_{h,i,w}$ is the $h$-week-ahead change in case rate, $\bm{Z}_{i,w} = (\mathbf{1}_{\text{Red}}, \mathbf{1}_{\text{Orange}}, \mathbf{1}_{\text{Yellow}})$ are zone dummies, $\alpha_i$ and $\delta_w$ are MODZCTA and week fixed effects, and standard errors are clustered at the MODZCTA level. Coefficients on exposure are standardized: cases per 100k per one-standard-deviation increase in exposure. The zone dummies absorb the treatment effect of the policy itself---so the exposure coefficients measure predictive power after controlling for the policy response.

Table~\ref{tab:predictive_comparison} reports the degree-matched horse-race. Individually, both networks predict case growth (Specs~1--2), but in the horse-race (Spec~3), $\hat\beta_G = 19.2$--$20.0$ (all $p < 0.01$) while $\hat\beta_L$ is statistically indistinguishable from zero at every horizon. Since the two networks have identical per-node degree, the gap reflects \emph{which} connections matter, not how many. The geo-superset comparison tells the same story: even though $G_{\text{super}}$ includes every geographic edge, $\hat\beta_{G_s} = 20.6$--$23.8$ (all $p < 0.01$) while $\hat\beta_L \approx 0$ (Appendix Table~\ref{tab:superset_comparison}). The predictive signal is entirely in the mobility edges. 

Zone dummies confirm that Red and Orange restrictions reduced case growth at 2--4 week horizons (Appendix~\ref{sec:nyc-appendix}). MODZCTA and week fixed effects control for neighborhood characteristics and city-wide trends, though we cannot rule out all confounders and interpret the regression as predictive rather than causal.

\begin{table}[htbp]
\centering
\caption{Predictive Comparison: Geographic vs.\ Mobility Exposure}
\label{tab:predictive_comparison}
\small
\begin{tabular}{r cc cc ccc}
\toprule
 & \multicolumn{2}{c}{Spec 1: $L$ only} & \multicolumn{2}{c}{Spec 2: $G$ only} & \multicolumn{3}{c}{Spec 3: Horse race}  \\
\cmidrule(lr){2-3} \cmidrule(lr){4-5} \cmidrule(lr){6-8}
$h$ & $\hat\beta_L$ & $R^2$ & $\hat\beta_G$ & $R^2$ & $\hat\beta_L$ & $\hat\beta_G$ & $R^2$ \\
\midrule
  1 & $13.632$ (3.297) & 0.119 & $20.749$ (2.763) & 0.176 & $2.002$ (2.973) & $19.226$ (3.822) & 0.183 \\
  2 & $12.793$ (3.428) & 0.233 & $20.532$ (3.527) & 0.252 & $0.690$ (3.715) & $20.007$ (5.083) & 0.260 \\
  3 & $10.668$ (3.684) & 0.340 & $18.904$ (4.164) & 0.349 & $-1.422$ (3.882) & $19.986$ (5.564) & 0.352 \\
  4 & $6.598$ (3.195) & 0.456 & $15.367$ (4.161) & 0.460 & $-4.999$ (5.212) & $19.171$ (5.944) & 0.461 \\
\bottomrule
\end{tabular}
\begin{tablenotes}\footnotesize
\item \textit{Notes:} Dependent variable is $\Delta_{h,i,w}$, the $h$-week-ahead change in case rate. Standardized coefficients shown; standard errors (clustered by MODZCTA) in parentheses. All specifications include MODZCTA and week fixed effects plus controls for policy tier and own case rate.
\end{tablenotes}
\end{table}

\begin{figure}[ht]
\centering
\includegraphics[width=\textwidth]{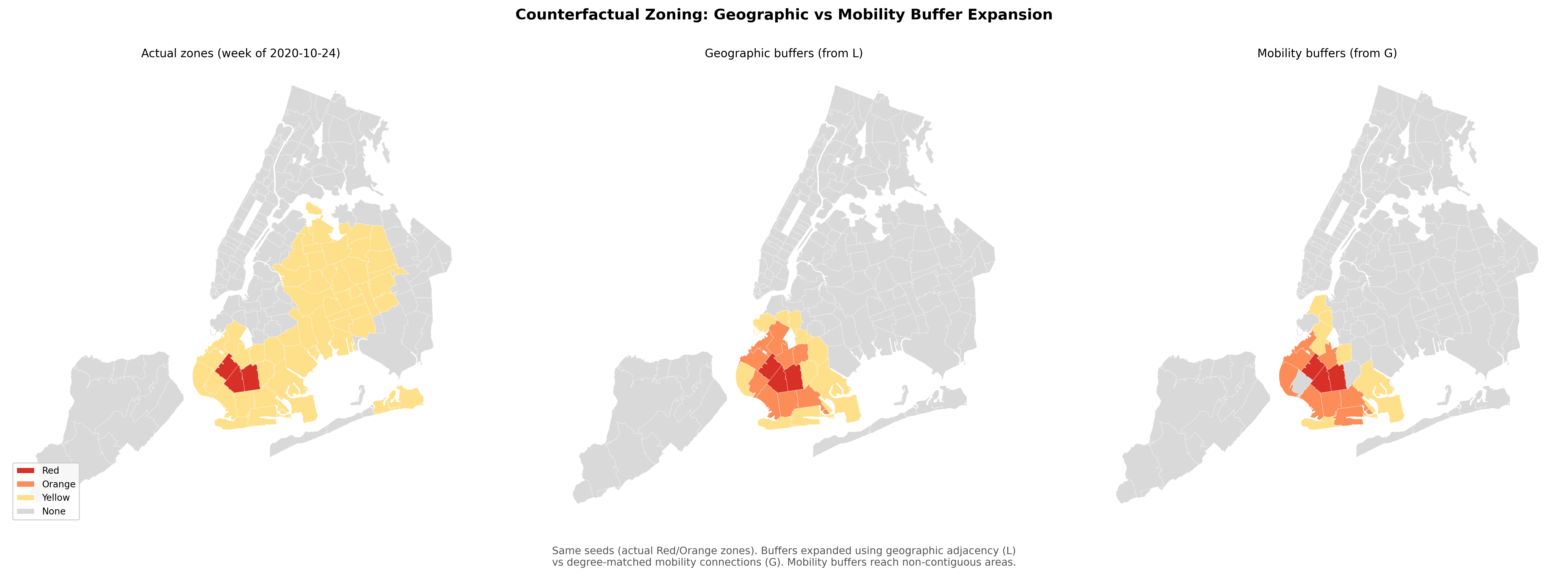}
\caption{Counterfactual zoning: geographic vs.\ mobility buffer expansion (week of 2020-10-24). Left: actual zone assignments. Center: buffer zones expanded using geographic adjacency ($L$). Right: buffer zones expanded using degree-matched mobility connections ($G_{\text{match}}$). Same seeds (Red/Orange zones) held fixed; mobility buffers reach non-contiguous areas across the city.}
\label{fig:nyc-counterfactual}
\end{figure}

We replicate the actual zoning rule---seeds designated Red, then buffer zones expanded via 1-hop (Orange) and 2-hop (Yellow) neighbors---but vary the buffer network, holding seeds fixed each week. Figure~\ref{fig:nyc-counterfactual} maps the results for a representative week. Replacing geography with the degree-matched mobility network ($G_{\text{match}}$) reaches only 187 buffer neighborhoods versus $L$'s 242 over the full program---a 23\% reduction---with 79 zone-assignment changes. The mobility buffers reach non-contiguous areas across the city, including parts of Brooklyn and Queens that are geographically distant from the epicenter but strongly connected through commuting flows. Augmenting geography with additional mobility links ($G_{\text{super}}$) covers all 242 neighborhoods in $L$'s buffer plus 26 additional neighborhoods reached by mobility connections that do not share a border with the hotspot, for a total of 268. The Jaccard similarity between $L$ and $G_{\text{match}}$ buffers averages 0.71, indicating roughly 30\% divergence in which neighborhoods receive containment resources (Appendix~\ref{sec:nyc-appendix} reports the full week-by-week comparison). The practical recommendation is augmentation: keep every geographic neighbor and add mobility connections on top.

The degree-matched comparison maps cleanly to our framework: $G_{\text{match}}$ has the same edges per node as $L$, so $G_{\text{match}} \setminus L$ consists precisely of unaligned jump links connecting MODZCTAs far apart in adjacency distance. The Yellow buffer zones are analogous to diffusion balls---regions the policymaker believes bound the outbreak. The regression evidence matches the theory: mobility exposure (capturing jump links) predicts case growth; geographic exposure does not. Two caveats: the theory targets asymptotic regimes while this is a finite setting with 177 MODZCTAs, and the ${\sim}30\%$ edge difference is not vanishingly small in the sense of $\beta_n \to 0$. Nonetheless, the theoretical mechanism is visible in the data.

\subsection{Optimal Targeting under Measurement Error}\label{sec:beaman-application}

\cite{beaman2021can} design a targeting strategy for information diffusion in Malawi that identifies optimal seed nodes to maximize technology adoption. Their strategy significantly outperforming alternatives like selecting village leaders. However, their strategy depends on the network being correctly measured -- here, we examine how robust the targeting strategy is to small measurement error in the network. Our theory predicts that the answer depends on expansion properties: in well-mixed networks, seed choice matters little; in structured networks, seed choice matters enormously but is most fragile.

We take the 225 village networks from the Malawi data used by \cite{beaman2021can} and calibrate an SIR percolation model to match their threshold diffusion model.\footnote{%
The \cite{beaman2021can} model uses a threshold rule: each agent draws a threshold $\tau$ from a truncated normal $\text{TruncN}(\lambda, 0.5; \tau > 0)$ and adopts when the number of informed neighbors exceeds $\tau$. We calibrate the per-edge infection probability in our SIR model by computing $p(\lambda) = \Pr(\tau \leq 1 \mid \tau \sim \text{TruncN}(\lambda, 0.5; \tau > 0))$, yielding $p \approx 0.46$ for $\lambda = 1$ (``simple'' contagion).}
For each village, we enumerate all pairs of seed nodes and run 2{,}000 SIR simulations per pair over $T = 4$ time steps to identify the optimal seed pair on the true network. We then introduce measurement error by adding false edges with probability $p_{\text{false}} = 1/n_{\text{households}}$ per non-edge (a vanishingly small rate for typical village sizes of 50--200 households), generate 1{,}000 perturbed networks per village, and re-evaluate the same seed pairs on each perturbed network.

To index the expansion properties of each village network, we compute the spectral gap---the difference between the first and second eigenvalues of the adjacency matrix. A large spectral gap indicates a well-connected, expander-like network with fast mixing; a small spectral gap indicates a more structured network with bottlenecks and slower expansion, closer to the polynomial-growth regime of our theoretical framework.

To measure sensitivity of the network to optimal seeds, we consider the change in diffusion from targeting the optimal seed (Top1 Seed) and the fifth best seed (Top5 Seed). In networks where the optimal choice matters quite a bit, the percentage difference will be large. In places where it is unimportant, this difference will be small.

\begin{figure}[ht]
\centering
\begin{tabular}{ccc}
\includegraphics[width=0.32\textwidth]{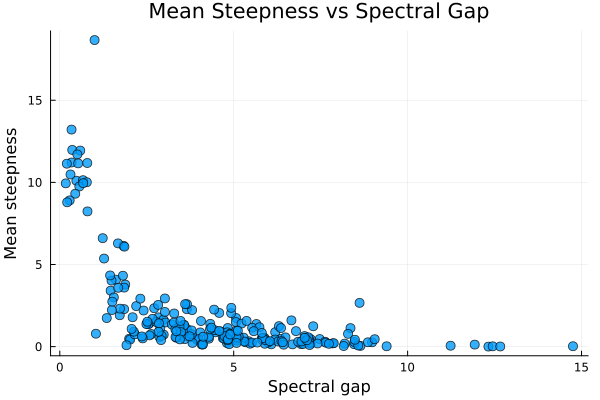} &
\includegraphics[width=0.32\textwidth]{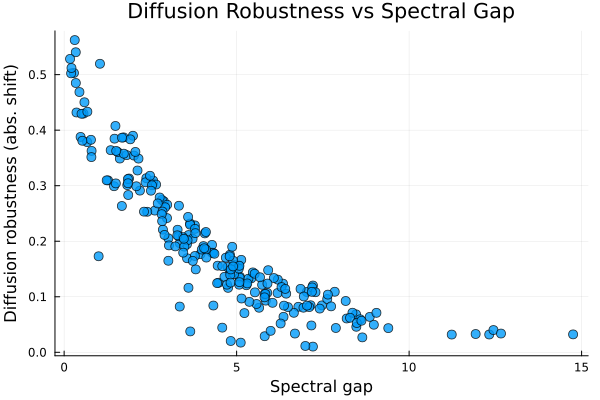} &
\includegraphics[width=0.32\textwidth]{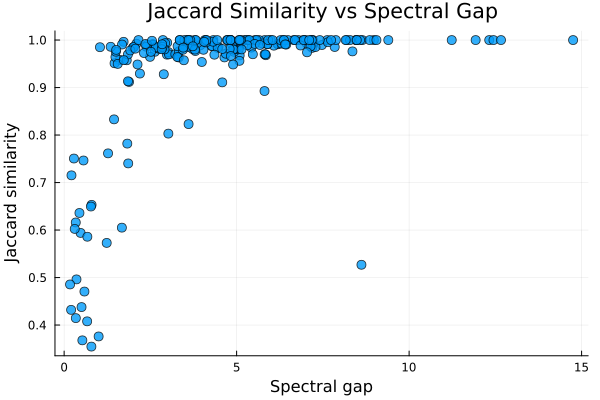} \\
{\small (a) Mean steepness} & {\small (b) Diffusion robustness} & {\small (c) Jaccard similarity} \\[6pt]
\includegraphics[width=0.32\textwidth]{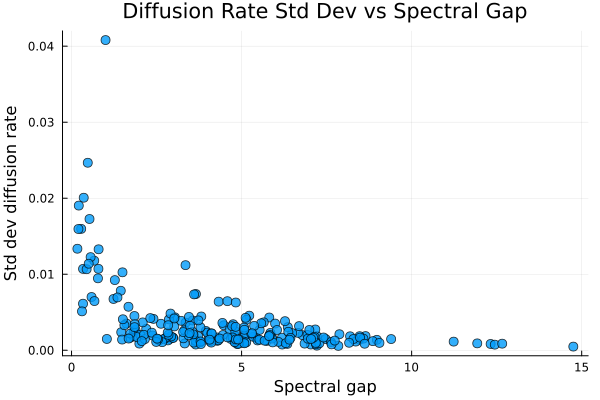} &
\includegraphics[width=0.32\textwidth]{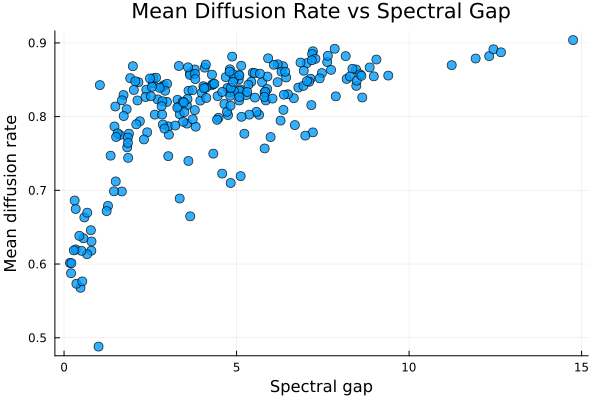} & \\
{\small (d) Std.\ dev.\ of diffusion rate} & {\small (e) Mean diffusion rate} &
\end{tabular}
\caption{Sensitivity of optimal targeting to measurement error across 225 Malawi villages from \cite{beaman2021can}. Each dot is a village; the $x$-axis is the spectral gap of the village network. Panel (a): steepness of the seed-quality objective function. Panel (b): absolute shift in diffusion rate between true and perturbed networks. Panel (c): Jaccard similarity of reached nodes (true vs.\ perturbed). Panel (d): standard deviation of diffusion rate across 1{,}000 error draws. Panel (e): mean diffusion rate (fraction of village reached). Measurement error: false edges added with probability $1/n_{\text{households}}$ per non-edge; $T = 4$; $p = 0.46$; 2{,}000 SIR simulations per seed pair per error draw.}
\label{fig:beaman-results}
\end{figure}

\paragraph*{Results.} Figure~\ref{fig:beaman-results} presents five outcomes across the 225 villages, each plotted against the village's spectral gap. The results reveal a tension at the core of network-based policy. Panel~(a) shows that the steepness of the objective function ($100 \times (\text{Top1 Seed} - \text{Top5 Seed})/\text{Top1 Seed}$) is 10--19\% for low-spectral-gap villages and near zero for high-spectral-gap villages. Targeting \emph{matters most} in bottlenecked networks---precisely where our theory predicts measurement error is most damaging.

Panels~(b)--(c) confirm this. Diffusion robustness (the absolute shift in the diffusion rate, panel~b) is 0.3--0.6 for low-spectral-gap villages but ${\sim}0.03$ for high-spectral-gap villages. Jaccard similarity (panel~c) drops to 0.4--0.6 for low spectral gap villages---measurement error changes not just \emph{how much} diffusion occurs but \emph{who} is reached. The standard deviation across error draws (panel~d) is large for structured villages and near zero for well-mixed ones. Panel~(e) shows well-mixed villages achieve 85--90\% adoption from any seed, while structured villages reach only 60\%.

Low-spectral-gap villages have bottlenecked networks where the optimal seed sits at a critical bridge; a single false edge can redirect diffusion entirely. High-spectral-gap villages are near-expanders where diffusion reaches most nodes from any seed. A policymaker would invest in targeting precisely for the villages where it appears most valuable---the structured, low-spectral-gap villages. But these are exactly where targeting is most fragile: the networks with the highest potential payoff from targeting are those where measurement error most changes that payoff.

Our exercise uses i.i.d.\ false edges---the dispersed, unaligned structure our theory identifies as most damaging. More broadly, these simulations demonstrate that unaligned measurement error is particularly damaging for targeting applications, where seed choice interacts with network structure. Real survey error is plausibly unaligned and may be \emph{worse}. First, non-responding households tend to differ in degree and centrality \citep{chandrasekharl2010}, and may disproportionately bridge subgroups. Second, respondents are more likely to name strong, salient ties than weak ones---but weak ties bridge clusters \citep{granovetter1973}. Third, surveys typically enumerate links within a defined boundary, excluding connections to neighboring villages and visiting traders.

All three channels systematically miss bridging links while retaining local ones. Our i.i.d.\ exercise may therefore \emph{understate} targeting fragility, since it adds shortcuts uniformly rather than targeting bottleneck points. Some real error may be closer to aligned (random omission within well-surveyed categories), which Theorem~\ref{thm:no-failure-mar} predicts would be harmless. Researchers deploying network-based targeting should invest in understanding which error channel dominates.

\section{Discussion}
Our results establish that forecast failure in network diffusion is governed by the geometric relationship between missing and observed links. When the observed network has polynomial expansion, even vanishingly small geometrically unaligned missing links transform expansion properties, generating catastrophic forecast errors. The key mechanism is that missing links function as jump links---even those connecting geographically or socially proximate nodes---serving as escape hatches in graph distance on the observed network $L_n$, bridging regions that are macroscopically separated in the metric that governs diffusion dynamics. The small-worlds phenomenon of \cite{watts1998collective}---demonstrated numerically on ring lattices---is a manifestation of this principle.

Theorem~\ref{thm:no-failure-mar} provides sufficient conditions for robustness that complement the failure conditions in Theorem~\ref{thm:main-polynomial}: forecast failure arises when missing links disrupt the expansion structure of the observed network, but not when they merely thin it. The contrast between aligned and unaligned measurement error has direct practical implications, though intermediate cases remain open (Remark~\ref{rem:scope-contrast}). Common data-collection practices---geographic proximity proxies, interaction frequency thresholds, survey boundaries---systematically produce unaligned measurement error, because they disproportionately miss links that bridge different network regions. The most valuable investment for applied researchers may therefore be understanding the \emph{geometry} of measurement error in their data, rather than estimating the \emph{rate} of missing links. That said, geometry is not the only consideration: even under aligned error, Theorem~\ref{thm:no-failure-mar} requires the censoring rate to satisfy $\varepsilon_n T_n^{q+1} \to 0$, so a non-trivial volume of missing links remains problematic regardless of geometry.

Importantly, ``long-range'' in our framework means far apart in \emph{graph distance on $L_n$}---not necessarily in geographic or social distance. A link between geographically nearby locations is a long-range shortcut if $L_n$ provides no short path between them.

Aggregates like $\mathcal{R}_0$ may be better used as descriptive than prescriptive tools. The failure of robustness would almost certainly propagate to welfare calculations that rely on the extent or location of diffusion \citep{acemoglu2021optimal,fajgelbaum2021optimal}. The susceptibility to small measurement error may argue for earlier, more aggressive policy responses \citep{barnett2023epidemic}. The full decision theory exercise is beyond our scope, but the statistical force we document pushes strongly in that direction.

\begin{remark}[Directed Networks]\label{rem:directed}
Many applied settings involve directed interactions. Sensitive dependence may be \emph{stronger} in directed settings, since cones of influence from nearby seeds diverge more readily when paths are asymmetric. Volume growth assumptions would apply to out-neighborhoods. We conjecture qualitatively similar results hold under analogous directed expansion conditions.
\end{remark}

Our results complement \cite{lewbel2024estimating}, who study link misclassification in peer effects regressions. Both papers show that network measurement error has more severe consequences than standard intuitions suggest, but the mechanisms differ: misclassification biases peer-effects coefficients, while in our setting vanishingly small error causes diffusion \emph{trajectories} to diverge catastrophically. 

Our results are specific to SIR models, but the same perturbation robustness failure may affect general models of treatment effects with spillovers \citep{aronow2017estimating, hardy2019estimating, athey2018exact}---a direction we leave to future work.


\bibliographystyle{plainnat}
\bibliography{model_failures}


\appendix

\renewcommand\thefigure{\thesection.\arabic{figure}}    
\setcounter{figure}{0}  
\renewcommand\thetable{\thesection.\arabic{table}} 
\setcounter{table}{0}

\setcounter{figure}{0}  
\setcounter{table}{0}  

\section{Proofs}\label{sec:proofs}

\begin{proof}[Proof of Lemma \ref{lem:diffusion}]
We establish matching upper and lower bounds on $\calE_t$.

\textit{Upper bound.} When $p_n = 1$, every edge in the percolation transmits the diffusion with certainty, providing an upper bound for any $p_n \in (0,1]$. In this case, the diffusion volume is determined entirely by the graph structure of $L_n$. By part 1 of Assumption \ref{ass:disease}, any ball of radius $t$ contains at most $a_{max} t^{q+1}$ nodes, so $\calE_t \leq a_{max} t^{q+1} = O(t^{q+1})$.

\textit{Lower bound.} We track activations in a specific region to establish the lower bound. Fix a node $z$ at distance at most $(1-\alpha)t$ from the seed $i_0$. Consider the ball $B_z(\alpha t)$ of radius $\alpha t$ centered at $z$; note that $B_z(\alpha t) \subseteq B_{i_0}(t)$ by the triangle inequality. By part 1 of Assumption \ref{ass:disease}, this ball contains at least $a_{min}(\alpha t)^{q+1} = a_{min}\alpha^{q+1} t^{q+1}$ nodes. 

By part 2 of Assumption \ref{ass:disease} (uniform spread), with probability at least $\varepsilon > 0$, at least a fraction $\theta > 0$ of these nodes are activated by time $t$. Therefore, even considering only this single ball and ignoring all other activated nodes,
\begin{align*}
    \calE_t \geq \varepsilon \cdot \theta \cdot a_{min}\alpha^{q+1} t^{q+1} = \Omega(t^{q+1}),
\end{align*}
where the final equality uses that $\varepsilon, \theta, \alpha > 0$ by Assumption \ref{ass:disease} and $a_{min} \geq 1$.

Combining the upper and lower bounds yields $\calE_t = \Theta(t^{q+1})$.
\end{proof}

\begin{proof}[Proof of Theorem \ref{thm:sensitive-dep}]
We construct a set of alternative seeds $J_{i_0}$ near $i_0$ and show that for any $j_0 \in J_{i_0}$, the diffusions from $i_0$ and $j_0$ reach substantially disjoint sets of nodes by a suitable time horizon $T_n$. Throughout, distances are measured with respect to the base network $L_n$. For notational simplicity, we suppress floor and ceiling functions when discretizing continuous quantities.

\medskip
\noindent\textit{Step 1: Construction of the alternative seed set $J_{i_0}$.}
Fix a node $k$ at distance $r_n$ from $i_0$, where $r_n \in [\overline{T}_n/2, \overline{T}_n]$ is even. Such a node exists because $|B_{i_0}(\overline{T}_n)| - |B_{i_0}(\overline{T}_n/2)| = \Theta(\overline{T}_n^{q+1}) > 0$ by Assumption~\ref{ass:disease}, part~1, so the shell between radii $\overline{T}_n/2$ and $\overline{T}_n$ is nonempty and contains nodes at some even distance $r_n$. Set $a_n := r_n/(1+\lambda)$ for some constant $\lambda \in (0,1)$ to be specified. Define the lens-shaped region
\[
J_{i_0} := B(i_0, a_n) \cap B(k, a_n),
\]
the intersection of balls of radius $a_n$ centered at $i_0$ and $k$, respectively.

We now verify that $J_{i_0}$ contains a constant fraction of $B(i_0, a_n)$. Let $m$ be the midpoint of a shortest path from $i_0$ to $k$, so $d(i_0, m) = d(k, m) = r_n/2$. Consider a ball $B(m, w)$ of radius $w := a_n - r_n/2$ around $m$. By the triangle inequality, for any node $x \in B(m, w)$,
\[
d(i_0, x) \leq d(i_0, m) + d(m, x) \leq r_n/2 + w = a_n,
\]
and symmetrically $d(k, x) \leq a_n$. Thus $B(m, w) \subseteq B(i_0, a_n) \cap B(k, a_n) = J_{i_0}$. 

Substituting $r_n = (1+\lambda)a_n$ yields $w = a_n - (1+\lambda)a_n/2 = (1-\lambda)a_n/2$. By Assumption~\ref{ass:disease}, part 1,
\begin{align*}
    \frac{|J_{i_0}|}{|B(i_0, a_n)|} 
    &\geq \frac{|B(m, w)|}{|B(i_0, a_n)|}
    \geq \frac{a_{\min} w^{q+1}}{a_{\max} a_n^{q+1}}\\
    &= \frac{a_{\min}}{a_{\max}} \left(\frac{1-\lambda}{2}\right)^{q+1} =: C > 0,
\end{align*}
where $C > 0$ since $a_{\min}, a_{\max} \geq 1$, $q > 0$, and $\lambda \in (0,1)$.

\medskip
\noindent\textit{Step 2: Divergence of diffusion paths.}
Fix an arbitrary alternative seed $j_0 \in J_{i_0}$ and set the forecast horizon $T_n := a_n + \lambda a_n/2$. By choosing $r_n$ appropriately, $T_n$ can be placed in $[\underline{T}_n, \overline{T}_n]$, which is feasible since both endpoints diverge with $n$. We identify a region reachable from $j_0$ within $T_n$ steps but not from $i_0$.

Consider the ball $B(k, \lambda a_n(1-c)/2)$ for some $c\in(0,1-3\alpha)$, noting that the interval is non-empty since $\alpha < 1/3$. For any node $v$ in this ball, the reverse triangle inequality gives
\begin{align*}
    d(i_0, v) 
    &\geq |d(i_0, k) - d(k, v)| 
    \geq (1+\lambda)a_n - \lambda a_n(1-c)/2 
    = a_n + \frac{\lambda(1+c)}{2} a_n
    > T_n.
\end{align*}
Conversely, by the triangle inequality,
\[
d(j_0, v) \leq d(j_0, k) + d(k, v) \leq a_n + \lambda a_n(1-c)/2 < T_n,
\]
where the first inequality uses $j_0 \in J_{i_0} \subseteq B(k, a_n)$. Thus every node in $B(k, \lambda a_n(1-c)/2)$ is reachable from $j_0$ in $T_n$ steps but not from $i_0$.

\medskip
\noindent\textit{Step 3: Positive probability of divergence.} We then apply Assumption \ref{ass:disease}. The ball $B(k, \lambda a_n(1-c)/2)$ contains at least $a_{\min}(\lambda a_n(1-c)/2)^{q+1}$ nodes by Assumption~\ref{ass:disease}, part 1. We then apply part 2 (uniform spread). To apply the assumption, it must be that $d(j_0, k)\leq (1-\alpha)T_n$. Given that $j_0 \in B(k, a_n)$, the worst case is that $d(j_0, k) = a_n$. To ensure the conditions are concurrent, $a_n \leq (1-\alpha)(1+\lambda/2)a_n$, or that $\alpha \leq \frac{\lambda}{2+\lambda}$. Given that $\alpha < 1/3$, we can always find some $\lambda \in (0,1)$ such that this condition holds. By the assumption, with probability $\varepsilon > 0$, at least a fraction $\theta > 0$ of nodes in $B(k, \alpha T_n)$ are activated. To apply the assumption directly, we choose $\lambda = \frac{2\alpha}{1-\alpha-c}$ so that $\lambda a_n(1-c)/2 = \alpha a_n(1+\lambda/2) = \alpha T_n$. Note that as $\alpha < 1/3$ and $c \in (0,1 - 3\alpha)$, it follows $\lambda \in (0,1)$.

Since $|B(k, \lambda a_n(1-c)/2)| = \Theta(a_n^{q+1})$ and these nodes are unreachable from $i_0$ in $T_n$ steps, the symmetric difference $|I_{\mathcal{P}}(i_0, T_n) \triangle I_{\mathcal{P}}(j_0, T_n)|$ includes $\Theta(a_n^{q+1})$ nodes with probability bounded away from zero. Therefore,
\[
\Delta_n(i_0, j_0) = \frac{|I_{\mathcal{P}}(i_0, T_n) \cap I_{\mathcal{P}}(j_0, T_n)|}{|I_{\mathcal{P}}(i_0, T_n) \cup I_{\mathcal{P}}(j_0, T_n)|} \leq 1 - c'
\]
for some constant $c' > 0$, with probability bounded away from zero.
\end{proof}

\begin{lemma}[Tiling Construction]\label{lem:tiles}
There exists an algorithm that constructs $K_n$ disjoint tiles $\{k_0, k_1,\ldots, k_{K_n-1}\}$, covering $\Theta(n)$ nodes (some nodes may lie outside all tiles), with the following properties:
\begin{enumerate}
    \item Each tile is contiguous (forms a connected subgraph).
    \item Tiles are pairwise disjoint: $k_i \cap k_j = \emptyset$ for all $i \neq j$.
    \item Each tile contains a ``core'' of nodes at distance at least $\overline{T}_n$ from the tile boundary, comprising a constant fraction of the tile's volume.
    \item The union of all tile cores contains $\Theta(n)$ nodes.
\end{enumerate}

\medskip
\noindent\textbf{Tiling Algorithm.}
The algorithm takes as input $\overline{T}_n$, a graph $L_n$ satisfying Assumption~\ref{ass:disease}, and an initial seed $i_0$. All distances are measured in $L_n$.

\begin{enumerate}
    \item \textbf{Initialization.} Set the tile radius $R_{\text{tile}} := 2\overline{T}_n$ and exclusion radius $R_{\text{excl}} := 4\overline{T}_n$. Initialize the set of tile centers $\calC = \emptyset$ and the set of available nodes $\calV_{\text{avail}} = V$.

    \item \textbf{Seed tile.} Place the first tile center at the seed: $\calC \leftarrow \{i_0\}$. Remove its exclusion zone: $\calV_{\text{avail}} \leftarrow \calV_{\text{avail}} \setminus B(i_0, R_{\text{excl}})$.

    \item \textbf{Greedy placement.} While $\calV_{\text{avail}} \neq \emptyset$:
    \begin{enumerate}
        \item Select an arbitrary node $c_j \in \calV_{\text{avail}}$.
        \item Add $c_j$ to the centers: $\calC \leftarrow \calC \cup \{c_j\}$.
        \item Remove its exclusion zone: $\calV_{\text{avail}} \leftarrow \calV_{\text{avail}} \setminus B(c_j, R_{\text{excl}})$.
    \end{enumerate}

    \item \textbf{Tile definition.} For each center $c_j$ in the final set $\calC = \{i_0, c_1,\ldots, c_{K_n}\}$, define tile $k_j := B(c_j, R_{\text{tile}})$.
\end{enumerate}
\end{lemma}

\begin{proof}
The algorithm terminates in finite steps for a given $n$ since each iteration removes at least one node (the chosen center $c_j$) from $\calV_{\text{avail}}$, which is initially finite. We verify each claimed property in turn.

\medskip
\noindent\textit{Property 1: Contiguity.} 
Each tile $k_j = B(c_j, R_{\text{tile}})$ is a ball in the graph metric of $L_n$, hence consists of all nodes reachable from $c_j$ within $R_{\text{tile}}$ steps. Any such ball is connected by definition.

\medskip
\noindent\textit{Property 2: Disjointness.} 
Consider distinct tiles $k_i$ and $k_j$ with centers $v_i$ and $v_j$. By the algorithm's exclusion mechanism, $d(v_i, v_j) > R_{\text{excl}} = 4\overline{T}_n$ (otherwise $v_j$ would have been removed when $v_i$ was selected).

Suppose for contradiction that $k_i \cap k_j \neq \emptyset$, and let $v_k \in k_i \cap k_j$. Then $d(v_k, v_i) \leq R_{\text{tile}} = 2\overline{T}_n$ and $d(v_k, v_j) \leq 2\overline{T}_n$. By the triangle inequality,
\[
d(v_i, v_j) \leq d(v_i, v_k) + d(v_k, v_j) \leq 4\overline{T}_n,
\]
contradicting $d(v_i, v_j) > 4\overline{T}_n$. Therefore tiles are pairwise disjoint.

\medskip
\noindent\textit{Property 3: Constant fraction in tile cores.} 
Each tile $k_j = B(c_j, 2\overline{T}_n)$ has radius $2\overline{T}_n$. Define the \emph{core} of tile $k_j$ as the inner ball $B(c_j, \overline{T}_n)$, consisting of nodes at distance at least $\overline{T}_n$ from the tile boundary. By Assumption~\ref{ass:disease}, part 1,
\[
\frac{|B(c_j, \overline{T}_n)|}{|B(c_j, 2\overline{T}_n)|} 
\geq \frac{a_{\min} \overline{T}_n^{q+1}}{a_{\max} (2\overline{T}_n)^{q+1}} 
= \frac{a_{\min}}{2^{q+1} a_{\max}} =: \rho > 0,
\]
where $\rho > 0$ is a constant since $q > 0$ is fixed and $1 \leq a_{\min} \leq a_{\max} < \infty$.

\medskip
\noindent\textit{Property 4: Linear number of core nodes.} 
We establish that the total volume of core nodes is $\Theta(n)$ by proving matching bounds.

\textit{Upper bound:} Core nodes are a subset of all nodes, so $|V_{\text{core}}| \leq n = O(n)$.

\textit{Lower bound:} We first bound the number of tiles. The algorithm terminates when the exclusion zones cover all nodes: $V \subseteq \bigcup_{j=1}^{K_n} B(c_j, R_{\text{excl}})$. Taking volumes and applying Assumption~\ref{ass:disease}, part 1,
\[
n = |V| \leq \sum_{j=1}^{K_n} |B(c_j, R_{\text{excl}})| 
\leq \sum_{j=1}^{K_n} a_{\max}(4\overline{T}_n)^{q+1} 
= K_n \cdot a_{\max} \cdot 4^{q+1} \overline{T}_n^{q+1}.
\]
Rearranging yields
\[
K_n \geq \frac{n}{4^{q+1} a_{\max} \overline{T}_n^{q+1}}.
\]

Each tile core contains at least $a_{\min} \overline{T}_n^{q+1}$ nodes (all nodes within distance $\overline{T}_n$ of the center). Therefore,
\[
|V_{\text{core}}| \geq K_n \cdot a_{\min} \overline{T}_n^{q+1} 
\geq \frac{n}{4^{q+1} a_{\max} \overline{T}_n^{q+1}} \cdot a_{\min} \overline{T}_n^{q+1} 
= \frac{a_{\min}}{4^{q+1} a_{\max}} \cdot n = \Theta(n),
\]
where the final equality uses that the leading constant is strictly positive.
\end{proof}

We now extend our analysis beyond the i.i.d. link structure of Assumption~\ref{ass:beta} to accommodate more general error patterns. The key modification allows each node's missing links to concentrate on a subset of potential partners, while maintaining sufficient dispersion across the network to enable our results.

\begin{ass}[Heterogeneous Missing Links]\label{ass:beta-corr}
Each node $i$ has an \emph{affinity set} $A_i \subseteq V_n$ with $|A_i| = \delta_n n$, where $\delta_n \in (0,1]$ may depend on $n$. Affinity is \emph{symmetric}: $j \in A_i$ if and only if $i \in A_j$. Missing links form independently according to
\[
E_{ij} \sim \begin{cases}
\mathrm{Ber}(\beta_n) & \text{if } j \in A_i \ (\text{equivalently, } i \in A_j)\\
0 & \text{otherwise},
\end{cases}
\]
with all potential links drawn independently.

After constructing tiles via Lemma~\ref{lem:tiles}, denote the resulting tile set by $\calT$. We impose the following \emph{dispersion condition}: a constant fraction of nodes $\eta_1$ in each tile have affinity sets that cover at least a constant fraction $\eta_2$ of other tiles, where $\eta_1, \eta_2 > 0$ are universal constants independent of $n$. 

Additionally, we require:
\begin{enumerate}
    \item \textbf{Sufficient affinity volume:} $\delta_n = \omega\left(\frac{1}{ \underline{T}_n^q}\right)$.
    \item \textbf{Sparse error rate:} $\beta_n = \omega\left(\frac{1}{p_n\underline{T}_n^{q+1} \delta_n n}\right), \quad \beta_n = o\left(\frac{1}{n}\right).$
\end{enumerate}
\end{ass}

This assumption generalizes the i.i.d. structure in three respects. First, nodes need not be able to link to all others—each node $i$ can only form missing links within its affinity set $A_i$, which may comprise only a vanishing fraction $\delta_n$ of the population. Second, the affinity structure may be arbitrary, subject only to the dispersion condition. Third, the link formation rate $\beta_n$ is rescaled to account for the restricted affinity range.

The dispersion condition ensures that missing links from any collection of tiles have sufficiently broad reach across the network. When $\delta_n = 1$ (full affinity), this holds trivially: every node can potentially link to every other node, so any set of nodes can reach any tile with positive probability. For $\delta_n < 1$, the condition requires that affinity sets are not pathologically concentrated---they must provide coverage across tiles. The symmetry requirement ($j \in A_i \Leftrightarrow i \in A_j$) ensures that cross-tile coverage by source nodes translates into feasible link formation: if node $i$ in one tile has node $j$ in a distant tile in its affinity set, the link $(i,j)$ is feasible. Without symmetry, the dispersion condition on $A_i$ alone would not guarantee that cross-tile links can form. Despite the symmetry constraint, the assumption allows for substantial heterogeneity: different nodes may have vastly different affinity sets, provided each set is symmetric and dispersed.

The parametric restrictions ensure our results remain meaningful in this generalized setting. Condition (1) requires that affinity sets are not too restrictive: each node must be able to reach enough potential partners that the range $\left(\frac{1}{p_n \underline{T}_n^{q+1} \delta_n n}, \frac{1}{n}\right)$ for $\beta_n$ is nonempty. Condition (2) maintains sparsity: the upper bound $\beta_n < 1/n$ ensures that each node has $o(1)$ missing links in expectation, so $E_n$ remains a collection of isolated edges rather than forming dense clusters or giant components. The lower bound on $\beta_n$ is scaled by $\delta_n$ to account for the reduced support—nodes can only link within their affinity sets, so a higher per-pair probability is needed to maintain the same expected number of cross-tile jumps.

Assumption~\ref{ass:beta} corresponds to the special case where $\delta_n = 1$ and $A_i = V_n \setminus \{i\}$ for all $i$, recovering global i.i.d. link formation. We first state and prove an auxiliary lemma which demonstrates that most links are across, rather than within tiles. 

\begin{lemma}\label{lem:intratile}
    Suppose Assumptions~\ref{ass:forecast-time}, \ref{ass:disease}, and \ref{ass:beta-corr} hold. Then the following holds: 
    \begin{itemize}
        \item Fix a tile, denoted $k$. Then $\PP(E_n \text{ has a link that starts and ends in }k)\rightarrow 0$.
        \item The fraction of tiles containing at least one within-tile link is $O_p\left(n^{-\frac{1}{2q+3}}\right)$.
    \end{itemize}
    As $n\rightarrow\infty$.
\end{lemma}
\begin{proof}
Let $V_k = |k|$ denote the number of nodes in an arbitrary tile $k$. By Lemma \ref{lem:tiles}, each tile is a ball of radius $R_{\text{tile}} = 2\overline{T}_n$ in the graph $L_n$. By Assumption \ref{ass:disease}, part 1, the volume of any such ball is deterministically bounded by $a_{\max} (2\overline{T}_n)^{q+1} := V_{\max}$. The number of potential node pairs within tile $k$ is $\binom{V_k}{2}$. Since this is monotonic in $V_k$, we can bound the number of potential intra-tile links by the worst-case volume:
\[
N_{\text{pairs}} = \binom{V_k}{2} \leq \frac{1}{2} V_{\max}^2 = O\left(\overline{T}_n^{2q+2}\right).
\]
The probability that at least one such link forms is bounded by the union bound over these pairs. Using the upper bound $\beta_n < 1/n$ from Assumption \ref{ass:beta-corr}:
\[
\PP(E_n \text{ has a link that starts and ends in }k) 
\leq N_{\text{pairs}} \cdot \beta_n 
\leq O\left(\overline{T}_n^{2q+2}\right) \cdot \frac{1}{n}.
\]
Substituting the time horizon constraint $\overline{T}_n = o(n^{\frac{1}{2q+3}})$ from Assumption \ref{ass:forecast-time}:
\[
\PP(E_n \text{ has a link that starts and ends in }k) 
\leq O\left( \frac{n^{\frac{2q+2}{2q+3}}}{n} \right) 
= O\left( n^{\frac{2q+2}{2q+3} - 1} \right) 
= O\left( n^{-\frac{1}{2q+3}} \right).
\]
We can then consider the fraction of tiles with within tile links. By linearity of expectation, the expected number of tiles containing at least one internal link is at most $K_n \times O(n^{-\frac{1}{2q+3}})$. Dividing by $K_n$:
\[
\frac{\mathbb{E}[\#\text{ tiles with internal links}]}{K_n} \leq \frac{K_n \cdot O\left(n^{-\frac{1}{2q+3}}\right)}{K_n} = O\left(n^{-\frac{1}{2q+3}}\right) \to 0.
\]
By Markov's inequality, the fraction of tiles containing at least one within-tile link is $O_p\left(n^{-\frac{1}{2q+3}}\right)$.
\end{proof}

Define a tile to be activated if any of the node in the core of a tile are activated. We then construct a lower bound on the number of total activated tiles. 

\begin{lemma}[Tile Activation Growth]\label{lem:regions-corr}
Suppose Assumptions~\ref{ass:disease}, \ref{ass:forecast-time}, and \ref{ass:beta-corr} hold. Let $X_s$ denote the number of tiles activated for the first time at time step $s$. Then there exists a constant $C > 0$ (depending only on $q, a_{\min}, a_{\max}, \varepsilon, \theta, \alpha, \eta_1, \eta_2$) such that
\[
\sum_{s=1}^{T}\EE_{P_n(G_n), E_n}[X_s] \geq C \cdot n\beta_n\delta_n p_n \cdot (T-1)^{q+1}.
\]
\end{lemma}

\begin{proof}
We establish a lower bound on tile activations by tracking only the \emph{first} activation in each tile, ignoring subsequent within-tile spread. Apply Lemma~\ref{lem:tiles} to split $L_n$ into tiles $\calT = \{k_1,\ldots,k_{K_n}\}$. Throughout, we focus on tile cores—nodes at distance at least $\overline{T}_n$ from tile boundaries—as these are the nodes whose activations cannot escape their tiles via $L_n$ alone within $\overline{T}_n$ steps. We consider a tile activated if any node in the core of the tile has been activated by the diffusion process. 

Define $\calX_t := \EE_{P_n(G_n), E_n}[X_t]$ as the expected number of newly activated tiles at time $t$, where the expectation is over both the percolation $P_n(G_n)$ and the realization of missing links $E_n$.

\medskip
\noindent\textit{Step 1: A positive fraction of tiles remain available.}
We show that the expected number of tiles activated by \emph{first-generation} jump links from the primary wavefront on $L_n$ is $o(K_n)$, so a positive fraction of tiles remain available as targets throughout the time horizon. This conservative count ignores secondary wavefronts seeded by jump links; since secondary wavefronts can only \emph{increase} the true diffusion on $G_n$, restricting to first-generation links yields a valid lower bound in Step~2.

At each time step $t$, the newly activated nodes on $L_n$ (the wavefront $\calS_t$) can generate cross-tile transmissions. Since each node transmits for exactly one period under SIR dynamics, the expected number of first-generation cross-tile transmissions by time $\overline{T}_n$ is at most
\[
\sum_{t=1}^{\overline{T}_n} \calS_t \cdot \beta_n \delta_n n \cdot p_n 
\leq \sum_{t=1}^{\overline{T}_n} O(t^{q}) \cdot \beta_n \delta_n n \cdot p_n
= O(\overline{T}_n^{q+1}) \cdot \beta_n \delta_n n p_n,
\]
where we used $\calS_t = O(t^q)$ (since $\calE_t = \Theta(t^{q+1})$ and $\calS_t = \calE_t - \calE_{t-1}$) and summed over $\overline{T}_n$ time steps. Since $\beta_n < 1/n$ and $p_n\delta_n \leq 1$, this is $O(\overline{T}_n^{q+1})$.

By Lemma~\ref{lem:tiles}, there are at least $K_n \geq \frac{n}{4^{q+1}a_{\max}\overline{T}_n^{q+1}}$ tiles. The expected fraction of tiles activated by first-generation links is therefore at most
\[
\frac{1 + O(\overline{T}_n^{q+1})}{K_n} 
= O\left(\frac{\overline{T}_n^{q+1} \cdot 4^{q+1}a_{\max}\overline{T}_n^{q+1}}{n}\right) 
= O\left(\frac{\overline{T}_n^{2q+2}}{n}\right) = o(1),
\]
where the final equality uses $\overline{T}_n = o(n^{1/(2q+3)})$ from Assumption~\ref{ass:forecast-time}, which gives $\overline{T}_n^{2q+2} = o(n^{(2q+2)/(2q+3)}) = o(n)$. Thus a positive fraction of tiles remain available as targets for the lower bound construction in Step~2.\footnote{The total number of tiles activated (including by secondary wavefronts) may be larger, but this does not affect our lower bound: Step~2 counts only first-generation activations, and each such activation contributes independently to the lower bound on $\hat{Y}_T(G_n)$.}

\medskip
\noindent\textit{Step 2: Lower bound on expected growth.}
We use the Law of Iterated Expectations to handle the randomness of the wavefront. Let $\mathcal{S}_{s-1}$ be the set of nodes newly activated at time $s-1$ (the wavefront). We condition on $\mathcal{S}_{s-1}$ to calculate the expected number of new tiles activated at step $s$:
\[
\mathbb{E}[X_s] = \mathbb{E}_{P_n(G_n)} \left[ \mathbb{E}_{E_n} [ X_s \mid \mathcal{S}_{s-1} ] \right].
\]

Consider a fixed realization of $\mathcal{S}_{s-1}$. For any node $u \in \mathcal{S}_{s-1}$, a "successful activation" event occurs if $u$ forms a link to a node $v$ such that:
\begin{enumerate}
    \item The link exists in $E_n$ and transmits the diffusion (probability $\approx \beta_n p_n$).
    \item $v$ is in the core of a tile $k_j$ distinct from $u$'s tile.
    \item $k_j$ has not been previously activated.
\end{enumerate}

By the dispersion condition in Assumption \ref{ass:beta-corr}, a constant fraction $\eta_1$ of nodes in $\mathcal{S}_{s-1}$ have affinity sets that cover at least a fraction $\eta_2$ of other tiles. For such a "dispersed" node $u$, the number of valid targets is proportional to the size of its affinity set restricted to these tiles. Since core nodes comprise a constant fraction $\rho$ of each tile (Lemma \ref{lem:tiles}), and a positive fraction of tiles remain available as targets (Step 1), the probability that a random link from $u$ hits a valid, available core is lower bounded by a constant $c > 0$. Finally, we know that by Lemma \ref{lem:intratile} within tile links are rare -- therefore, almost all links in $E_n$ are to other tiles. Collisions (multiple first-generation links hitting the same tile simultaneously) are negligible because $\beta_n < 1/n$ ensures the graph of missing links is sparse, by the same computation as in Lemma \ref{lem:intratile}. 

Thus, the expected number of new tiles is lower bounded by the sum of expected successful links:
\[
\mathbb{E}[X_s \mid \mathcal{S}_{s-1}] \geq C \cdot |\mathcal{S}_{s-1}| \cdot (\delta_n n \beta_n p_n),
\]
where $C$ absorbs the constants $\eta_1, \eta_2, \rho$ and the probability that the tile is unactivated. This lower bound holds pointwise for every realization of $\mathcal{S}_{s-1}$: conditional on the wavefront, each node independently probes for missing links via i.i.d.\ Bernoulli draws, and the dispersion condition guarantees that a constant fraction $\eta_1$ of nodes in any tile have affinity sets covering $\eta_2$ of other tiles.

Summing over $s = 1$ to $T$ and applying the tower property to the pointwise inequality:
\[
\sum_{s=1}^{T} \mathbb{E}[X_s]
= \sum_{s=1}^{T} \mathbb{E}\bigl[\mathbb{E}[X_s \mid \mathcal{S}_{s-1}]\bigr]
\geq C \cdot \delta_n n \beta_n p_n \cdot \sum_{s=1}^{T} \mathbb{E}[|\mathcal{S}_{s-1}|]
\geq C \cdot \delta_n n \beta_n p_n \cdot \mathcal{E}_{T-1},
\]
where we used $\sum_{s=1}^{T} \mathbb{E}[|\mathcal{S}_{s-1}|] = \sum_{s=0}^{T-1} \mathbb{E}[|\mathcal{S}_s|] = \mathcal{E}_{T-1}$. By Lemma~\ref{lem:diffusion}, $\mathcal{E}_{T-1} = \Theta((T-1)^{q+1})$, which completes the proof.
\end{proof}

\begin{cor}[Global Seed Sensitivity]\label{cor:sensitive}
Suppose Assumptions~\ref{ass:forecast-time}, \ref{ass:disease}, and \ref{ass:beta-corr} hold. Consider a local perturbation from seed $i_0$ to alternative seed $j_0$ as in Theorem~\ref{thm:sensitive-dep}. Then with strictly positive probability, the following events occur jointly:
\begin{enumerate}
    \item \textbf{Shortcut activation.} Both diffusions utilize missing links in $E_n$: there exist nodes $e_1, e_2 \in V_n$ such that the diffusion from $i_0$ reaches $e_1$ via $L_n$ and then transmits through a missing link to some node outside $B_{i_0}(T_n)$, and similarly the diffusion from $j_0$ reaches $e_2$ and transmits outside $B_{j_0}(T_n)$.
    
    \item \textbf{Macroscopic divergence.} Let $s := \max\{d_{L_n}(i_0, e_1), d_{L_n}(j_0, e_2)\}$ denote the maximum graph distance to the shortcut nodes. Conditional on the shortcut activations in part (1), the expected cumulative number of disjoint tiles reached by the two diffusions over the remaining $T_n - s - 1$ steps is at least
    \[
    2C\,n\beta_n\delta_n p_n(T_n - s - 2)^{q+1},
    \]
    where $C > 0$ is the constant from Lemma~\ref{lem:regions-corr}.
\end{enumerate}
\end{cor}

\begin{proof}
We establish that both diffusions escape their local neighborhoods via missing links with positive probability, then quantify the divergence using the tile activation dynamics from Lemma~\ref{lem:regions-corr}.

\medskip
\noindent\textit{Step 1: Diffusion from $j_0$ escapes via $E_n$.}
Recall from Theorem~\ref{thm:sensitive-dep} the ball $B(k, \alpha T_n)$ reachable from $j_0$ in $T_n$ steps but not from $i_0$ in $T_n$ steps, where $k$ is a node at distance $(1+\lambda)a_n$ from $i_0$.

By Assumption~\ref{ass:disease}, part~1, $|B(k, \alpha T_n)| \geq a_{\min}(\alpha T_n)^{q+1}$. Define $V := \theta \cdot a_{\min}(\alpha T_n)^{q+1}$. By part~2 (uniform spread), with probability at least $\varepsilon > 0$, at least $V$ nodes in $B(k, \alpha T_n)$ are activated from $j_0$. Call this event $\calA$.

Conditional on $\calA$, consider the $V$ activated source nodes. By the dispersion condition in Assumption~\ref{ass:beta-corr}, a constant fraction $\eta_1$ of these sources have affinity sets covering at least $\eta_2 \delta_n n$ nodes outside $B_{j_0}(T_n)$. For each such source node $u$, the probability that $u$ has \emph{no} missing link in $E_n$ to any of its $\delta_n n$ feasible targets that also transmits is $(1 - \beta_n p_n)^{\delta_n n}$. Since missing links are independent across pairs (Assumption~\ref{ass:beta-corr}), the probability that \emph{none} of the $\eta_1 V$ dispersed source nodes generates a successful escape transmission is at most
\[
\bigl(1 - \beta_n p_n\bigr)^{\eta_1 V \cdot \delta_n n}
\;\leq\; \exp\!\bigl(-\beta_n p_n \cdot \eta_1 V \cdot \delta_n n\bigr).
\]
Since $V = \Theta(T_n^{q+1})$ and $\beta_n p_n \delta_n n \cdot T_n^{q+1} \to \infty$ by Assumption~\ref{ass:beta-corr}, the exponent diverges to $-\infty$. Therefore,
\[
\PP(\text{at least one successful escape} \mid \calA) \;\geq\; 1 - \exp\!\bigl(-\eta_1 \beta_n p_n \delta_n n \cdot V\bigr) \;\to\; 1.
\]
Combining with $\PP(\calA) \geq \varepsilon > 0$, the unconditional probability of escape is at least $\varepsilon \cdot (1 - o(1)) > 0$ for all $n$ sufficiently large.

\medskip
\noindent\textit{Step 2: Diffusion from $i_0$ escapes via $E_n$.}
By an analogous construction, there exists a ball $B(k', \alpha T_n)$ reachable from $i_0$ in $T_n$ steps but not from $j_0$, where $k'$ is at distance $(1+\lambda)a_n$ from $j_0$. The same argument as in Step~1 applies: conditional on at least $V$ nodes being activated in $B(k', \alpha T_n)$ (probability $\geq \varepsilon$), the probability of at least one successful escape transmission converges to one. Hence the diffusion from $i_0$ escapes via $E_n$ with positive probability.

The two escape events are independent because the source nodes probing $E_n$ lie in disjoint subsets of $V_n$: nodes in $B(k, \alpha T_n)$ are unreachable from $i_0$ within $T_n$ steps (established in Step~1), and nodes in $B(k', \alpha T_n)$ are unreachable from $j_0$. Since missing links in $E_n$ are drawn i.i.d.\ across all pairs (Assumption~\ref{ass:beta-corr}), the Bernoulli draws at disjoint source sets are independent. Therefore,
\[
\Pr[\text{both escapes occur}]  > 0.
\]
By Lemma~\ref{lem:intratile}, the probability that an escape link lands within the initial tile vanishes, so the escapes reach distinct remote tiles with probability $1 - o(1)$.

\medskip
\noindent\textit{Step 3: Quantifying divergence via tile activations.}
Conditional on the diffusion from $i_0$ reaching node $e_1$ at time $s_1 := d(i_0, e_1)$ and escaping via $E_n$, the diffusion seeds a new wavefront with $T' := T_n - s_1 - 1$ remaining steps. The Tile Activation Growth lemma (Lemma~\ref{lem:regions-corr}) applies at the landing node because Assumption~\ref{ass:disease} holds for every node $j \in V_n$, so volume growth and uniform spread are satisfied regardless of where the jump link lands. By Property~4 of the Tiling Lemma, core nodes comprise a $\Theta(1)$ fraction of $V_n$; under i.i.d.\ link formation, the landing node falls in an unactivated tile core with probability $1-o(1)$. Applying Lemma~\ref{lem:regions-corr} with time horizon $T'$, the expected cumulative number of tiles activated is at least $C\,n\beta_n\delta_n p_n (T')^{q+1}$. Similarly for the diffusion from $j_0$ via $e_2$, with $s_2 := d(j_0, e_2)$.

These tile sets are disjoint with high probability: only a vanishing fraction $o(1)$ of tiles are ever activated from any given seed (Step~1 of the proof of Lemma~\ref{lem:regions-corr}), so the overlap is $o(K_n)$ tiles. Setting $s := \max\{s_1, s_2\}$,
\[
\EE[\text{disjoint tiles}] \geq 2C\,n\beta_n\delta_n p_n(T_n - s - 2)^{q+1}. 
\]
\end{proof}

\begin{proof}[Proof of Theorem~\ref{thm:main-polynomial}]
We establish asymptotic forecast failure by deriving an upper bound on $\hat{Y}_T(L_n)$ and a lower bound on $\hat{Y}_T(G_n)$, showing their ratio vanishes as $n \to \infty$. We work under Assumptions \ref{ass:forecast-time}, \ref{ass:disease}, and \ref{ass:beta-corr}.

\medskip
\noindent\textit{Step 1: Upper bound on the observed forecast.}
The econometrician's forecast $\hat{Y}_T(L_n)$ equals the expected number of nodes activated by time $T$ on the base network $L_n$ alone:
\[
\hat{Y}_T(L_n) = \calE_T = \EE_{P_n(L_n)}[|I(i_0, T, P_n(L_n))|].
\]
By Lemma~\ref{lem:diffusion}, $\calE_T = \Theta(T^{q+1})$, so in particular $\hat{Y}_T(L_n) = O(T^{q+1})$.

\medskip
\noindent\textit{Step 2: Lower bound on the true diffusion.}
The true diffusion count $\hat{Y}_T(G_n)$ accounts for both within-tile spread on $L_n$ and cross-tile jumps via missing links $E_n$. Apply Lemma~\ref{lem:tiles} to partition $L_n$ into tiles $\{k_1,\ldots,k_{K_n}\}$. Let $X_s$ denote the number of tiles activated for the first time at step $s$.

A tile activated at time $s$ subsequently spreads within itself via $L_n$. By time $T$, the diffusion has $(T-s)$ additional steps to propagate, activating an expected $\calE_{T-s} = \Theta((T-s)^{q+1})$ nodes by Lemma~\ref{lem:diffusion}. Summing over all tiles activated at all times $s \in \{0,1,\ldots,T-1\}$ (with $X_0 = 1$ for the initial seed tile),
\begin{align*}
\hat{Y}_T(G_n) 
&\geq \EE\left[\sum_{s=0}^{T-1} X_s \cdot \calE_{T-s}\right]\\
&\geq T^{q+1} + \sum_{s=1}^{T-1} \EE[X_s] \cdot (T-s)^{q+1},
\end{align*}
where the first term accounts for the seed tile and subsequent terms account for tiles activated via $E_n$.

\medskip
\noindent\textit{Step 3: Simplifying the lower bound via Abel summation.}
We bound the weighted sum $\sum_{s=1}^{T-1} \EE[X_s] \cdot (T-s)^{q+1}$ using summation by parts (Abel summation) and the cumulative bound from Lemma~\ref{lem:diffusion}. From the proof of Lemma~\ref{lem:regions-corr}, the conditional bound $\EE[X_s \mid \calS_{s-1}] \geq C |\calS_{s-1}| \delta_n n \beta_n p_n$ holds pointwise, so
\[
\sum_{s=1}^{T-1} \EE[X_s] \cdot (T-s)^{q+1}
\;\geq\; C\,\delta_n n \beta_n p_n \cdot \sum_{s=1}^{T-1} \EE[|\calS_{s-1}|] \cdot (T-s)^{q+1}.
\]
Define $A(t) := \sum_{s=1}^{t} \EE[|\calS_{s-1}|] = \calE_{t-1}$ and $w(s) := (T-s)^{q+1}$. By summation by parts,
\[
\sum_{s=1}^{T-1} \EE[|\calS_{s-1}|] \cdot w(s) 
= A(T{-}1)\,w(T{-}1) + \sum_{t=1}^{T-2} A(t)\bigl[w(t) - w(t{+}1)\bigr].
\]
Since $w(t) - w(t+1) = (T-t)^{q+1} - (T-t-1)^{q+1} \geq (q+1)(T-t-1)^q \geq 0$ for $t \leq T-2$, and $A(t) = \calE_{t-1} \geq c\, (t-1)^{q+1}$ by Lemma~\ref{lem:diffusion}, the sum is bounded below by
\[
c\sum_{t=1}^{T-2} (t-1)^{q+1} \cdot (q+1)(T-t-1)^q
\;\geq\; c'  T^{2q+2}
\]
for a constant $c' > 0$, where the final inequality follows from the standard Beta-integral approximation $\sum_{t=1}^{T} t^{a}(T-t)^{b} = \Theta(T^{a+b+1})$. Therefore,
\[
\hat{Y}_T(G_n) 
\geq T^{q+1} + C'\, n\beta_n\delta_n p_n \cdot T^{2q+2}
= T^{q+1}\left(1 + C'\, n\beta_n\delta_n p_n\, T^{q+1}\right).
\]

\medskip
\noindent\textit{Step 4: Ratio diverges to zero.}
Combining the upper and lower bounds,
\[
\frac{\hat{Y}_T(L_n)}{\hat{Y}_T(G_n)} 
\leq \frac{C_1 T^{q+1}}{T^{q+1}\left(1 + C'\, n\beta_n\delta_n p_n\, T^{q+1}\right)}
= \frac{C_1}{1 + C'\, n\beta_n\delta_n p_n\, T^{q+1}}
\]
for a constant $C_1 > 0$ from the upper bound on $\calE_T$.

By Assumption~\ref{ass:beta-corr}, $\beta_n = \omega\!\left(\frac{1}{p_n \underline{T}_n^{q+1} \delta_n n}\right)$, so for any $T \geq \underline{T}_n$,
\[
n\beta_n\delta_n p_n T^{q+1} 
\geq n\beta_n\delta_n p_n \underline{T}_n^{q+1}
\to \infty
\]
as $n \to \infty$, directly by the $\omega$ condition. Therefore,
\[
\frac{\hat{Y}_T(L_n)}{\hat{Y}_T(G_n)} \to 0
\]
as $n \to \infty$ for any forecast horizon $T \in [\underline{T}_n, \overline{T}_n]$, establishing asymptotic forecast failure.
\end{proof}

\textit{Coupling convention.}
To compare diffusions on $G_n$ and $L_n$ on a common probability space,
we use the standard monotone coupling: for each undirected edge $\{u,v\}\in G_n$
draw i.i.d.\ uniforms $U_{u\to v},U_{v\to u}\sim \mathrm{Unif}(0,1)$ and declare
the directed edge $u\to v$ open iff $U_{u\to v}\le p_n$ (and similarly for $v\to u$).
This defines a percolation $P^\star:=P_n(G_n)$. After edge censoring produces
$L_n\subseteq G_n$, define $P^{\mathrm{obs}}$ as the restriction of $P^\star$
to edges in $L_n$. Under this coupling, the observed diffusion is always a sub-diffusion
of the true one, and events like $\{I(i_0,T_n, L_n)=I(i_0,T_n, G_n)\}$
are well-defined conditional on $P_n$.

\begin{proof}[Proof of Theorem~\ref{thm:no-failure-mar}]
Fix $i_0$ and $T_n$. For the sake of compact notation, we use $^\star$ to denote objects on $G_n$. Write $B^\star := B_{i_0}^\star(T_n)$ for the radius-$T_n$
ball around $i_0$ in the true graph $G_n$ (graph distance in $G_n$).
Let $H^\star$ denote the induced subgraph of $G_n$ on vertex set $B^\star$,
and let $E(H^\star)$ be its edge set.

\medskip\noindent
\textit{Step 1: A high-probability event on which $L_n$ and $G_n$ coincide
on the relevant region.}
Define the event
\[
\mathsf{Good}_n := \{ \text{no edge in }E(H^\star)\text{ is censored} \}.
\]
Under Assumption~\ref{ass:mar}, each edge of $G_n$ is deleted independently
with probability $\varepsilon_n$, hence
\[
\mathbb{P}(\mathsf{Good}_n^c)
= 1-(1-\varepsilon_n)^{|E(H^\star)|}
\le \varepsilon_n |E(H^\star)|,
\]
where the inequality is the union bound. By Assumption~\ref{ass:disease} part~1, $|B^\star| \leq a_{\max} T^{q+1}$ nodes and the maximum degree is at most $a_{\max} - 1$ (taking $t=1$ in the volume bound), so $|E(H^\star)| \leq \frac{a_{\max}-1}{2} \cdot a_{\max} T^{q+1}$.

Combining (and absorbing $\frac{a_{\max}-1}{2}$ into the constant):
\[
\mathbb{P}(\mathsf{Good}_n^c)\le C_E\,\varepsilon_n T_n^{q+1}\to 0
\]
by Assumption~\ref{ass:mar}. Thus $\mathbb{P}(\mathsf{Good}_n)\to 1$.

\medskip\noindent
\textit{Step 2: On $\mathsf{Good}_n$, the activated sets coincide.}
Under the coupling convention stated above, the observed percolation
$P^{\mathrm{obs}}$ is the restriction of the true percolation $P^\star$ to
edges that survive censoring. On $\mathsf{Good}_n$ we have
$L_n[ B^\star ] = G_n[ B^\star ]$ (same induced subgraph on $B^\star$),
hence also $P^{\mathrm{obs}}[B^\star]=P^\star[B^\star]$.

Moreover, any node infected by time $T_n$ in the true diffusion must lie in $B^\star$:
if $v\in I(i_0,T_n, G_n)$, then there exists a directed open path of length at most $T_n$
from $i_0$ to $v$ in $P^\star$, hence an undirected path of length at most $T_n$ in
$G_n$, so $v\in B_{i_0}^\star(T_n)=B^\star$. The same containment holds for the
observed diffusion since $L_n\subseteq G_n$.

Therefore, on $\mathsf{Good}_n$ both diffusions are completely determined by the
same percolation restricted to the same induced subgraph on $B^\star$, implying
\[
I(i_0,T_n, G_n)=I(i_0,T_n, L_n)\quad\text{on }\mathsf{Good}_n.
\]
Consequently,
\[
\mathbb{P}\bigl(I(i_0,T_n, G_n)=I(i_0,T_n, L_n)\mid P_n\bigr)
\ge \mathbb{P}(\mathsf{Good}_n)\to 1.
\]

\medskip\noindent
\textit{Step 3: Symmetric-difference and expectation ratios.}
For the sake of compact notation, let $I^\star = I(i_0,T_n, G_n)$ and $I^{\mathrm{obs}} = I(i_0,T_n, L_n)$, both using the same percolation $P_n$. Since the two sets are equal with probability tending to one, we have
\[
\frac{|I^\star(i_0,T_n)\triangle I^{\mathrm{obs}}(i_0,T_n)|}{|I^\star(i_0,T_n)|}
\xrightarrow{\mathbb{P}} 0.
\]

For the expectation ratio, note that on $\mathsf{Good}_n$ the difference is zero, and
always $|I^\star(i_0,T_n)|\le |B^\star|\le a_{\max}T_n^{q+1}$. Hence
\[
0\le \mathbb{E}\!\left[|I^\star|-|I^{\mathrm{obs}}|\right]
= \mathbb{E}\!\left[(|I^\star|-|I^{\mathrm{obs}}|)\mathbf{1}\{\mathsf{Good}_n^c\}\right]
\le a_{\max}T_n^{q+1}\,\mathbb{P}(\mathsf{Good}_n^c).
\]
Using the bound from Step 1 gives
\[
\mathbb{E}\!\left[|I^\star|-|I^{\mathrm{obs}}|\right]
\le a_{\max}T_n^{q+1}\cdot C_E\varepsilon_nT_n^{q+1}.
\]
By Lemma~\ref{lem:diffusion} applied to $G_n$ (which satisfies
Assumption~\ref{ass:disease}), we have $\mathbb{E}[|I^\star(i_0,T_n)|]=\Theta(T_n^{q+1})$,
so dividing by $\mathbb{E}[|I^\star|]$ yields
\[
0\le 1-\frac{\mathbb{E}[|I^{\mathrm{obs}}|]}{\mathbb{E}[|I^\star|]}
= \frac{\mathbb{E}[|I^\star|-|I^{\mathrm{obs}}|]}{\mathbb{E}[|I^\star|]}
= O(\varepsilon_nT_n^{q+1})\to 0.
\]
Therefore,
\[
\frac{\mathbb{E}[|I^{\mathrm{obs}}(i_0,T_n)|]}{\mathbb{E}[|I^\star(i_0,T_n)|]}\to 1.
\]
This completes the proof.
\end{proof}


\begin{proof}[Proof of Proposition~\ref{prop:sampling_beta}]
We work under Assumptions~\ref{ass:forecast-time},~\ref{ass:disease}, and~\ref{ass:beta}, so that missing links form independently across all pairs with probability $\beta_n \in (1/(p_n n T_n^{q+1}), 1/n)$. The i.i.d. structure provides a best-case scenario for detection, as the econometrician need not account for heterogeneity in where missing links are likely to occur.

\textit{Part (i): Detection failure.} 
Suppose $m_n = o(\sqrt{n})$. Since $\beta_n < 1/n$ by Assumption~\ref{ass:beta}, we have $\beta_n m_n = o(1/\sqrt{n})$ and $\beta_n m_n^2 = o(1)$. The probability of finding no missing links among the $\binom{m_n}{2}$ sampled pairs is
\begin{align*}
    \mathbb{P}\left(\text{No links from } E_n \text{ found}\right) 
    &= (1-\beta_n)^{\binom{m_n}{2}} \\
    &\approx \exp\left(-\beta_n \binom{m_n}{2}\right) \\
    &= \exp\left(-\beta_n \frac{m_n^2 - m_n}{2}\right) \\
    &= \exp\left(-\frac{\beta_n m_n^2}{2} + O(\beta_n m_n)\right) \\
    &\to \exp(0) = 1,
\end{align*}
where the first approximation uses $(1-x)^k \approx \exp(-kx)$ for small $x$, valid since $\beta_n \binom{m_n}{2} = o(1)$, and the final limit follows because $\beta_n m_n^2 \to 0$ and $\beta_n m_n \to 0$. 

Note that this result holds even when $\beta_n$ takes its maximum admissible value of $1/n$ under Assumption~\ref{ass:beta}, demonstrating that detection failure is not an artifact of considering unrealistically small error rates.

\textit{Part (ii): Estimation failure.} 
Suppose $m_n = O(1/\sqrt{\beta_n})$, so that $m_n^2 \beta_n = O(1)$. We establish that no consistent estimator can exist by showing that the sample variance fails to vanish, violating a necessary condition for the law of large numbers.

Consider the normalized indicator $z_{ij}^n := e_{ij}^n / \beta_n$ for each potential link $(i,j)$, where $e_{ij}^n \in \{0,1\}$ indicates whether the link exists in $E_n$. Note that $\mathbb{E}[z_{ij}^n] = 1$ and $\text{Var}(z_{ij}^n) = (1-\beta_n)/\beta_n$. The natural estimator of the average $\mathbb{E}[z_{ij}^n]$ based on the sampled pairs is
\begin{equation*}
    \bar{z}_n := \frac{2}{m_n(m_n-1)} \sum_{\substack{i,j : s_{ij} = 1 \\ i < j}} z_{ij}^n,
\end{equation*}
where $s_{ij} = 1$ indicates that both nodes $i$ and $j$ were sampled. Since the indicators $\{e_{ij}^n\}$ are independent across pairs, the variance is
\begin{align*}
    \text{Var}(\bar{z}_n) 
    &= \left(\frac{2}{m_n(m_n-1)}\right)^2 \binom{m_n}{2} \cdot \frac{1-\beta_n}{\beta_n} \\
    &= \frac{2(1-\beta_n)}{\beta_n \cdot m_n(m_n-1)} \\
    &= \frac{2(1-\beta_n)}{m_n^2 \beta_n - m_n \beta_n}.
\end{align*}
For consistency, we require $\text{Var}(\bar{z}_n) \to 0$, which holds if and only if $m_n^2 \beta_n \to \infty$. However, by hypothesis $m_n^2 \beta_n = O(1)$, so the variance remains bounded away from zero.

To extend beyond the sample mean to arbitrary estimators, we apply Le Cam's two-point method. Consider hypotheses $H_0\colon \beta = \beta_n$ and $H_1\colon \beta = (1+3\epsilon)\beta_n$. Under $H_0$ the expected link count is $\lambda_0 := \binom{m_n}{2}\beta_n = O(1)$; under $H_1$ it is $\lambda_1 = (1+3\epsilon)\lambda_0 = O(1)$. Since $\lambda_0 = O(1)$, the squared Hellinger distance $H^2(P_0, P_1)$ between the two Binomial laws is bounded by a constant strictly less than 1. By Le Cam's lemma, for any estimator $\hat\beta_n$,
\begin{equation*}
\max\!\left\{\Pr_{H_0}\!\left(\left|\frac{\hat\beta_n}{\beta_n} - 1\right| \geq \epsilon\right),\; \Pr_{H_1}\!\left(\left|\frac{\hat\beta_n}{(1{+}3\epsilon)\beta_n} - 1\right| \geq \epsilon\right)\right\} \geq \frac{1 - H(P_0, P_1)}{2} > 0.
\end{equation*}
The accuracy intervals $[(1-\epsilon)\beta_n, (1+\epsilon)\beta_n]$ and $[(1+2\epsilon)\beta_n, (1+4\epsilon)\beta_n]$ are disjoint for small $\epsilon$, so no estimator can be $\epsilon$-accurate under both hypotheses. In particular, $\Pr_{H_0}(|\hat\beta_n/\beta_n - 1| \geq \epsilon) \geq 1 - c$ for some $c < 1$ independent of $n$.
\end{proof}

\begin{proof}[Proof of Theorem~\ref{thm:regionalDetection}]
We work under Assumptions~\ref{ass:forecast-time},~\ref{ass:disease}, and~\ref{ass:beta}. Apply the tiling construction from Lemma~\ref{lem:tiles} to partition the network into disjoint spatial regions (tiles). Under the testing regime, the policymaker observes each activated node independently with probability $\gamma_n$.

\textit{Step 1: Detection probability for a single tile.}
Consider a tile containing $x$ activated nodes at time $T$. The probability that at least one activation is detected is
\begin{equation*}
    1 - (1-\gamma_n)^x \leq \gamma_n x,
\end{equation*}
by the union bound.

\textit{Step 2: Activated nodes per tile.}
A tile first activated at time $s \leq T$ has had $T - s$ periods of local spread. By Lemma~\ref{lem:diffusion} and the tiling construction, such a tile contains at most $a_{\max}(T-s)^{q+1}$ activated nodes at time $T$. By Step~1, its detection probability is therefore at most $\gamma_n \cdot a_{\max}(T-s)^{q+1}$.

\textit{Step 3: Expected number of detected tiles.}
Let $X_s$ denote the number of tiles newly activated at time $s$. Since $(T-s)^{q+1} \leq T^{q+1}$ for all $s \geq 0$, and $K_T^\star = \sum_{s=1}^T \EE[X_s]$,
\begin{align*}
    \EE[\hat{K}_T]
    &\leq \sum_{s=1}^{T} \EE[X_s] \cdot \gamma_n \cdot a_{\max}(T-s)^{q+1} \\
    &\leq \gamma_n \cdot a_{\max} \cdot T^{q+1} \cdot \sum_{s=1}^{T} \EE[X_s] \\
    &= \gamma_n \cdot a_{\max} \cdot T^{q+1} \cdot K_T^\star.
\end{align*}

\textit{Step 4: Detection ratio.}
Dividing both sides by $K_T^\star$,
\begin{equation*}
    \frac{\hat{K}_T}{K_T^\star} \leq \gamma_n \cdot a_{\max} \cdot T^{q+1} = \gamma_n \cdot O(T^{q+1}) < 1,
\end{equation*}
where the final inequality uses $\gamma_n = O(T^{-(q+1)})$ with a sufficiently small implied constant.
\end{proof}

\subsection{Random Geometric Graph Example}\label{sec:rgg-proof}

We prove that the random geometric graph of Example~\ref{ex:rgg} satisfies Assumption~\ref{ass:disease}. The argument proceeds in two parts: polynomial volume growth (Assumption~\ref{ass:disease}.1), established in Theorem~\ref{thm:rgg-growth}, and supercritical uniform spread (Assumption~\ref{ass:disease}.2), established in Theorem~\ref{thm:rgg-spread}.

\subsubsection*{Setup}
Fix dimension $d \geq 2$, intensity $\lambda > 0$, and connection radius $r > 0$ in the supercritical regime of continuum percolation.
For each $n$, let $W_n$ be a $d$-dimensional torus of volume $|W_n| = n/\lambda$.
Let $\calP_n$ be a homogeneous Poisson point process of intensity $\lambda$ on $W_n$, and define
\[
G_n^{\mathrm{rgg}} = (V_n, E_n^{\mathrm{rgg}}),
\qquad V_n := \calP_n, \qquad
E_n^{\mathrm{rgg}} := \bigl\{\{u,v\} \subset V_n : u \neq v,\ \|u-v\|_2 \leq r\bigr\},
\]
where $\|\cdot\|_2$ denotes toroidal Euclidean distance. Let $\calC_n$ be the unique giant component (which exists with probability tending to one under supercriticality), and let $d_{G_n}(\cdot,\cdot)$ denote graph distance. For $v \in \calC_n$ and integer $t \geq 0$, define $B_n(v,t) := \{u \in \calC_n : d_{G_n}(u,v) \leq t\}$.

\begin{theorem}[Typical polynomial volume growth]\label{thm:rgg-growth}
For every fixed $C < \infty$ and every $\varepsilon > 0$, there exist constants $0 < a_{\min} \leq a_{\max} < \infty$ such that, with probability tending to one,
\[
a_{\min}\, t^d \;\leq\; |B_n(v,t)| \;\leq\; a_{\max}\, t^d
\]
for all $1 \leq t \leq (\log |V_n|)^C$, for at least a $(1-\varepsilon)$ fraction of vertices $v \in \calC_n$.
\end{theorem}

\begin{theorem}[Supercritical uniform spread]\label{thm:rgg-spread}
Fix $p_{\mathrm{tr}} \in (0,1)$ such that $p := p_{\mathrm{tr}}^2 > p_c(\lambda,r)$, where $p_c$ is the critical percolation threshold for the random connection model. For every fixed $C < \infty$ and every $\varepsilon > 0$, there exist constants $\alpha, \theta_{\mathrm{spr}}, \rho \in (0,1)$ such that, with probability tending to one, for at least a $(1-\varepsilon)$ fraction of seeds $v \in \calC_n$ and all $1 \leq t \leq (\log |V_n|)^C$,
\[
\PP\bigl(\,|I_n(v,t) \cap B_n(z,\alpha t)| \;\geq\; \theta_{\mathrm{spr}}\,|B_n(z,\alpha t)|\;\bigm|\; G_n^{\mathrm{rgg}}\bigr) \;\geq\; \rho
\]
for all $z \in B_n\bigl(v,(1-\alpha)t\bigr)$, where $I_n(v,t)$ is the SIR activation set.
\end{theorem}

\subsubsection*{Geometric discretization}

\begin{lemma}[Geometric tessellation]\label{lem:rgg-tessellation}
There exists $K < \infty$ such that, for each $n$, setting $s_n := \max\{s' \leq r/K : (n/\lambda)^{1/d}/s' \in \mathbb{Z}\}$ and partitioning $W_n$ into cubes of side length $s_n$ yields (for $n$ sufficiently large, so that $s_n \geq r/(K+1)$):
\begin{description}
\item[(P1)] If two cubes share a face, then any point in one is within Euclidean distance $r$ of any point in the other.
\item[(P2)] If $\{u,v\} \in E_n^{\mathrm{rgg}}$, then the cubes containing $u$ and $v$ differ by at most $K+1$ in $\ell_\infty$ distance.
\item[(P3)] The induced subgraph on any single cube is a clique.
\end{description}
\end{lemma}
\begin{proof}
Choose $K \geq 2\sqrt{d}$ and set $s = r/K$.
If two cubes share a face, their maximum coordinate-wise separation is at most $2s$, hence Euclidean separation is at most $2s\sqrt{d} \leq r$, giving (P1).
If $\{u,v\} \in E_n^{\mathrm{rgg}}$ then $\|u-v\|_2 \leq r = Ks$, so their cube indices differ by at most $K+1$ in $\ell_\infty$ distance, giving (P2).
Any two points in a cube are at distance at most $\sqrt{d}\,s \leq r/2 < r$, giving (P3).
\end{proof}

\subsubsection*{Block renormalization}

Partition $W_n$ into \textbf{blocks} of $L^d$ micro-cubes from Lemma~\ref{lem:rgg-tessellation}, each of side length $\ell := Ls$. Here $L = L(n)$ grows slowly with $n$ (e.g., $L = (\log\log n)^{C_1}$); see the proof of Theorem~\ref{thm:rgg-spread} for the precise choice. For a block $B$, let $B^+ := \{x \in W_n : \mathrm{dist}_\infty(x,B) \leq r\}$ denote its $r$-enlargement.

A block $B$ is \emph{Good} if, in the induced subgraph of $G_n^{\mathrm{rgg}}$ on $V_n \cap B^+$: (i) there exists a unique crossing component $\calK(B)$ intersecting every boundary slab of $B$; (ii) $|\calK(B) \cap B| \geq (\theta - \delta)\lambda\,\mathrm{Vol}(B)$ for a parameter $\delta \in (0,\theta/4)$, where $\theta = \theta(\lambda,r)$ is the infinite-cluster density; (iii) intrinsic diameter of $\calK(B) \cap B$ is at most $\beta L$; and (iv) $\calK(B)$ intersects at least $\delta L^{d-1}$ micro-cubes in each boundary slab. These are standard finite-box regularity properties in the supercritical regime \citep{Penrose2003}.

For face-adjacent blocks $B_1, B_2$, the bond $(B_1, B_2)$ is \emph{Open} if both are Good and their crossing components are connected within $(B_1 \cup B_2)^+$ by a path of length at most $\beta L$.

\begin{lemma}[Supercritical domination]\label{lem:rgg-lss}
For every $\varepsilon > 0$ there exists $L_0 < \infty$ such that if $L \geq L_0$, then the Open-bond field stochastically dominates i.i.d.\ Bernoulli bond percolation on the block lattice with parameter $q_{\mathrm{LSS}} \geq 1 - \varepsilon$.
\end{lemma}
\begin{proof}
By the supercritical finite-box theory for continuum percolation \citep{PenrosePisztora1996}, for any $\eta > 0$ one can choose $\delta$ small and $L$ large so that $\inf_e \PP(\omega(e) = 1) \geq 1 - \eta$ uniformly in $n$. The Open-bond field is $k$-dependent (for fixed $k = k(d)$) since $\omega(e)$ depends only on the Poisson configuration in $(B_1 \cup B_2)^+$. The Liggett--Schonmann--Stacey domination theorem \citep{LSS1997} then yields stochastic domination by i.i.d.\ Bernoulli-$q_{\mathrm{LSS}}$ percolation with $q_{\mathrm{LSS}} \uparrow 1$ as $\eta \downarrow 0$.
\end{proof}

Define the \emph{strong core} $V_n^{\mathrm{core}} := \bigcup_{B \in \calG_n^{\mathrm{strong}}} (\calK(B) \cap B)$, where $\calG_n^{\mathrm{strong}}$ is the giant component of the Open-bond block lattice.

\begin{lemma}[Core density]\label{lem:rgg-core}
For any $\eta > 0$, with suitable choice of $\delta$ and $L$, with high probability $|V_n^{\mathrm{core}}| \geq (1-\eta)|\calC_n|$.
\end{lemma}
\begin{proof}
By Lemma~\ref{lem:rgg-lss}, for $L$ large the Open-bond process dominates i.i.d.\ bond percolation with parameter close to $1$. The giant bond-percolation cluster occupies a $(1-\eta/4)$ fraction of blocks with high probability. Each such block contributes at least $(\theta - \delta)\lambda\,\mathrm{Vol}(B)$ vertices to the core. Since $|\calC_n|/|V_n| \to \theta$, choosing $\delta$ and $\eta$ small yields $|V_n^{\mathrm{core}}| \geq (1-\eta)|\calC_n|$ w.h.p.
\end{proof}

\subsubsection*{Proof of Theorem~\ref{thm:rgg-growth}}

\begin{proof}
\textit{Upper bound.}
A path of $t$ edges has Euclidean length at most $rt$, so $B_n(v,t) \subseteq V_n \cap B_{\RR^d}(v,rt)$. By Palm theory for Poisson processes, $|V_n \cap B_{\RR^d}(v,rt)| - 1 \sim \mathrm{Poisson}(\mu_t)$ with $\mu_t = \lambda\,\mathrm{Vol}(B_{\RR^d}(0,rt)) = \Theta(t^d)$. Standard Chernoff bounds and a second-moment argument (see \citealt{Penrose2003}, Ch.~3) yield that for $a_{\max}$ large enough, $|B_n(v,t)| \leq a_{\max} t^d$ for all $t \leq (\log n)^C$ and at least a $(1-\varepsilon/2)$ fraction of $v \in \calC_n$, w.h.p.

\medskip
\textit{Lower bound.}
By Lemma~\ref{lem:rgg-lss}, graph-distance balls in $G_n^{\mathrm{rgg}}$ contain block-percolation balls: if $v \in V_n^{\mathrm{core}}$ and $d_{\mathrm{strong}}(B_v, B') \leq m$, then $\calK(B') \cap B' \subseteq B_n(v, \beta L + 2\beta L m)$. To see this, consider a path of $m$ Open bonds from $B_v$ to $B'$ in the block lattice. Each Open bond $(B_i, B_j)$ certifies that $\calK(B_i)$ and $\calK(B_j)$ are connected within $(B_i \cup B_j)^+$ by a path of at most $\beta L$ edges (by definition of Open). Within each Good block $B_i$, the crossing component $\calK(B_i)$ has intrinsic diameter at most $\beta L$ (by definition of Good, property~(iii)). Concatenating: from $v$ to $\calK(B_v)$'s boundary costs $\leq \beta L$; each of the $m$ block transitions costs $\leq \beta L$ (crossing path) $+ \beta L$ (traversal within the next block) $= 2\beta L$; total $\leq \beta L + 2\beta L m$. Setting $m(t) := \max\{0, \lfloor(t - \beta L)/(2\beta L)\rfloor\}$, we get
\[
|B_n(v,t)| \;\geq\; \sum_{B':\, d_{\mathrm{strong}}(B_v,B') \leq m(t)} |\calK(B') \cap B'|.
\]
By the LSS coupling (Lemma~\ref{lem:rgg-lss}), $|\mathsf{Ball}_\omega(B_v, m)| \geq |\mathsf{Ball}_{q_{\mathrm{LSS}}}(B_v, m)|$. Standard chemical-distance estimates for supercritical Bernoulli percolation on $\ZZ^d$ (\citealt{AntalPisztora1996}; large-deviation theory from \citealt{PenrosePisztora1996}) imply that for at least a $(1-\varepsilon/4)$ fraction of blocks, $|\mathsf{Ball}_{q_{\mathrm{LSS}}}(B_v, m)| \geq c_1 m^d$ for all $1 \leq m \leq (\log n)^C$. Since each Good block contributes $\geq (\theta-\delta)\lambda(Ls)^d$ vertices to its crossing component, we obtain $|B_n(v,t)| \geq a_{\min}\,t^d$ after absorbing constants. Combining with the core density (Lemma~\ref{lem:rgg-core}) gives the result for a $(1-\varepsilon)$ fraction of $\calC_n$.
\end{proof}

\subsubsection*{Proof of Theorem~\ref{thm:rgg-spread}}

\begin{proof}
Let $H_n^{(p)}$ be the bidirectional percolated subgraph of $G_n^{\mathrm{rgg}}$, retaining each edge independently with probability $p = p_{\mathrm{tr}}^2$. Since $p > p_c(\lambda,r)$, $H_n^{(p)}$ is supercritical, and the block renormalization of the previous subsection applies with $G_n^{\mathrm{rgg}}$ replaced by $H_n^{(p)}$.

\medskip\noindent
\textit{Step 1: Quenched supercriticality.} Define a block to be \emph{$p$-Good} and a bond to be \emph{$p$-Open} by the same criteria applied to $H_n^{(p)}$. Let $L = L(n)$ grow slowly with $n$ (e.g., $L = (\log\log n)^{C_1}$ for a large constant $C_1$), with $L \to \infty$ and $L = o(\log n)$. By the supercritical finite-box theory for continuum percolation \citep{PenrosePisztora1996}, the annealed per-bond failure probability decays exponentially in the block surface area:
\[
\PP(\omega_p(e) = 0) \;\leq\; \exp(-c\, L^{d-1})
\]
for a constant $c > 0$ depending on $p, \lambda, r$, and $d$ but not on $L$ or $n$.

The proof requires quenched supercriticality only in the \emph{local} neighborhood of a seed---a ball of polylogarithmic radius in the block lattice containing $N_{\mathrm{bonds}} = O((\log n)^{Cd})$ bonds. By Markov's inequality, the quenched conditional probability satisfies $\PP(\omega_p(e) = 0 \mid G_n^{\mathrm{rgg}}) \leq \exp(-c'\, L^{d-1})$ except on an event of probability $\leq \exp(-c'\, L^{d-1})$. A union bound over $N_{\mathrm{bonds}}$ bonds gives total failure probability
\[
N_{\mathrm{bonds}} \cdot \exp(-c'\, L^{d-1})
\;=\; O\!\left((\log n)^{Cd} \cdot \exp\bigl(-c'(\log\log n)^{C_1(d-1)}\bigr)\right)
\;\to\; 0,
\]
since exponential decay in $(\log\log n)^{C_1(d-1)}$ dominates polynomial growth in $\log n$. The $p$-Open field is $k$-dependent with $k = k(d)$ bounded independently of $L$ (the dependency range in block-lattice units is $2 + 2r/L \to 2$ as $L \to \infty$). The LSS domination theorem (Lemma~\ref{lem:rgg-lss}) therefore yields stochastic domination by i.i.d.\ Bernoulli-$q_{\mathrm{block}}$ with $q_{\mathrm{block}} > p_c(\ZZ^d)$. Chemical-distance estimates \citep{AntalPisztora1996} hold with constants depending on $q_{\mathrm{block}}$ and $d$ but not on $L$. This holds w.h.p.\ over $G_n^{\mathrm{rgg}}$ for a $(1-\varepsilon)$ fraction of seeds.

\medskip\noindent
\textit{Step 2: Block geometry implies spread.} Under i.i.d.\ Bernoulli-$q_{\mathrm{block}}$ percolation, chemical-distance control \citep{AntalPisztora1996} and the positive cluster density ensure that, with probability $\geq \rho_0 > 0$, the intrinsic ball of radius $m$ around the seed-block intersects a positive fraction of blocks in every sub-ball of relative radius $\alpha$. Since each $p$-Good block contributes a $(\theta_p - \delta)$ fraction of its vertices to its crossing component, the bound
\[
|I_n(v,t) \cap B_n(z,\alpha t)| \;\geq\; |B_{H_n^{(p)}}(v,t) \cap B_n(z,\alpha t)| \;\geq\; \theta_{\mathrm{spr}}\,|B_n(z,\alpha t)|
\]
holds with conditional probability $\geq \rho := \rho_0$ for all $z \in B_n(v, (1-\alpha)t)$.
\end{proof}

\subsubsection*{Mapping to Assumption~\ref{ass:disease}}
Take $L_n := \calC_n$ (the giant component), set $q = d-1$ so that $t^{q+1} = t^d$, and take the forecast horizon $\overline{T}_n = (\log n)^{C_0}$ for any fixed $C_0$. Then $\overline{T}_n = o(n^{1/(2q+3)})$ for any $q > 0$, satisfying Assumption~\ref{ass:forecast-time}. Theorem~\ref{thm:rgg-growth} verifies Assumption~\ref{ass:disease}.1 and Theorem~\ref{thm:rgg-spread} verifies Assumption~\ref{ass:disease}.2, both at polylogarithmic horizons and for a $(1-\varepsilon)$ fraction of seeds in $\calC_n$.

\begin{remark}[Inhomogeneous intensities]\label{rem:rgg-inhomogeneous}
The argument extends to inhomogeneous Poisson point processes with intensity $\lambda(x)$ satisfying $0 < \underline{\lambda} \leq \lambda(x) \leq \overline{\lambda} < \infty$ uniformly. All concentration bounds and the coarse-graining step hold with constants depending on $(\underline{\lambda}, \overline{\lambda})$.
\end{remark}

\newpage

\setcounter{figure}{0}  
\setcounter{table}{0}  
\begin{center}
    \Huge Online Appendix
\end{center}

\section{Extension to the Exponential Case}\label{sec:extensions}

We turn to the case of exponential expansion. When the base network $L_n$ itself has exponential volume growth, the forecast window is compressed: diffusion spreads rapidly even without measurement error. However, this does not make forecasting irrelevant---the COVID-19 pandemic exhibited approximately exponential growth yet unfolded over months, a window in which forecasts informed consequential policy decisions. We show that even within this compressed window, measurement error generates forecast failure, though under stronger conditions on $\beta_n$ than in the polynomial case. The comparison is itself informative: the polynomial case requires only $\beta_n = \omega(1/(p_n \underline{T}_n^{q+1} n))$, while the exponential case demands $\beta_n = \omega(1/(p_n n \underline{T}_n))$---a stronger lower bound (slower decay to zero) reflecting the fact that exponential expansion leaves less room for jump links to create qualitatively new behavior. This comparison quantifies how network structure modulates vulnerability to measurement error: networks with slower expansion are more fragile.

We make assumptions that correspond to Assumptions \ref{ass:forecast-time}, \ref{ass:disease} and \ref{ass:beta}, to account for the faster-moving diffusion process. As before, we assume that each node $i$ can link to a fraction of nodes $\delta_n$ of the graph through $E_n$. 

\begin{ass}\label{ass:disease-exp}
    Fix $q>1$. For each $n$, let $L_n$ be a graph with degree in $[d_{min}, d_{max}]$ with $d_{min}\geq 2$ and $d_{max} < \infty$. The following two conditions hold: 
    \begin{enumerate}
        \item \textbf{Exponential volume growth.} For every node $j$ and every integer $t\geq 1$: 
        \begin{align*}
            z_{min} q^t \leq |B_j(t)| \leq z_{max} q^{t}
        \end{align*}
        where $z_{min} = (d_{min} + 1)/q$ and $z_{max} = (d_{max} + 1)/q$, so that the bounds hold at $t = 1$: $z_{min}\, q = d_{min} + 1 = |B_j(1)|_{\min}$ and similarly for the upper bound.\footnote{The normalization by $q$ ensures consistency at $t=1$ when $q > 1$, which would be violated by the unnormalized choice $z_{\min} = d_{\min}+1$.}
        \item \textbf{Passing probability and global spread.} The passing probability $p_n > \frac{1}{d_{min}}$. Furthermore, there exist constants $\theta \in (0,1)$ and $\varepsilon > 0$ such that for any seed $j$ and for all $T \leq \overline{T}_n$:
\begin{align*}
    \PP\bigl(|I(j, T)| > \theta|B_j(T)|\bigr) \geq \varepsilon > 0.
\end{align*}
    \end{enumerate}
\end{ass}
The first part of Assumption \ref{ass:disease-exp} considers a graph with exponential expansion, in comparison to the polynomial expansion in Assumption \ref{ass:disease}. The second part strengthens the sub-ball filling condition used in the polynomial case to a \emph{global} filling condition: the diffusion starting from any seed $j$ activates a positive fraction $\theta$ of the entire ball $B_j(T)$ with probability bounded away from zero. This is strictly stronger than requiring only that sub-balls of radius $\alpha T$ are filled (Part~2 of Assumption~\ref{ass:disease}), but it is necessary to obtain the $\Theta(q^t)$ lower bound on expected activations under exponential volume growth, where sub-balls comprise an exponentially vanishing fraction of the full ball.\footnote{Under polynomial growth, the sub-ball condition suffices because $|B(\alpha t)|/|B(t)| = \Theta(\alpha^{q+1})$, a constant. Under exponential growth, $|B(\alpha t)|/|B(t)| = \Theta(q^{-(1-\alpha)t}) \to 0$, so the sub-ball condition alone yields only $\calE_t = \Omega(q^{\alpha t})$, not $\Theta(q^t)$.}

\begin{ass}\label{ass:forecast-time-exp} 
For every $n$, $T_n\in [\underline T_n, \overline T_n]$ where: 
 (1) $\overline T_n = (\log(n))^{1/q}$ and  (2) $\underline T_n =\omega(1)$.
\end{ass}

We then note the difference in the bounds on $T_{n}$: we impose a smaller upper bound on time than in the polynomial case. The upper bound determines how ``far'' and how saturated the diffusion process can be. The smaller upper bound on $T_{n}$ is intuitive: because the diffusion spreads more quickly, the seeds from idiosyncratic links can cause the diffusion to explode much more quickly. 

\begin{ass}\label{ass:beta-exp}
For all $n, i, j$: $E_{ij}\mathop{\sim}\limits^{\mathrm{iid}} \mathrm{Ber}(\beta_n)$ with $\beta_n  = \omega\left(\frac{1}{p_n n \underline T_n}\right), \beta_n = O\left(\frac{1}{n}\right) $. 
\end{ass}

This assumption is analogous to Assumption \ref{ass:beta}, but the rates change in the exponential case. Namely, the lower bound is less permissive, as it only contains a linear (rather than polynomial) function of $T$.

\begin{theorem}\label{thm:main-exponential}
Under Assumptions \ref{ass:disease-exp}, \ref{ass:forecast-time-exp}, and \ref{ass:beta-exp} as $n\rightarrow\infty$, $\frac{\hat Y_T(L_n)}{\hat Y_T(G_n)}\rightarrow 0.$
\end{theorem}

We make a few comparisons to our previous result. Relative to Theorem \ref{thm:main-polynomial}, we impose a stronger lower bound on $\beta_n$ -- in order for similar results to hold, we require a larger probability of idiosyncratic links. This change follows from the structure of the proof -- the key comparison is the expansion in all of the areas ``seeded" via the idiosyncratic links compared to the expansion of the original diffusion process. When the original diffusion process is faster moving, it means that more idiosyncratic links are needed to overwhelm the original diffusion. 

In proving Theorem \ref{thm:main-exponential}, the exponential geometry means we cannot construct the disjoint tiles we used in the polynomial case. Instead, we consider balls around nodes with links in $E_n$ that transmit the diffusion, which may overlap. We bound the expected overlap, and show that it is an order of magnitude smaller than the total volume of diffusion accounting for $E_n$.

\begin{lemma}\label{lem:diffusion-exp}
    Let Assumptions \ref{ass:disease-exp} and \ref{ass:forecast-time-exp} hold. Then: 
    \begin{align*}
        \calE_t = \Theta(q^t).
    \end{align*}
\end{lemma}

\begin{proof}[Proof of Lemma \ref{lem:diffusion-exp}]
\textit{Upper bound.} When $p_n = 1$, every edge transmits with certainty.  The activated set is contained in $B_j(t)$, so $\calE_t \leq |B_j(t)| \leq z_{\max}\, q^t = O(q^t)$ by Assumption~\ref{ass:disease-exp}, part~1.

\textit{Lower bound.} By Assumption~\ref{ass:disease-exp}, part~2 (global spread), for any seed $j$ and any $t \leq \overline{T}_n$:
\[
\PP\bigl(|I(j,t)| \geq \theta\, |B_j(t)|\bigr) \geq \varepsilon > 0.
\]
By Assumption~\ref{ass:disease-exp}, part~1, $|B_j(t)| \geq z_{\min}\, q^t$. Therefore,
\[
\calE_t = \EE[|I(j,t)|] \geq \PP\bigl(|I(j,t)| \geq \theta\, z_{\min}\, q^t\bigr) \cdot \theta\, z_{\min}\, q^t \geq \varepsilon\, \theta\, z_{\min}\, q^t = \Omega(q^t),
\]
where $\varepsilon, \theta, z_{\min} > 0$ are constants independent of $n$ and $t$. Combining the upper and lower bounds yields $\calE_t = \Theta(q^t)$.
\end{proof}

\begin{proof}[Proof of Theorem \ref{thm:main-exponential}]
We establish the result by deriving an upper bound on the forecast $\hat{Y}_T(L_n)$ and a lower bound on the true diffusion $\hat{Y}_T(G_n)$, showing their ratio vanishes as $n \to \infty$.

\medskip
\noindent\textit{Step 1: Upper bound on the observed forecast.}
The forecast $\hat{Y}_T(L_n)$ is the expected number of nodes activated by a single seed $i_0$ within time $T$ on $L_n$. By Assumption~\ref{ass:disease-exp} (Exponential Volume Growth), there exists a constant $z_{\max} > 0$ such that the volume of any ball of radius $T$ satisfies $|B(T)| \le z_{\max} q^T$. Thus:
\begin{equation}
\hat{Y}_T(L_n) = \calE_T \leq z_{\max} q^T.
\end{equation}

\medskip
\noindent\textit{Step 2: Lower bound on primary hub activation.}
We lower bound $\hat{Y}_T(G_n)$ by counting only the ``primary hubs'' activated directly by the initial wavefront of $i_0$ via links in $E_n$. Let $X_s$ denote the number of new primary hubs activated at time $s$.

Let $\mathcal{S}_{s-1}$ be the wavefront (newly activated nodes) at time $s-1$ on $L_n$. We condition on the wavefront size to determine the expected number of new hubs.
\begin{enumerate}
    \item Sources: The number of potential sources is $|\mathcal{S}_{s-1}|$.
    \item Targets: The number of susceptible nodes is $n - |A_{s-1}|$, where $A_{s-1}$ is the set of previously activated nodes. By Step 1, $|A_{s-1}| \leq z_{\max} q^{s-1}$. Since $T \leq (\log n)^{1/q}$, we have $q^T = o(n)$. Thus, for sufficiently large $n$, $n - |A_{s-1}| \geq n/2$.
    \item Success Rate: The probability of a link existing in $E_n$ and transmitting is $\beta_n p_n$.
\end{enumerate}
Let $\lambda_n = n \beta_n p_n$. Then via the law of iterated expectations, the conditional bound $\EE_{E_n}[X_s \mid \mathcal{S}_{s-1}] \geq (n/2) \beta_n p_n |\mathcal{S}_{s-1}|$ holds pointwise. Summing over $s$ and applying the tower property:
\begin{align*}
    \sum_{s=1}^{T}\EE[X_s] &\geq \frac{\lambda_n}{2} \sum_{s=1}^{T} \EE[|\mathcal{S}_{s-1}|] = \frac{\lambda_n}{2} \calE_{T-1} \geq C' \lambda_n q^{T-1},
\end{align*}
where $C' > 0$ uses $\calE_{T-1} = \Theta(q^{T-1})$ from Lemma~\ref{lem:diffusion-exp}. For per-step bounds needed in Step~4, we apply Abel summation (summation by parts): with $A(t) = \calE_{t-1} \geq c\, q^{t-1}$ and decreasing weights $w(s) = q^{T-s}$, the differences $w(s) - w(s+1) = q^{T-s-1}(q-1) \geq 0$ yield
\[
\sum_{s=1}^{T} \EE[|\calS_{s-1}|] \cdot q^{T-s} \geq c(q-1)\sum_{t=1}^{T-1} q^{t-1} q^{T-t-1} = c(q-1)(T-1)q^{T-2} = \Theta(Tq^{T-1}).
\]
Therefore $\sum_{s=1}^T \EE[X_s] \cdot \calE_{T-s} \geq C\lambda_n T q^{T-1}$ for a constant $C > 0$.

\medskip
\noindent\textit{Step 3: Upper bound on second moment of hubs.}
To validate the volume sum in the next step, we require an upper bound on the variance of the hub count. Let $M_T = \sum_{s=1}^T X_s$ be the total number of primary hubs.
The random variable $M_T$ is stochastically dominated by a Binomial distribution where the number of trials is (total Activated on $L_n$) $\times$ (total Nodes $n$), and success probability is $\beta_n p_n$.
\[
\EE[M_T] \leq (z_{\max} q^T) \cdot n \cdot (\beta_n p_n) = z_{\max} \lambda_n q^T.
\]
For a Binomial variable $Z$, $\mathbb{E}[Z^2] = \text{Var}(Z) + \mathbb{E}[Z]^2 \leq \mathbb{E}[Z] + \mathbb{E}[Z]^2$. The squared term dominates:
\[
\EE[M_T^2] \leq 2(z_{\max} \lambda_n q^T)^2.
\]

\medskip
\noindent\textit{Step 4: Volume aggregation and overlap control.}
Each hub activated at time $s$ acts as a seed for a diffusion on $L_n$ for $T-s$ periods, with expected volume $\calE_{T-s} \ge z_{\min} q^{T-s}$. By the Abel summation result in Step~2,
\[
V_{sum} := \sum_{s=1}^{T} \EE[X_s]\, \calE_{T-s} \;\geq\; C \lambda_n T q^{T-1}.
\]
To show this is a valid lower bound for $\hat{Y}_T(G_n)$, we apply the Bonferroni inequality: $\EE[\text{Vol}] \geq V_{sum} - \EE[\text{Overlap}]$.

The overlap is the sum of intersections over all pairs of hubs. Let $S$ be the random set of hubs.
\[
\EE[\text{Overlap}] = \EE\left[ \sum_{u \neq v \in S} |B_u(T) \cap B_v(T)| \right].
\]
Since edges in $E_n$ are formed i.i.d., hub locations are independent and approximately uniform on $V_n$. For any pair of hubs, the expected intersection of their radius-$T$ balls is bounded by $(z_{\max} q^T)^2 / n = z_{\max}^2 q^{2T}/n$, since two uniformly placed balls of volume $\leq z_{\max} q^T$ in a population of $n$ nodes overlap in at most $z_{\max}^2 q^{2T}/n$ nodes in expectation. Summing over all pairs:
\begin{align*}
    \EE[\text{Overlap}] &\leq \frac{1}{2} \EE[M_T^2] \cdot \frac{z_{\max}^2 q^{2T}}{n} \\
    &\leq (z_{\max} \lambda_n q^T)^2 \cdot \frac{z_{\max}^2 q^{2T}}{n} \\
    &= \frac{z_{\max}^4 \lambda_n^2 q^{4T}}{n}.
\end{align*}

We now compare the overlap to the sum of volumes.
\begin{align*}
    \frac{\EE[\text{Overlap}]}{V_{sum}} &\leq \frac{z_{\max}^4 \lambda_n^2 q^{4T} / n}{C \lambda_n T q^{T-1}}
    = \frac{z_{\max}^4}{C} \cdot \frac{\lambda_n q^{3T+1}}{n T}.
\end{align*}
Since $\lambda_n = n \beta_n p_n < 1$ (by Assumption~\ref{ass:beta-exp}), the ratio is dominated by $q^{3T}/(nT)$. Since $T \leq (\log n)^{1/q}$, we have $q^{3T} = \exp(3T\log q) = \exp(3(\log n)^{1/q}\log q)$. Because $(\log n)^{1/q} = o(\log n)$ when $q > 1$, it follows that $q^{3T} = o(n^{\epsilon})$ for any $\epsilon > 0$, and in particular $q^{3T}/n \to 0$.

Thus the overlap is negligible, and
\[
\hat{Y}_T(G_n) \geq \frac{1}{2} V_{sum} = \frac{1}{2} C \lambda_n T q^{T-1}.
\]

\medskip
\noindent\textit{Step 5: Convergence.}
Combining the bounds from Steps~1 and~4:
\[
\frac{\hat{Y}_T(L_n)}{\hat{Y}_T(G_n)} \leq \frac{z_{\max}\, q^T}{\tfrac{1}{2}\, C\, \lambda_n\, T\, q^{T-1}} = O\!\left( \frac{q}{\lambda_n T} \right) = O\!\left( \frac{1}{\lambda_n T} \right).
\]
Substituting $\lambda_n = n \beta_n p_n$:
\[
\frac{\hat{Y}_T(L_n)}{\hat{Y}_T(G_n)} = O\!\left( \frac{1}{n \beta_n p_n T} \right).
\]
By Assumption~\ref{ass:beta-exp}, $\beta_n =\omega\!\left( \frac{1}{p_n n \underline T_n}\right)$, so $n \beta_n p_n T \geq n \beta_n p_n \underline{T}_n \rightarrow \infty$ for any $T \geq \underline{T}_n$. Thus the ratio converges to $0$.
\end{proof}

\section{Simulations}\label{sec:sims}

We simulate SIR diffusion on synthetic networks to illustrate the finite-sample behavior of our results.

We generate $L_n$ by placing $n = 4{,}000$ nodes on a $q$-dimensional lattice in $[0,1]^q$ with additional uniformly placed nodes, linked to nearby nodes with a radius ensuring connectivity. We simulate two networks ($q=4$ and $q=2$) with $\calR_0 = 2.5$ and $p_n = \calR_0 / \bar{d}$. Summary statistics appear in Table~\ref{tab:mc-graph-stats}. We set $T$ to twice the diameter of $L_n$ ($T = 36$ for $q = 4$; $T = 186$ for $q = 2$), which extends beyond the intermediate-horizon regime of Assumption~\ref{ass:forecast-time} into the saturation phase.\footnote{Recall the time period bounds from Assumption~\ref{ass:forecast-time}. We use horizons beyond $\overline{T}_n$ to show that the qualitative patterns predicted by the theory---forecast underestimation during the intermediate regime, followed by convergence as the network saturates---persist in finite samples even when the formal asymptotic bounds are not binding.}

\paragraph*{Sensitive Dependence}
We fix $L_n$, draw a single $E_n$, set $i_0$ at the lattice center, and construct the alternate seed set $J_{i_0}$ at distance $2$ from $i_0$. For $q = 4$, this neighborhood covers 1.85\% of nodes; for $q = 2$, the neighborhood covers 0.45\% of nodes in the graph. We approximate the Jaccard index from Theorem \ref{thm:sensitive-dep}, $\calJ(i_{0}, j_{0})$, by fixing $E_n$ and averaging the Jaccard index $\calJ$ over 2,500 percolation draws.

\begin{figure}
  \centering
  \begin{subfigure}[b]{0.45\textwidth}
         \centering
         \includegraphics[width=\textwidth]{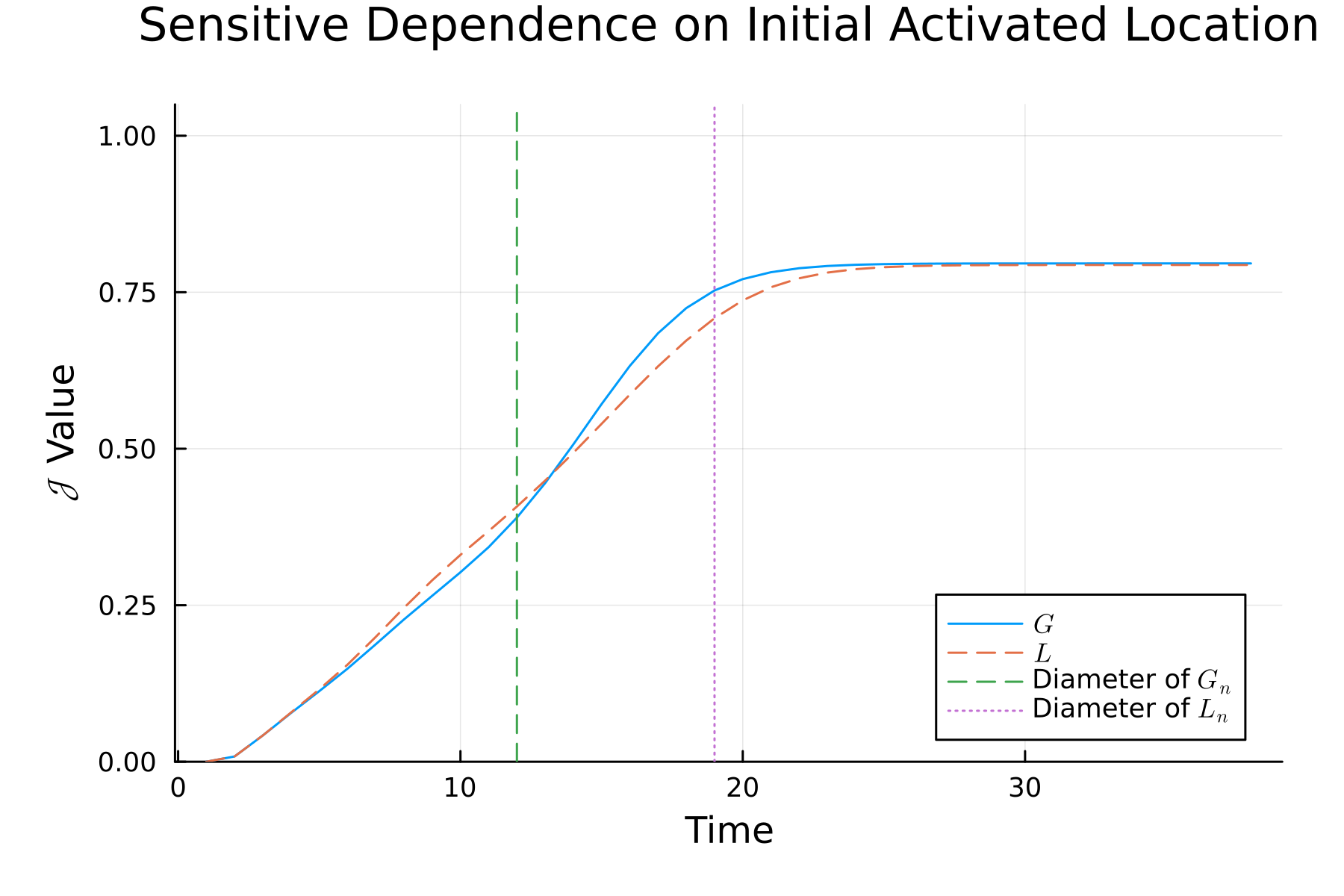}
         \caption{$ q= 4$}
         \label{fig:mc-sens-dep-4}
  \end{subfigure}
  \hfill
  \begin{subfigure}[b]{0.45\textwidth}
         \centering
         \includegraphics[width=\textwidth]{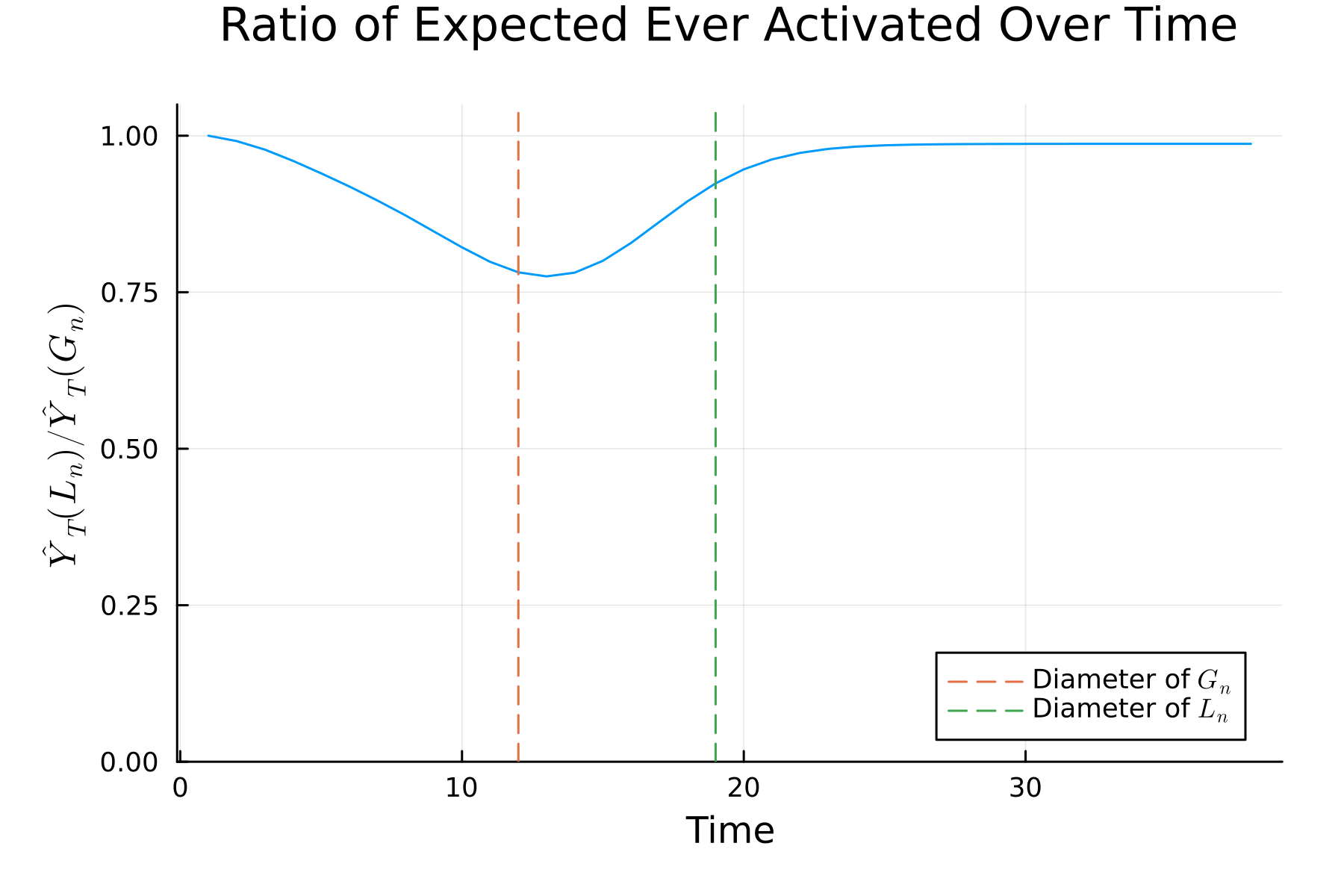}
         \caption{$q =4$}
         \label{fig:mc-ratio-4}
  \end{subfigure}

  \begin{subfigure}[b]{0.45\textwidth}
         \centering
         \includegraphics[width=\textwidth]{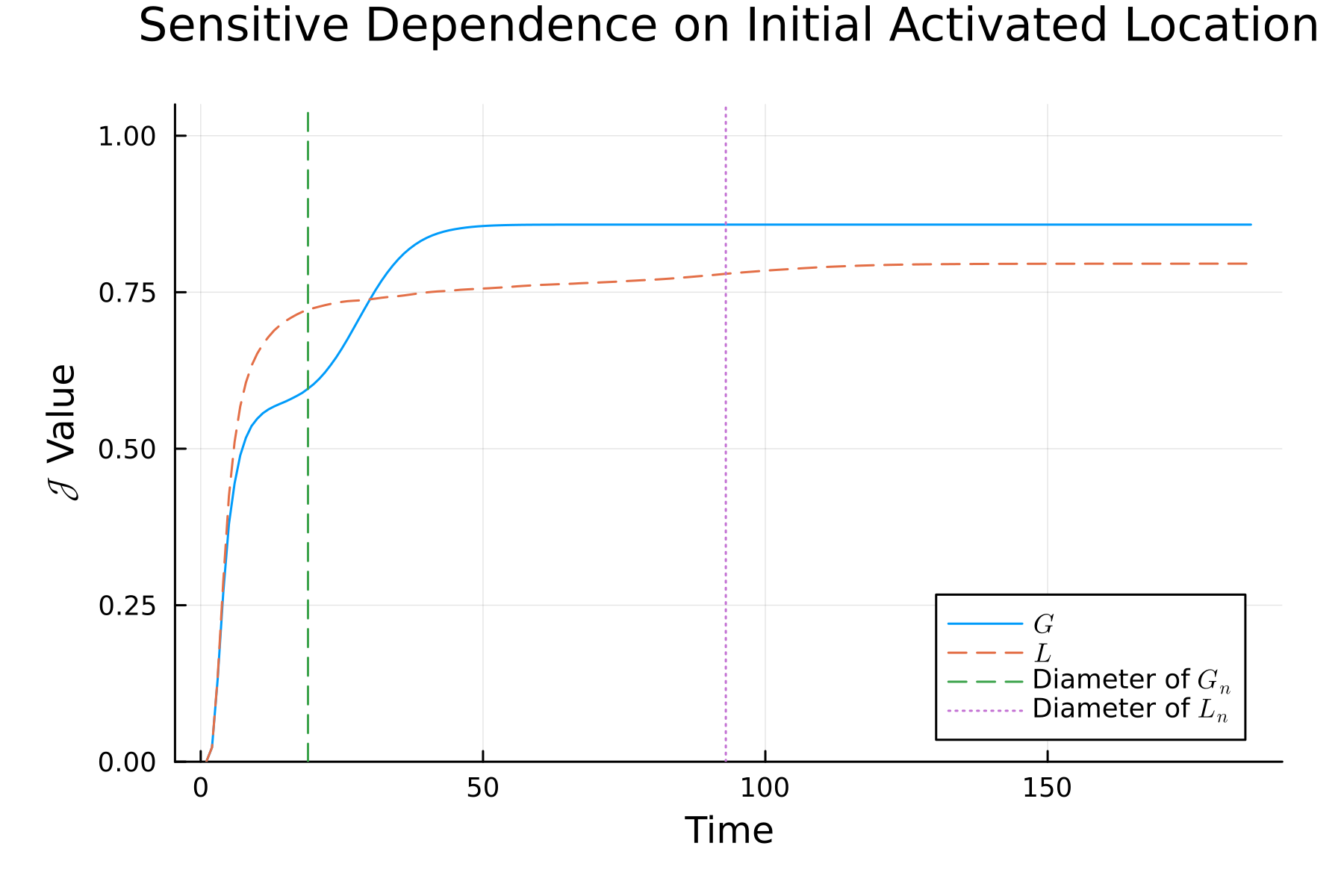}
         \caption{$q = 2$}
         \label{fig:mc-sens-dep-2}
  \end{subfigure}
  \hfill
  \begin{subfigure}[b]{0.45\textwidth}
         \centering
         \includegraphics[width=\textwidth]{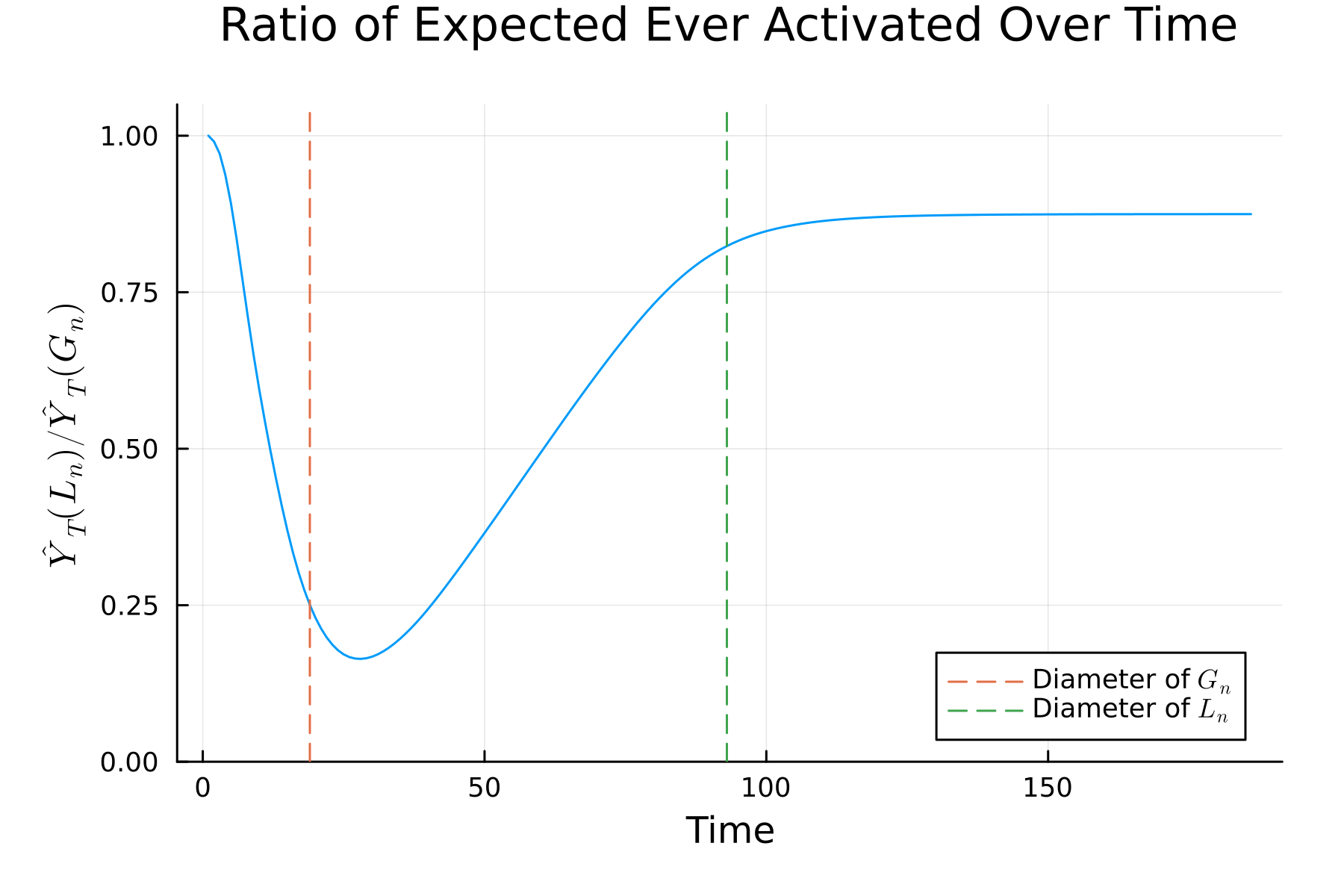}
         \caption{$q=2$}
         \label{fig:mc-ratio-2}
  \end{subfigure}

  \caption{Panels \ref{fig:mc-sens-dep-4} and \ref{fig:mc-sens-dep-2} show simulations of Theorem \ref{thm:sensitive-dep}, while panels \ref{fig:mc-ratio-4} and \ref{fig:mc-ratio-2} show simulations of Theorem \ref{thm:main-polynomial}. Panels \ref{fig:mc-sens-dep-4} and \ref{fig:mc-sens-dep-2} each fix a separate draw of $E_n$, then each choose a fixed $j_0$. We then simulate 2,500 diffusion processes while tracking the Jaccard index after perturbing the initial seed location.  In Panels \ref{fig:mc-ratio-4} and \ref{fig:mc-ratio-2}, we simulate 2,500 iterations of the diffusion process on both $L_n$ and $G_n$ for each value of $q$, re-drawing $E_n$ for each simulation. We then track the expected number of ever-activated nodes under each simulation at each time period, and then take the ratio.  
  }
  \label{fig:mc_thms}
\end{figure}

Figures~\ref{fig:mc-sens-dep-4} and~\ref{fig:mc-sens-dep-2} show little overlap between diffusions until the network saturates. For $q = 4$ at $T = 10$ (half the diameter of $L_n$), $\calJ = 0.29$---nearly disjoint processes; under $G_n$, this value is $\calJ = 0.27$. For $q = 2$ at $T = 47$ (half the diameter of $L_n$), $\calJ = 0.75$ under $L_n$ and $0.85$ under $G_n$. Comparing dimensions: lower $q$ produces more sensitivity in the \emph{extent} of diffusion (i.i.d.\ connections generate more activations), while higher $q$ produces more sensitivity in \emph{location}. Both show severe sensitive dependence early on.

\paragraph*{Forecast Errors}
We draw $E_n$ as an Erd\H{o}s--R\'{e}nyi graph with $\beta_n = 1/(10n)$, re-drawing $E_n$ each of 2,500 iterations. The average additional degree from $E_n$ is just 0.100, but the effect on global geometry is dramatic: for $q=2$, adding $E_n$ reduces average path length from 31.8 to 10.3 while mean degree rises only from 5.8 to 5.9---consistent with the path-length compression first observed numerically by \cite{watts1998collective}, and predicted by our framework for any polynomial-expansion base network with unaligned missing links.
We fix a random seed $i_0$ and compute $\hat{Y}_T(L_n)/\hat{Y}_T(G_n)$.

Figures~\ref{fig:mc-ratio-4} and~\ref{fig:mc-ratio-2} show $\hat{Y}_T(L_n)/\hat{Y}_T(G_n)$ over time. For $q = 4$, the minimum ratio is 0.775 at $T = 13$ (a 22\% underestimate); for $q = 2$, it drops to 0.164 at $T = 28$. Lower dimension produces greater sensitivity to additional links. Even with $\beta_n = 1/(100n)$ and $q = 2$, the minimum ratio remains far below the $q = 4$ values (Section~\ref{sec:extreme-sens}). The curves match the theoretical prediction: the ratio declines until diffusion reaches the diameter of $G_n$, then recovers toward one as the network saturates.\footnote{The ratio asymptotes just below 1, since $G_n$ permits slightly more activations in expectation than $L_n$.} Figure~\ref{fig:mc-thm1-rates} separates the curves for $\hat{Y}_T(L_n)$ and $\hat{Y}_T(G_n)$, confirming that divergence peaks near the diameter of $G_n$.

\subsection{Simulation Details}\label{sec:sim-details}
We now describe the simulation procedures in detail.

\subsection{Graph Generation}
Graph geometry plays a key role in our results. We build a network as follows, to generate an empirical analogue to the $L_n$ that we study theoretically. $L_n$ is generated as a graph of $n$ nodes in the following manner. 
\begin{enumerate}
    \item The base construction of the graph is a $q$-dimensional lattice, to mimic the properties of Assumption \ref{ass:disease}. We place $n_{side}$ nodes evenly spaced on $[0,1]^q$, meaning that there are $n_{side}^q$ nodes in the lattice portion of the graph. 
    \item The remainder of $n$ nodes are placed uniformly at random throughout $[0,1]^q$. 
    \item All nodes, regardless of whether they are in the lattice or placed randomly, link to all nodes within distance $r$. We set $r$ as:
    \begin{align*}
        r = \max\left\{\frac{1}{n_{side} - 1}, \frac{\sqrt{q}}{2}\frac{1}{n_{side} - 1}\right\}
    \end{align*}
    This ensures that the graph is connected, even when $q$ is large and thus nodes can be far apart. 
\end{enumerate}
We use the following parameters to generate $L_n$ in the graphs used in the main texts. In the first specification, we set $n = 4,000$, $q = 4$ and $n_{side} = 7$. In the second specification, we set $n = 4,000$, $q = 2$, and $n_{side} = 50$. To generate $G_n$, we add links with i.i.d. probability $\beta_n$. As a base rate, we use $\beta_n = \frac{1}{10n}$ -- in one variant of parameters, we set $\beta_n = \frac{1}{100n}$. Summary statistics are shown in Table \ref{tab:mc-graph-stats}, and for additional simulations in Table \ref{tab:mc-graph-stats-app2}.

\begin{table}[ht]
\caption{Graph statistics for $L_n$ with $n=4,000$ nodes}
\label{tab:mc-graph-stats}
    \begin{tabular}{c|cc|cc}
    \hline
    Statistic & $L_n$ & $G_n$ & $L_n$ & $G_n$\\
    \hline
Dimension & 4.0 & 4.0 & 2.0 & 2.0 \\
Diameter & 19.0 & 11.621 & 93.0 & 20.438 \\
Mean Degree & 10.164 & 10.263 & 5.826 & 5.926 \\
Min Degree & 3.0 & 3.091 & 2.0 & 2.0 \\
Max Degree & 24.0 & 24.107 & 16.0 & 16.126 \\
Mean Clustering Coefficient & 0.265 & 0.258 & 0.382 & 0.37 \\
Average Path Length & 7.592 & 6.017 & 31.942 & 10.304 \\
\hline
    \end{tabular}

{\small For $q = 4$, 60 percent of nodes are in the lattice, while with $q=2$ 62.5 percent are. Statistics for $G_n$ are the expectation over 2,500 draws of $E_n$, which is drawn Erd\H{o}s--R\'{e}nyi with $n=4,000$ and $\beta_n = \frac{1}{10n} = \frac{1}{40000}$.}
\end{table}

\subsection{Diffusion Process}
We use a Susceptible-Infected-Removed (SIR) diffusion process. Each node is infected (activated) for a single period, and has the opportunity to transmit the process with i.i.d. probability $p_n$ to each of its neighbors. After nodes are activated, they are removed and cannot be re-activated. We set the basic reproductive number to be $\calR_0 = 2.5$, and set $p_n = \calR_0/\bar d$, where $\bar d$ is the mean degree in $L_n$.

\subsection{Simulation of Theorem \ref{thm:sensitive-dep}}
As an analogue to Theorem \ref{thm:sensitive-dep}, we simulate SIR processes on a fixed $G_n$ with slightly perturbed starting points. We choose $i_0$ to be in the center of the lattice of $L_n$, that forms the backbone of $G_n$. Then, we build a set of alternative seeds $J_{i_0}$. All nodes at $2$ are included in $J_{i_0}$. We then choose a $j_0 \in J_{i_0}$ uniformly at random. 

The SIR process is then run, starting at both $i_0$ and $j_0$. We record which nodes are ever activated at each step of the process, under each simulation. To follow Theorem \ref{thm:sensitive-dep}, we fix the percolation across the simulation starting at $i_0$ and $j_0$. To do so, we use the fact that for a one-period SIR model, each link can transmit the disease at most one time. Therefore, we can simulate ex-ante which links will be able to transmit, which occurs with probability $p_n$, and intersect this with $G_n$ to get the realized percolation. 

We then compute a standard Jaccard index to track the intersection of the two epidemics. Let $I_P(i_0)$ be the set of ever-activated nodes under the epidemic from $i_0$, and $I_P(j_0)$ be the corresponding set from $j_0$. Then, we compute:
\begin{align*}
    \calJ := \EE_P\left[\frac{|I_P(i_0)\cap I_P(j_0)|}{|I_P(i_0)\cup I_P(j_0)|}\mid L_n, E_n\right]
\end{align*}
We work with the expected Jaccard index, rather than consider the probability the Jaccard index is bounded away from one. 

\subsection{Simulation of Theorem \ref{thm:main-polynomial}}
\begin{figure}
    \centering
    \begin{subfigure}[ht]{0.48\linewidth}
\includegraphics[width=\linewidth]{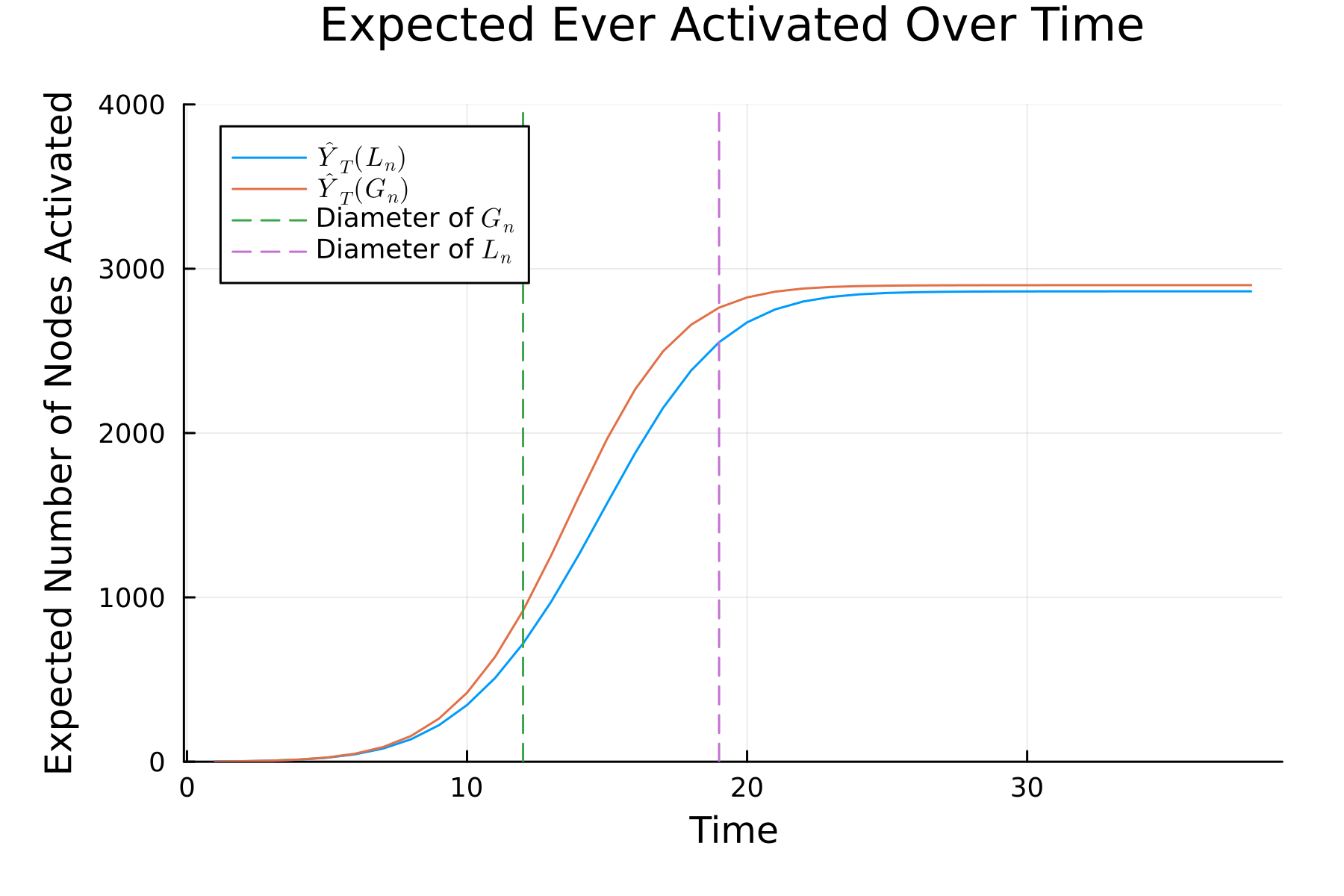}
\caption{$q=4$}
\end{subfigure}
\begin{subfigure}[ht]{0.48\linewidth}
\includegraphics[width=\linewidth]{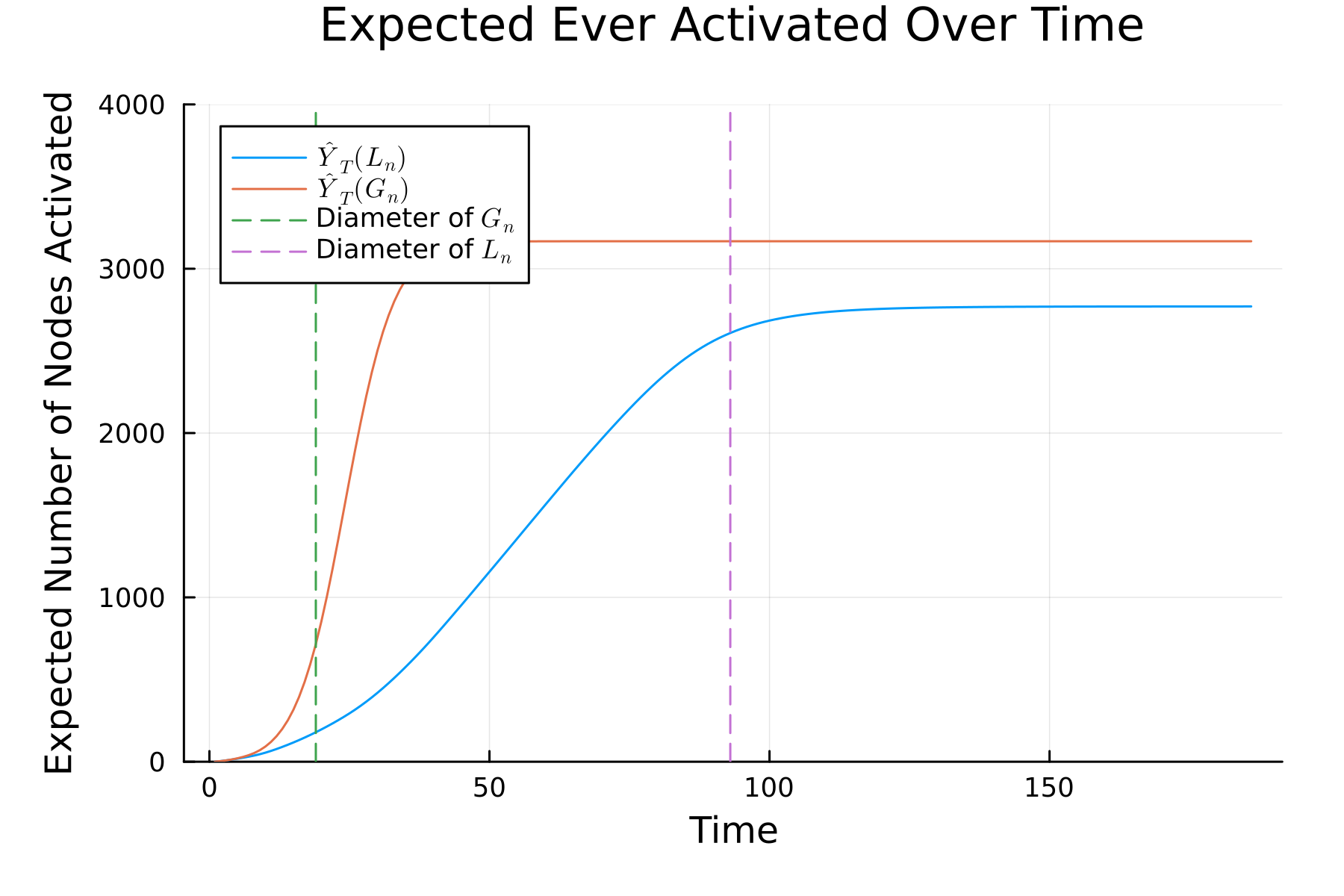}
\caption{$q=2$}
\end{subfigure}%
    \caption{This figure plots the same information as Figure \ref{fig:mc_thms}, but separated by graph for both $q = 4$ and $q=2$. The trajectory of $\hat Y_T(L_n)$ initially lags behind that of $\hat Y_T(G_n)$, leading to the decrease in the ratio shown in Figure~\ref{fig:mc_thms}. As $\hat Y_T(L_n)$ catches up, the ratio increases.}
    \label{fig:mc-thm1-rates}
\end{figure}

\begin{figure}
\centering
\begin{subfigure}[b]{0.45\linewidth}
\includegraphics[width=\linewidth]{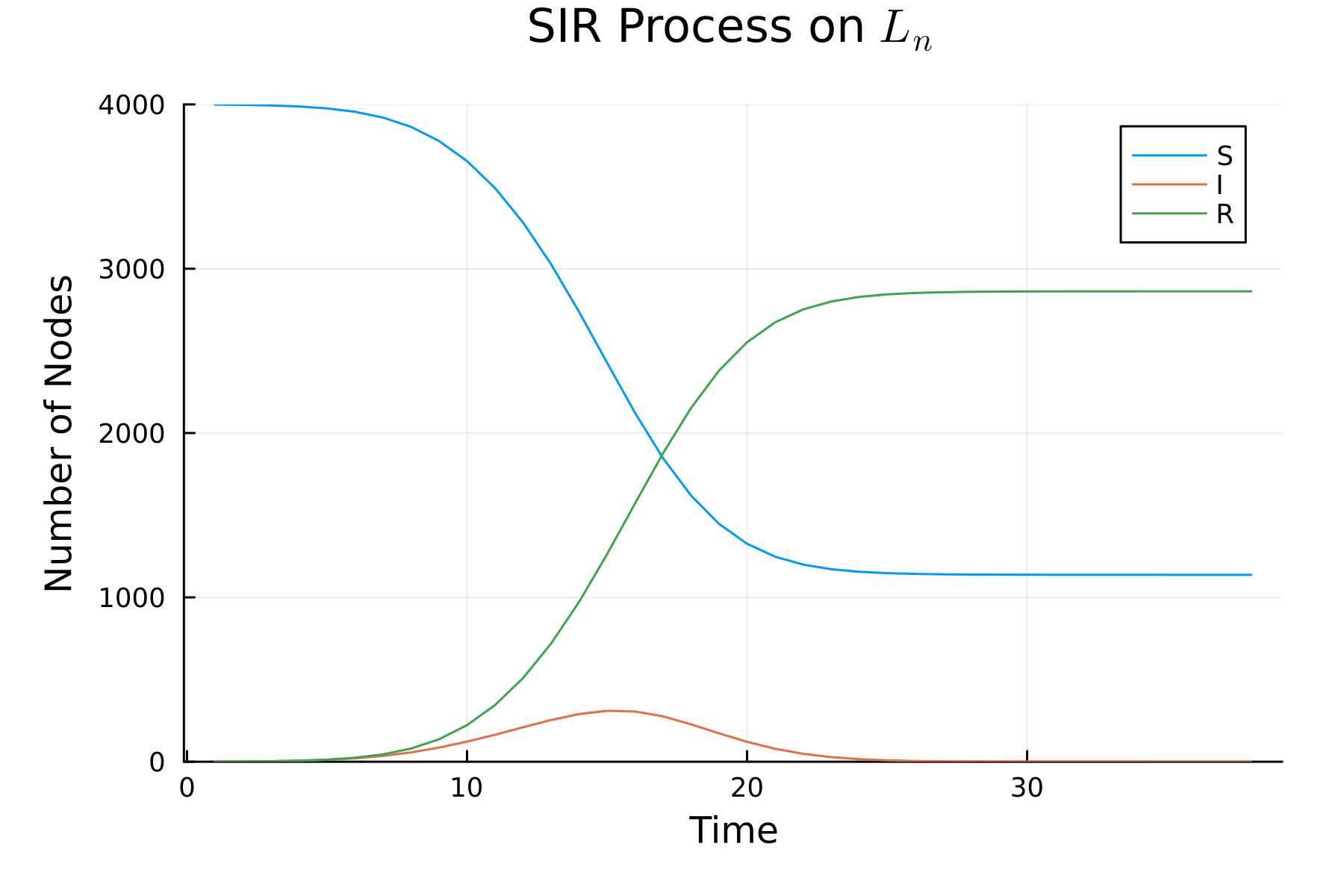}
\caption{$q=4$, $L_n$}
\end{subfigure}\hfill
\begin{subfigure}[b]{0.45\linewidth}
\includegraphics[width=\linewidth]{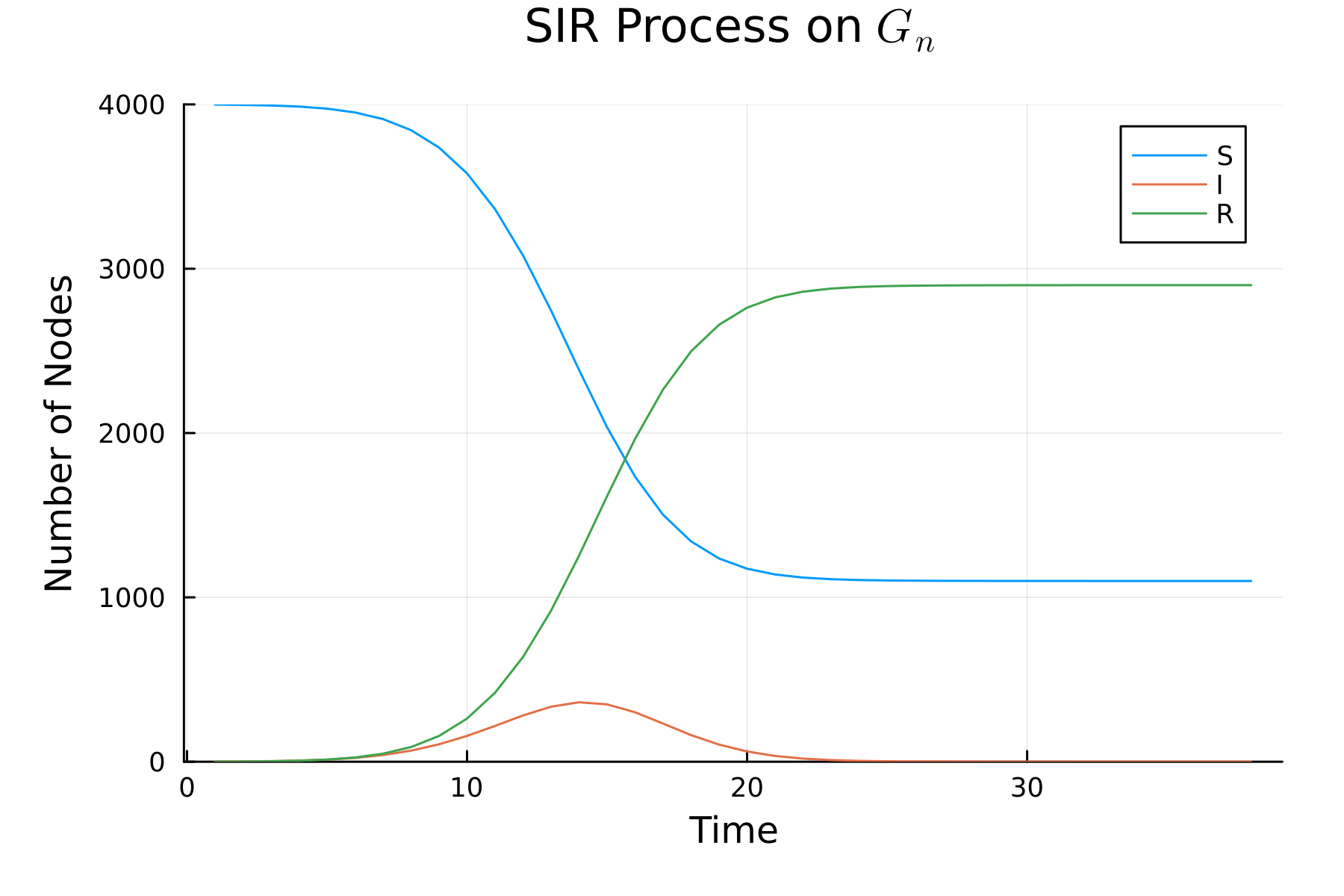}
\caption{$q=4$, $G_n$}
\end{subfigure}
\begin{subfigure}[b]{0.45\linewidth}
\includegraphics[width=\linewidth]{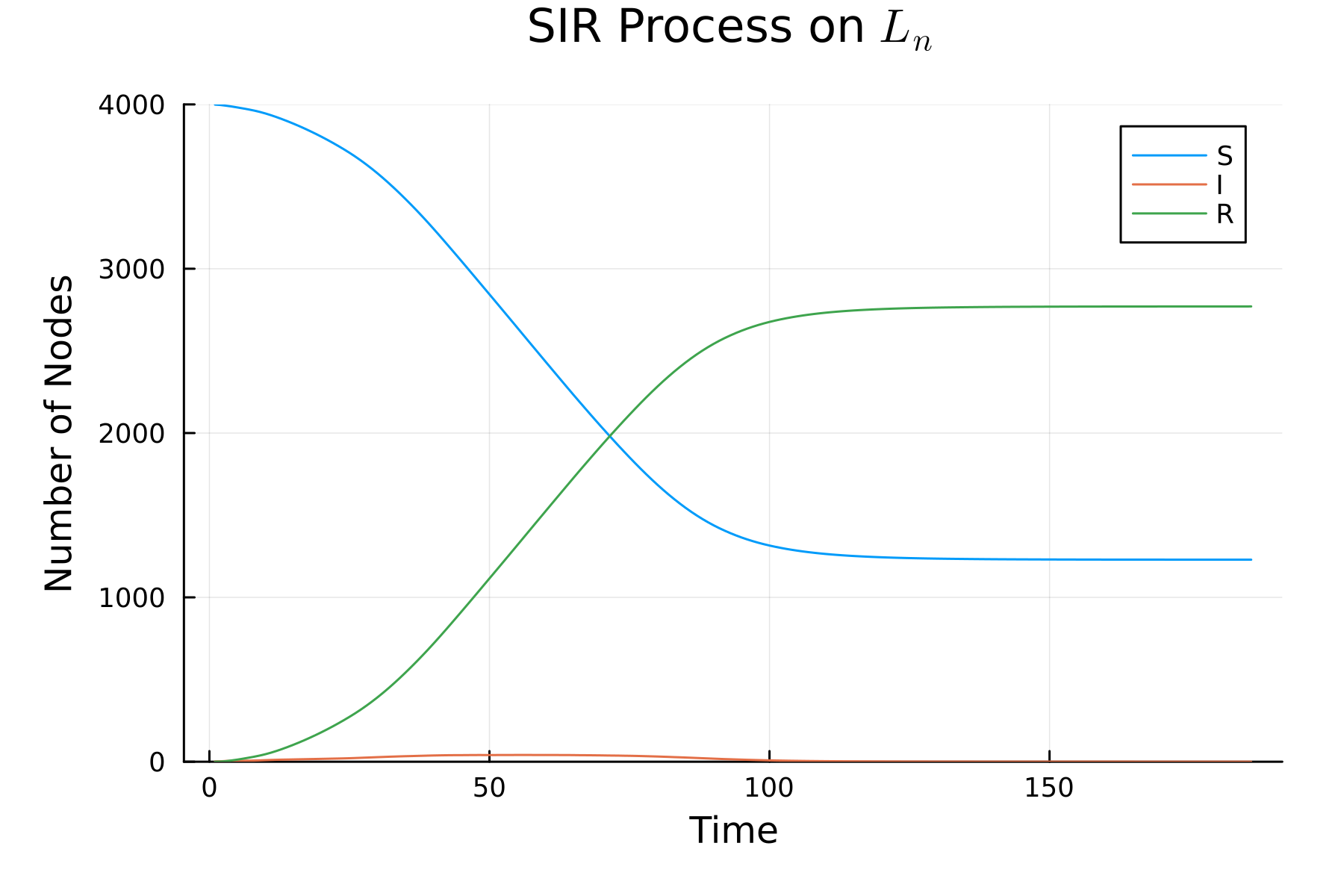}
\caption{$q=2$, $L_n$}
\end{subfigure}\hfill
\begin{subfigure}[b]{0.45\linewidth}
\includegraphics[width=\linewidth]{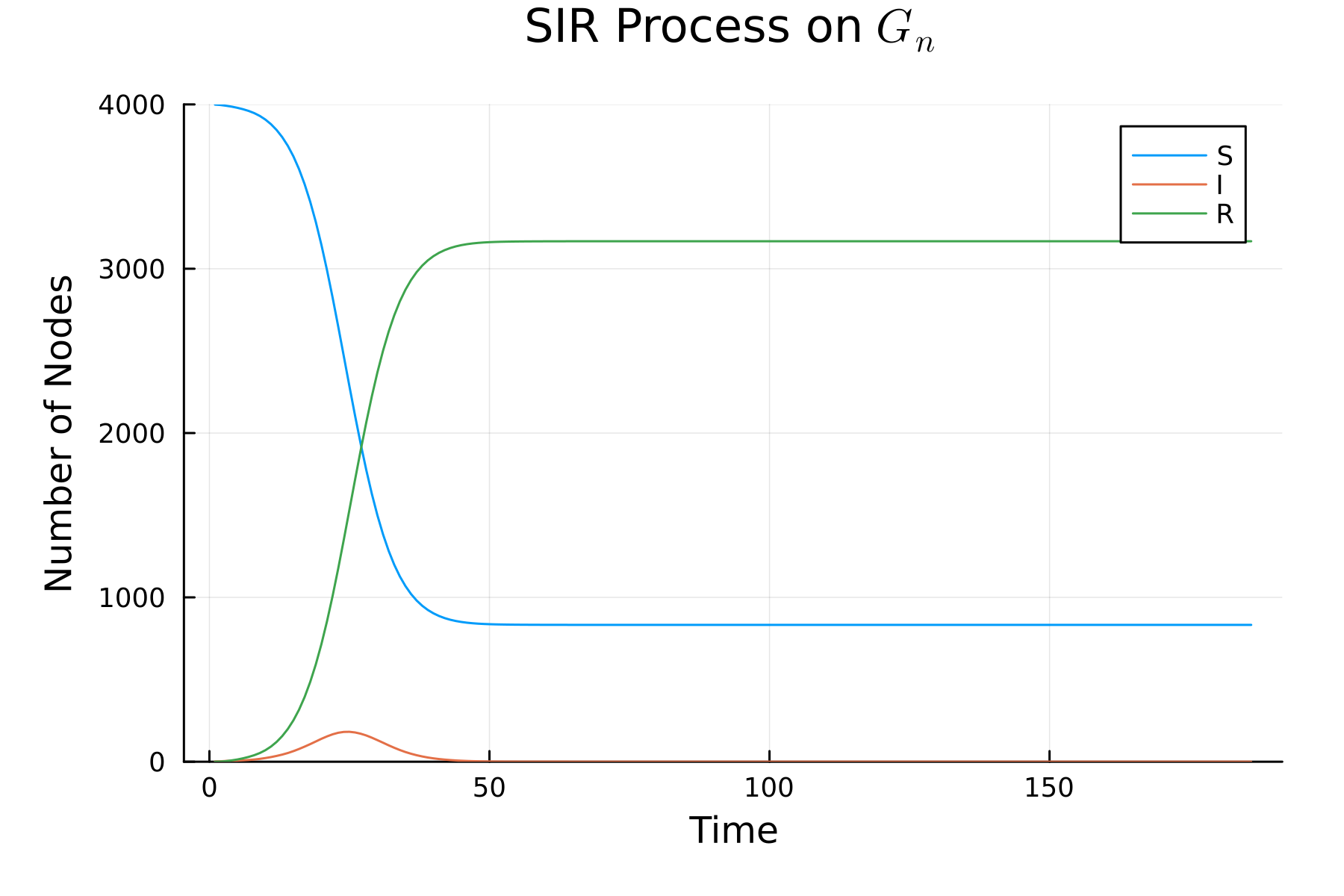}
\caption{$q=2$, $G_n$}
\end{subfigure}%
\caption{Simulations meant to emulate Theorem \ref{thm:main-polynomial}, disaggregated into the standard SIR framework. The figure is a result of averaging over simulation draws. Note that we see a larger spike in activations under $G_n$, which makes intuitive sense -- the additional links allow for more infections to occur. We show results for both $q = 4$ and $q=2$, both with $\beta_n = \frac{1}{10n}$. Note that the gap between total activations with $q=2$ is larger, as the additional links have a larger effect. }
\label{fig:mc-thm1-sir}
\end{figure}

To investigate the content of Theorem \ref{thm:main-polynomial}, we directly simulate the sample analogue. For 2,500 simulations, we do the following. We choose the initial seed $i_0$ uniformly at random, and fix it throughout the process. The SIR process is simulated for $T$ periods, where we set $T$ to be twice the diameter of $L_n$. 
\begin{enumerate}
    \item Simulate the SIR process on $L_n$. 
    \item Generate a draw of $E_n$, with links i.i.d. with probability $\beta_n$. 
    \item We define $G_n := L_n \cup E_n$, and simulate the SIR process on $G_n$.
\end{enumerate}
We track the number of ever-activated nodes in each simulation at each time step. We then take the average over simulations at each time step. Results are shown in Figure \ref{fig:mc_thms}. Additional results are shown in Figures \ref{fig:mc-thm1-rates} and \ref{fig:mc-thm1-sir}.

\subsection{Simulation of Theorem \ref{thm:no-failure-mar} (Aligned Error)}
\begin{figure}
\centering
\begin{subfigure}[b]{0.45\linewidth}
\includegraphics[width=\linewidth]{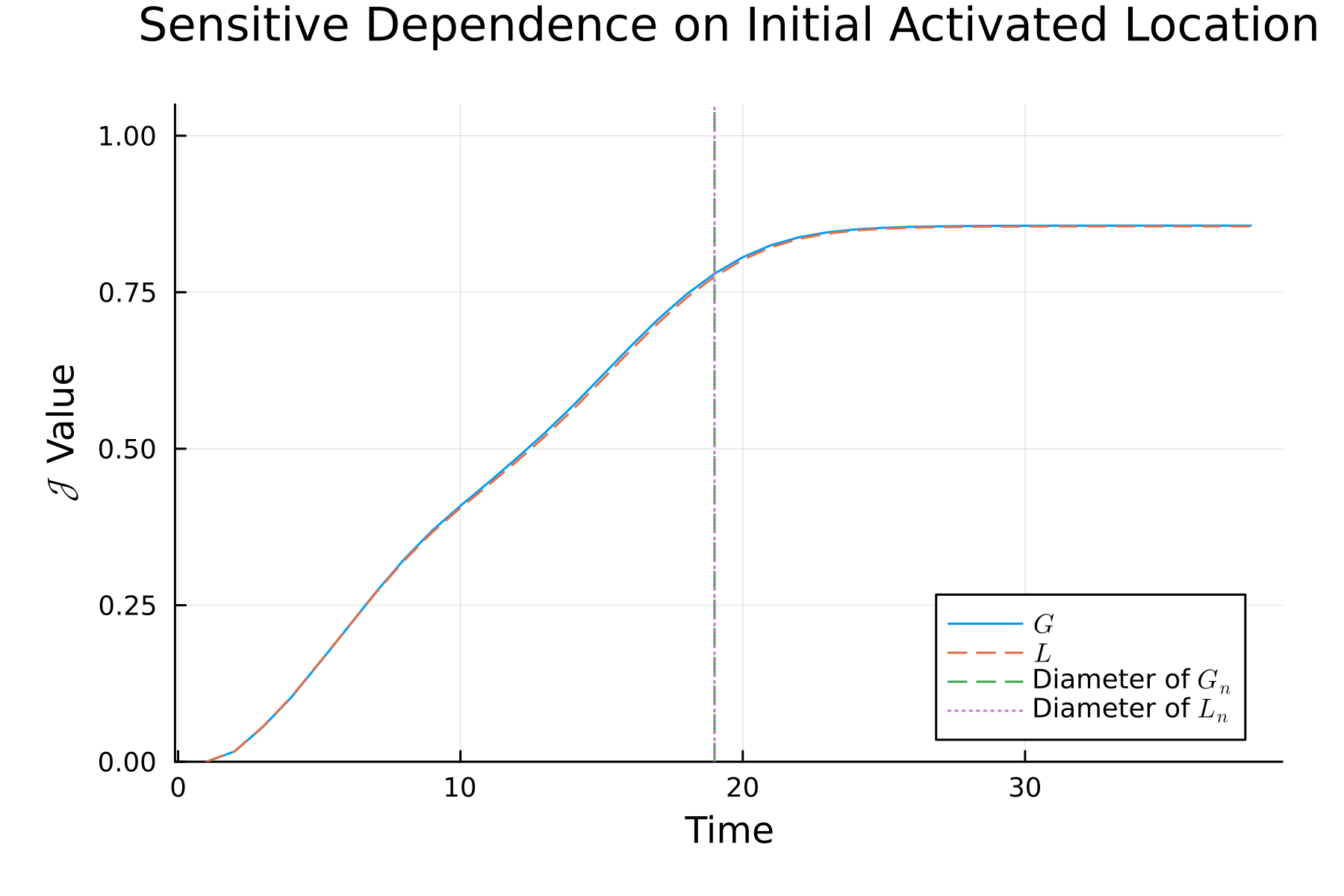}
\caption{$q=4$, sensitive dependence}
\end{subfigure}\hfill
\begin{subfigure}[b]{0.45\linewidth}
\includegraphics[width=\linewidth]{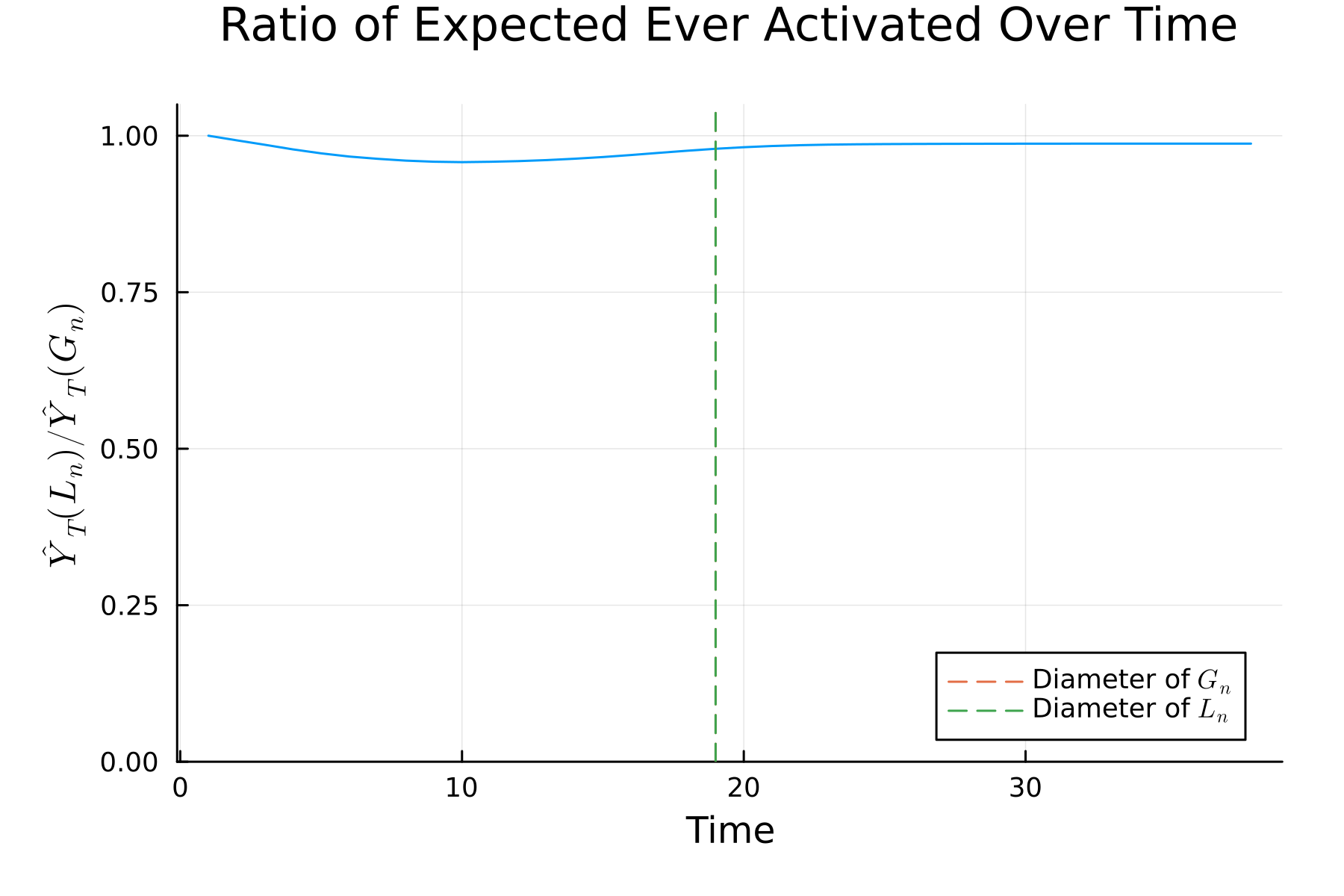}
\caption{$q=4$, diffusion ratio}
\end{subfigure}
\begin{subfigure}[b]{0.45\linewidth}
\includegraphics[width=\linewidth]{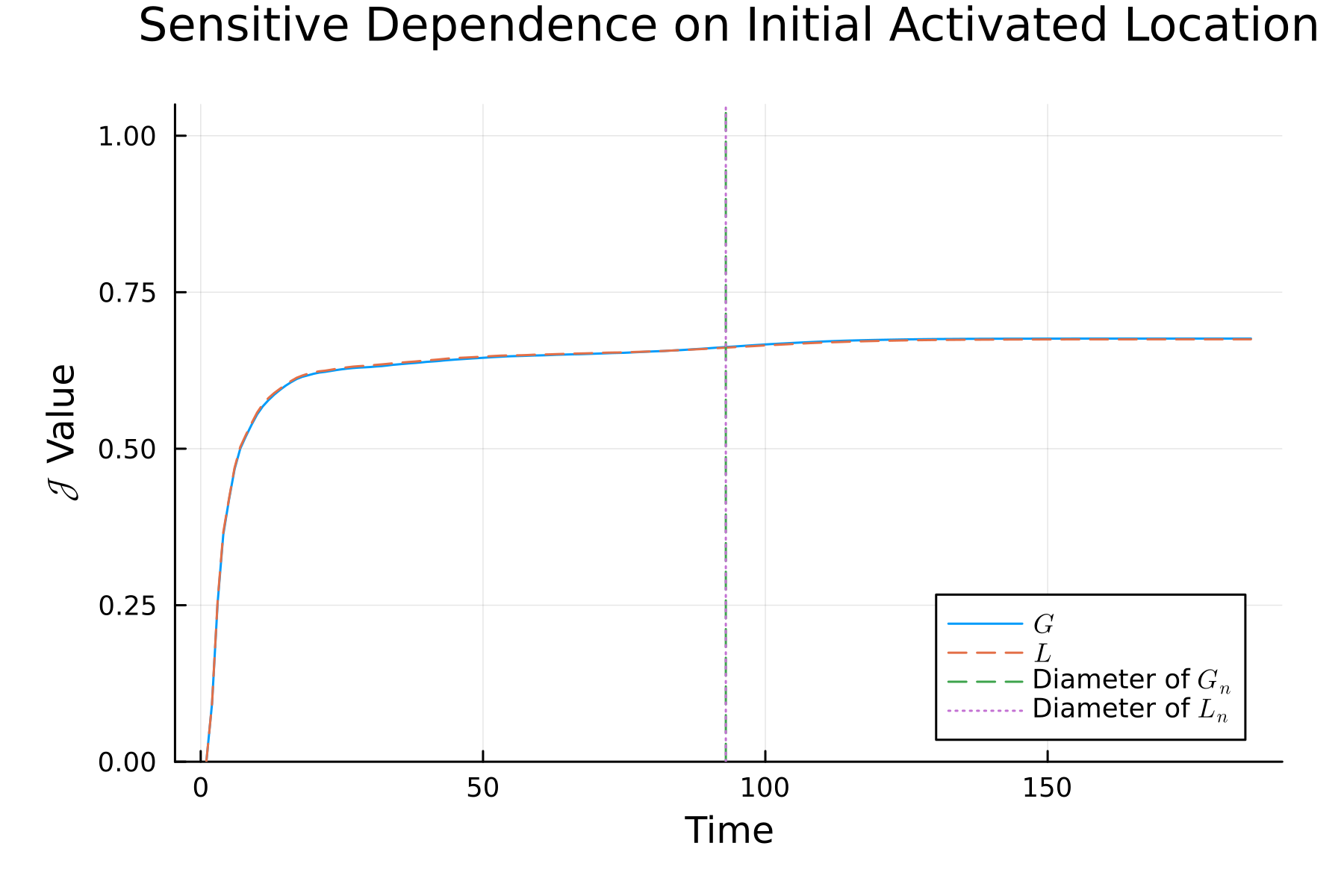}
\caption{$q=2$, sensitive dependence}
\end{subfigure}\hfill
\begin{subfigure}[b]{0.45\linewidth}
\includegraphics[width=\linewidth]{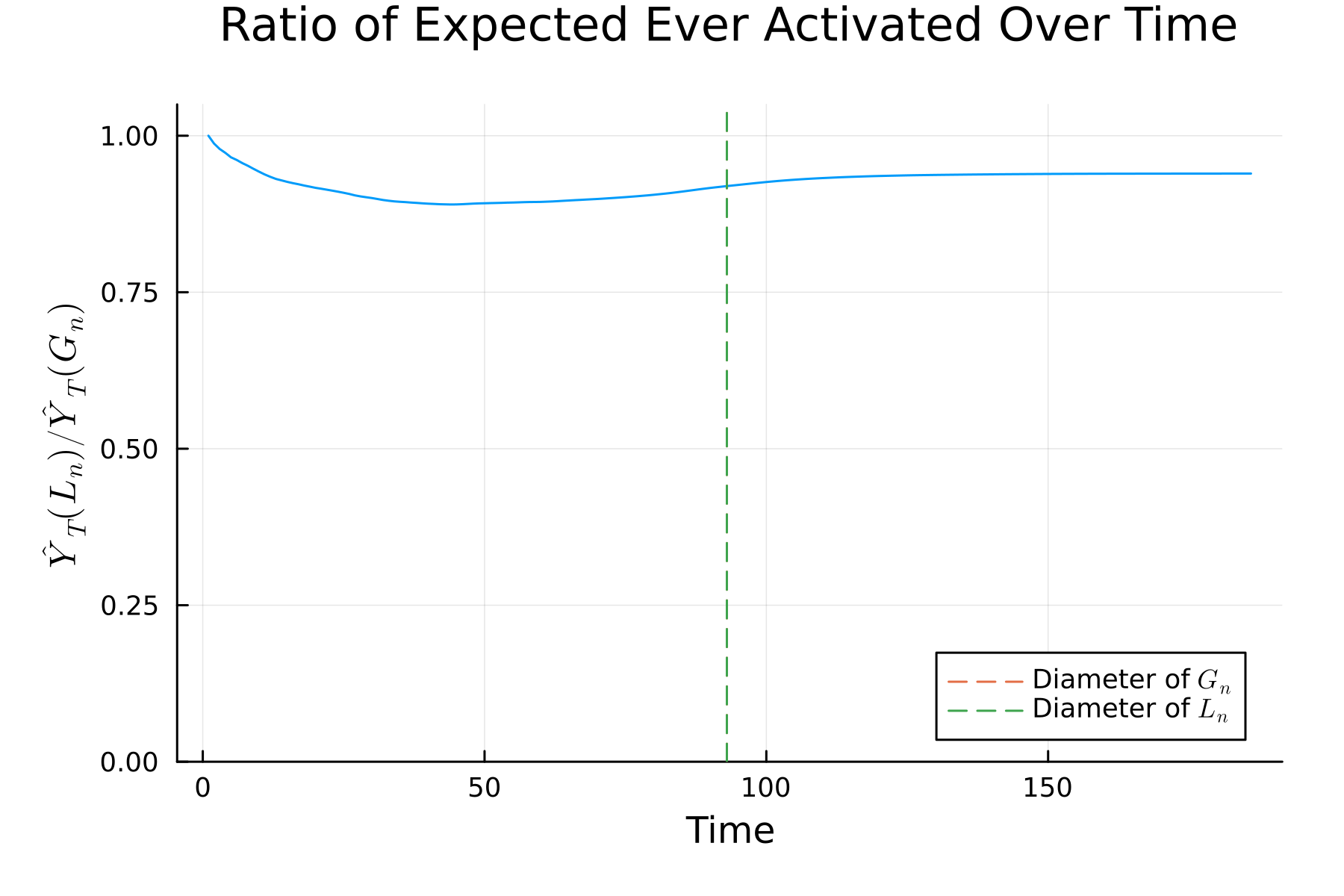}
\caption{$q=2$, diffusion ratio}
\end{subfigure}%
\caption{Simulations meant to emulate Theorem \ref{thm:no-failure-mar}, using the objectives from Theorems \ref{thm:sensitive-dep} and \ref{thm:main-polynomial}. We show results for both $q = 4$ and $q=2$, in both cases calibrating the probability of link deletion so that the same volume of links are deleted as in the case where we add links with probability $\frac{1}{10n}$.}
\label{fig:mc-thm3}
\end{figure}

We simulate a version of Theorem \ref{thm:no-failure-mar}. We begin with the base graph (what we used as $L_n$ in the prior two sections) and remove links with i.i.d. probability---producing aligned error, since missing links are a uniform thinning of the true network. Here, we choose the deletion probability $\varepsilon_n$ to delete the same number of links on average as we did when we added links with probability $\beta_n = \frac{1}{10n}$. We then investigate the objectives from Theorems \ref{thm:sensitive-dep} and \ref{thm:main-polynomial}, taking the base graph as $G_n$ and the thinned graph as $L_n$. Results are shown in Figure \ref{fig:mc-thm3}. We can see the system exhibits minimal additional sensitive dependence on the seed in both the $q=4$ and $q=2$ cases, beyond the base level of sensitivity from prior simulations. We also see reduced underestimation of the diffusion ratio $\hat Y(L_n)/\hat Y(G_n)$: for $q=4$, the minimal value is 0.96 at $T = 10$, while for $q=2$ it is 0.89 at $T = 44$. 

\begin{table}[ht]
\caption{Graph statistics for aligned-error simulations with $n=4,000$ nodes} 
\label{tab:mc-graph-stats-mar}
    \begin{tabular}{c|cc|cc}
    \hline
    Statistic & $L_n$ & $G_n$ & $L_n$ & $G_n$\\
    \hline
Dimension & 4.0 & 4.0 & 2.0 & 2.0 \\
Diameter & 18.805 & 19.0 & 92.858 & 93.0 \\
Mean Degree & 10.064 & 10.164 & 5.727 & 5.826 \\
Min Degree & 2.951 & 3.0 & 1.49 & 2.0 \\
Max Degree & 23.783 & 24.0 & 15.758 & 16.0 \\
Mean Clustering Coefficient & 0.261 & 0.269 & 0.375 & 0.383 \\
Average Path Length & 7.584 & 7.562 & 31.938 & 31.806 \\
\hline
    \end{tabular}
    
    {\small $L_n$ is generated from $G_n$ by thinning the network, removing links with i.i.d. probability $\varepsilon_n$. We calibrate $\varepsilon_n$ so that the expected difference in link volume between $L_n$ and $G_n$ is the same as in simulations where we generate $G_n$ by adding links to $L_n$. The diameter is computed over the largest connected component of the graph -- in some situations, the diameter decreases due to the thinning as pieces of the graph become disconnected.} 
\end{table}

\subsection{Extreme Sensitivity with \texorpdfstring{$q=2$}{ q = 2} }\label{sec:extreme-sens}
\begin{figure}[ht]
    \centering
    \begin{subfigure}[b]{0.45\textwidth}
         \centering
         \includegraphics[width=\textwidth]{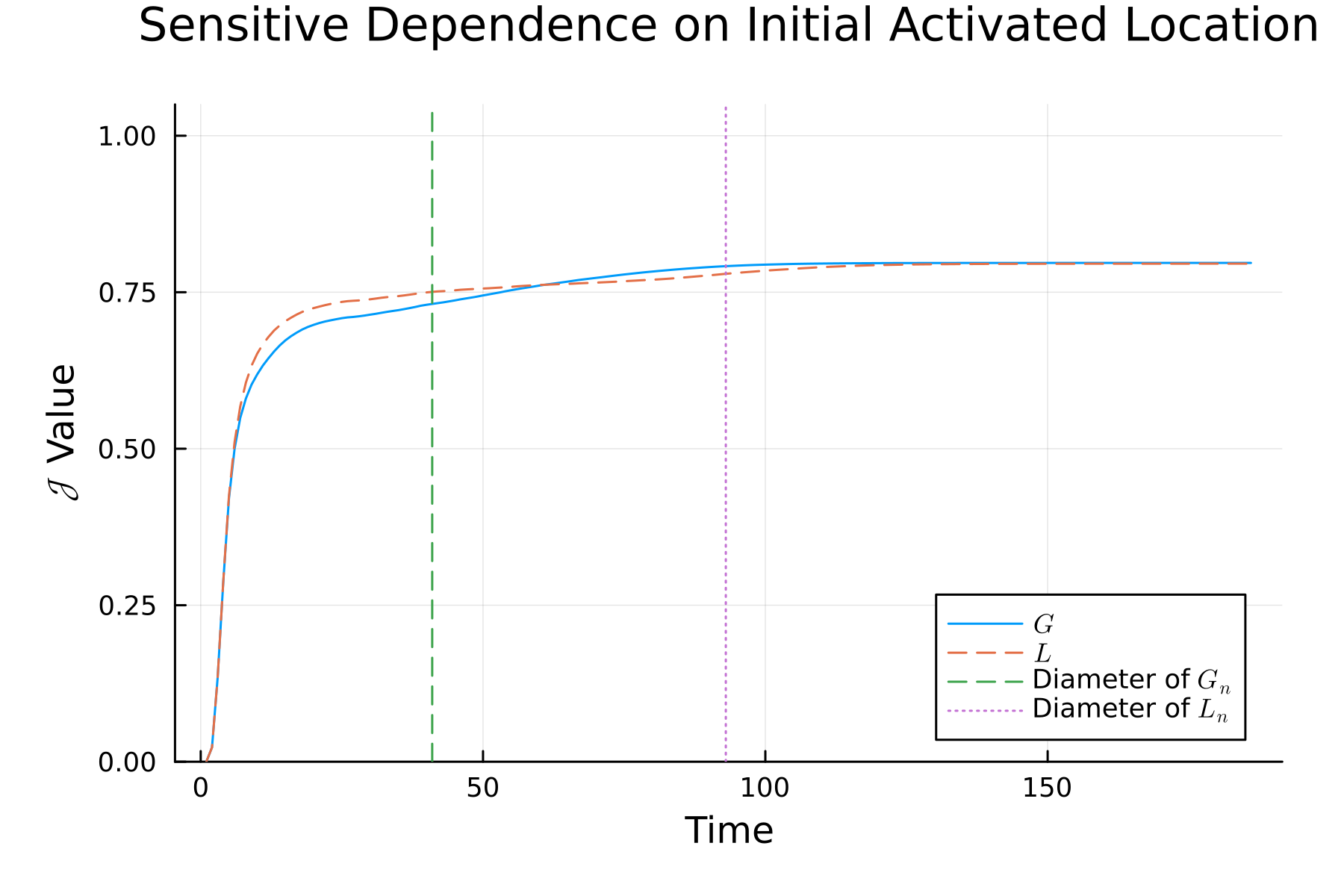}
         \caption{}
     \end{subfigure}
     \hfill
     \begin{subfigure}[b]{0.45\textwidth}
         \centering
         \includegraphics[width=\textwidth]{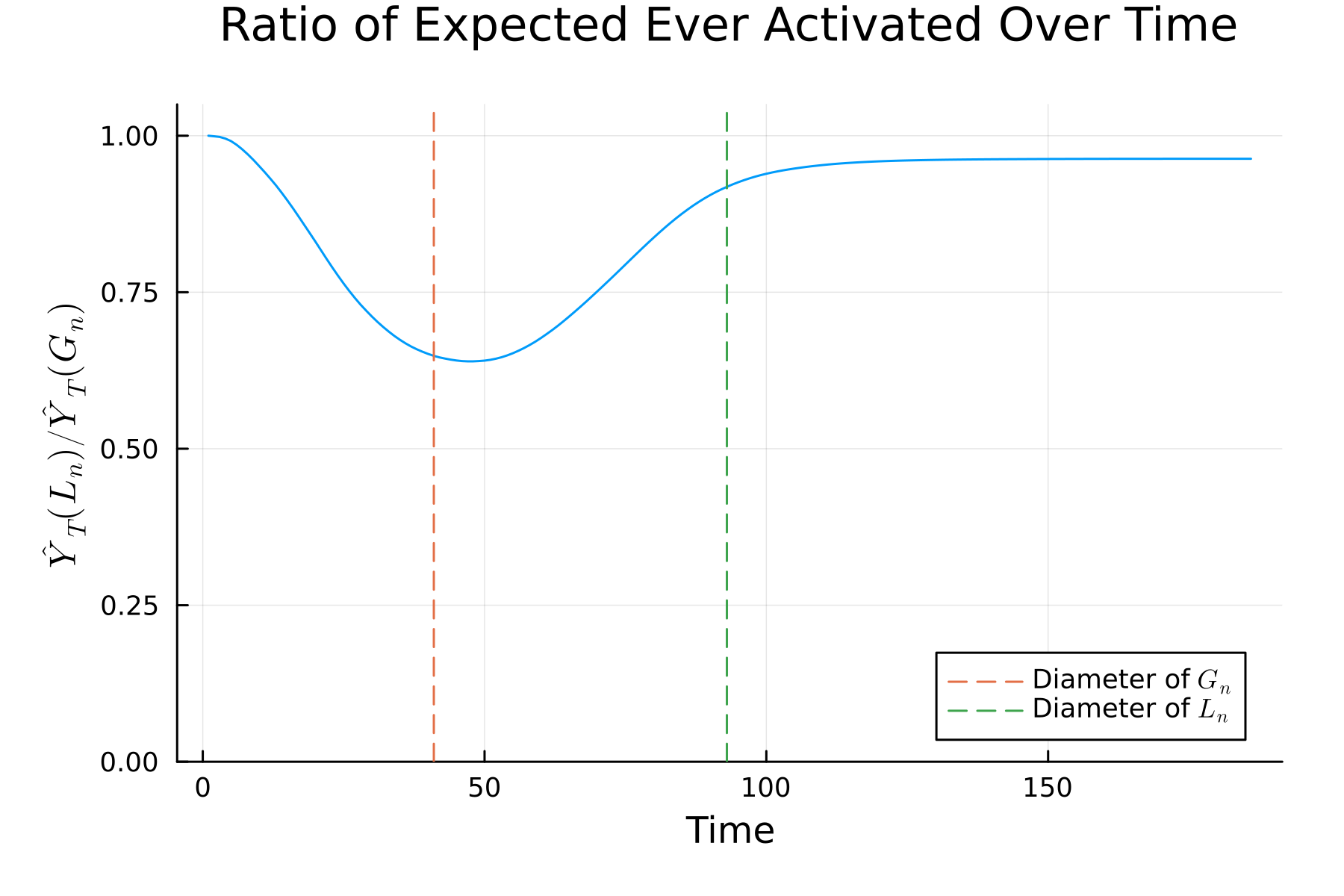}
         \caption{}
     \end{subfigure}
    \caption{Results with $q=2$ and $\beta_n = \frac{1}{100n}$. Panel (A) shows the Jaccard index $\calJ$, while Panel (B) shows the ratio $\hat Y_T(L_n)/\hat Y_T(G_n)$. 
    Averages are taken over 2,500 Monte Carlo simulations.}
    \label{fig:mc-low-beta}
\end{figure}

We explore an additional set of simulations in the case of $q=2$, this time using a much smaller value of $\beta_n = \frac{1}{100n}$. We show average graph statistics in Table \ref{tab:mc-graph-stats-app2}. Results are shown in Figures \ref{fig:mc-low-beta}. 

\begin{table}[ht]
\caption{Graph statistics for $L_n$ generated with $q=2$ and $G_n$ generated with $\beta_n=\frac{1}{100n}$ }
\label{tab:mc-graph-stats-app2}
    \begin{tabular}{c|cc}
    \hline
Statistic & $L_n$ & $G_n$\\
\hline
Dimension & 2.0 & 2.0 \\
Diameter & 93.0 & 45.059 \\
Mean Degree & 5.826 & 5.836 \\
Min Degree & 2.0 & 2.0 \\
Max Degree & 16.0 & 16.007 \\
Mean Clustering Coefficient & 0.379 & 0.38 \\
Average Path Length & 31.774 & 18.802 \\
\hline
    \end{tabular}

{\footnotesize Statistics for $G_n$ are taken as an average over 2,500 draws.}
\end{table}

As shown in Figure \ref{fig:mc-low-beta}, despite a much smaller value of $\beta_n$ forecasting issues persist. There remains meaningful sensitive dependence on the initial seed---the average $\calJ$ at half the diameter of $L_n$ is 0.74 on $L_n$ and $0.75$ on $G_n$, indicating that roughly 26\% of the activation set changes when the seed is perturbed by two hops. For forecasting diffusion volume, the minimum value of $\hat Y_T(L_n)/\hat Y_T(G_n)$ is achieved at $T = 47$, taking a value of 0.64. This value is still lower than the case with $q=4$ and $\beta_n  = \frac{1}{10n}$, showing the extreme sensitivity in the lower dimension.


\setcounter{figure}{0}  
\setcounter{table}{0}  
\section{Empirical Example: Location Data from the COVID-19 Epidemic}\label{sec:covid}
We give a detailed description of the data processing procedures, along with additional results using a graph constructed from location data. We build a network using visitor flows based on cell phone location data, provided by SafeGraph \citep{kang2020multiscale}. Our primary analysis studies the entirety of California and Nevada, with a small portion of Arizona included. 
Note that we only include areas in the United States. The region includes major cities including San Francisco, Los Angeles, and Las Vegas. 
We work with Census tracts as the unit of observation, which each contain approximately 4,000 people. Given privacy concerns, we focus on movement between tracts, rather than tracking individual people. We use tract-to-tract flows on March 1st, 2020. This date was before the WHO declared COVID-19 a pandemic, and before the United States government declared a national state of emergency. We construct graphs in the following manner. Fix a cutoff $c$. Then we take the following steps. 
\begin{enumerate}
    \item For each pair of Census tracts $a$ and $b$, we construct the average flow between tracts by taking the average of the flow from $a$ to $b$ and the flow from $b$ to $a$. Call this value $f_{ab}$. 

    \item Tracts $a$ and $b$ will be linked in the graph only if $f_{ab}>c$.
\end{enumerate}
We choose $c$ based on the empirical distribution of $f_{ab}$, the flows between tracts. We refer to this procedure as ``pruning." If the process results in a disconnected graph, we choose the largest connected subgraph. As before, we set $T$ as twice the diameter of $L_n$.

\subsection{Disease Process}
As with the simulated graphs, we fix $\calR_0 = 2.5$. We then compute $p_n= \calR_0/\bar d$, where $\bar d$ is the average degree in $L_n$.  Note that in this case, the meaning of $\calR_0$ is substantively different -- because nodes now refer to Census tracts, infecting 2.5 nodes in the disease free state on average means infected 2.5 tracts on average. 

\subsection{Errors Induced by Cutoff Choice}
We first study errors induced by choosing different cutoffs for pruning the graph. We construct $G_n$ by setting $c = 5$, which is at the 91st percentile of the empirical distribution of tract-to-tract flows. Then, we generate $L_n$ by choosing $c = 6$. Note that every link in $L_n$  will be in $G_n$, meaning that we can construct the implied error graph $E_n$.

We conduct the same three analyses that we did with the simulated graph. First, we study a version of Theorem \ref{thm:sensitive-dep}, comparing the overlap between epidemics after perturbing the starting point. Second, we study a version of Theorem \ref{thm:main-polynomial}, comparing the expected number of infections on each graph.  Finally, we consider the exercise of fitting a SIR differential equation model. 

For the sake of brevity, we only note differences unique to this section when compared to the procedures discussed in Section \ref{sec:sims}. When considering the simulation of Theorem \ref{thm:main-polynomial}, the key change is that we hold $G_n$ fixed: it is generated once from the data. When we take expectations, they are taken only over the disease process only. Otherwise, the process is identical. In addition, we conduct simulations with $E_n$ taken to be an Erd\H{o}s--R\'{e}nyi random graph, rather than via the pruning procedure. In the main text, we set $\beta_n$ so that the i.i.d. errors generate the same expected volume of links as the pruning procedure. As an additional set of results, we set $\beta_n = \frac{1}{10n}$, to compare with the Monte Carlo simulations. Summary statistics of the resulting graphs are shown in Table \ref{tab:graph_stats_sw_pruned}.

\begin{table}[ht]
    \caption{Graph statistics for $L_n$ and both hypothetical $G_n$s constructed from California, Nevada, and Arizona Census tract flow data}
    \label{tab:graph_stats_sw_pruned}
    \begin{tabular}{l|rrrr}
\toprule
    Statistic & $L_n$ & $G_n^{92}$ & $G_n^\beta$ & $L_n^{\varepsilon}$\\
\midrule
Error Type & --- & Pruned & i.i.d. Add & i.i.d. Remove \\
Diameter & 21.0 & 15.0 & 7.676 & 23.861 \\
Mean Degree & 12.962 & 15.486 & 16.172 & 9.982 \\
Min Degree & 1.0 & 1.0 & 1.843 & 0.0 \\
Max Degree & 298.0 & 329.0 & 301.044 & 229.663 \\
Mean Clustering Coefficient & 0.387 & 0.401 & 0.234 & 0.297 \\
Average Path Length & 7.277 & 5.861 & 4.03 & 8.139 \\
\bottomrule
    \end{tabular}

{\footnotesize  Statistics for $G_n^\beta$ and $G_n^{\varepsilon}$ with i.i.d. errors are averaged over 2,500 draws.}
\end{table}

\subsection{Lower Rates of I.I.D. Errors}
To make a more direct comparison to the Monte Carlo simulations, we repeat the simulation exercises using $E_n$ generated i.i.d. with $\beta_n = \frac{1}{10n}$. Graph statistics are shown in Table \ref{tab:sw-graph-stats-low}, again for $L_n$ and the average statistics for $G_n$ over 2,500 draws of $E_n$. Compared to $G_n$ in the main text (in Table \ref{tab:graph_stats_sw_pruned}), note that the change in degree, clustering, and average path length are all much smaller, as $E_n$ is much more sparse in this case. 

\begin{table}[ht]
    \caption{Average graph statistics with i.i.d. errors in the travel data for California, Nevada, and a small portion of Arizona}
    \label{tab:sw-graph-stats-low}
    \begin{tabular}{c|cc}
    \hline
    Statistic & $L_n$ & $G_n$\\
    \hline
Diameter & 21.0 & 16.914 \\
Mean Degree & 12.962 & 13.062 \\
Min Degree & 1.0 & 1.0 \\
Max Degree & 298.0 & 298.106 \\
Mean Clustering Coefficient & 0.39 & 0.381 \\
Average Path Length & 7.287 & 6.117 \\
\hline
    \end{tabular}

{\footnotesize $G_n$ is generated from $L_n$ using i.i.d. additional links, which occur with $\beta_n = \frac{1}{10n}$.}
\end{table}

\begin{figure}[ht]
    \centering
    \begin{subfigure}[b]{0.45\textwidth}
         \centering
         \includegraphics[width=\textwidth]{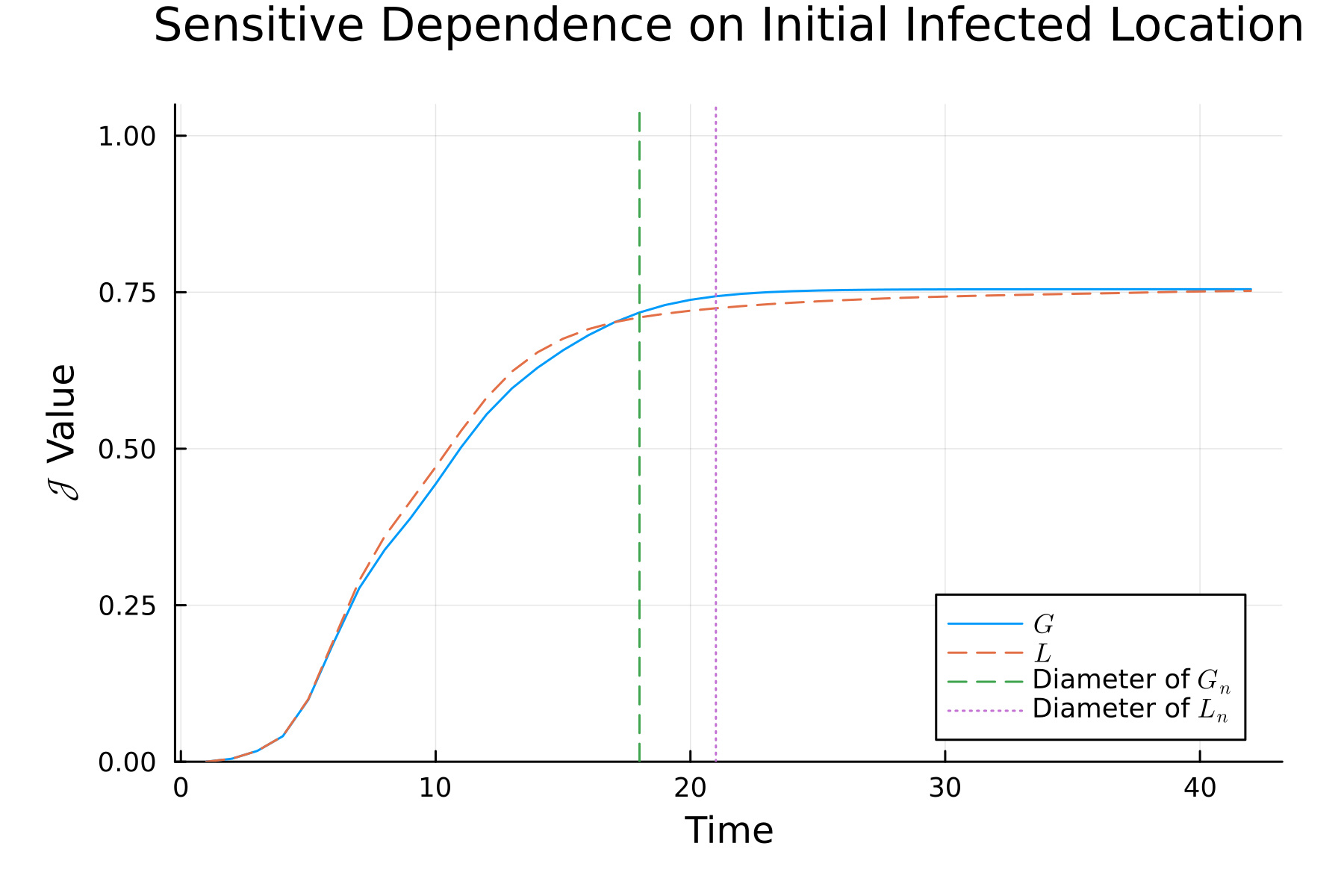}
         \caption{}
     \end{subfigure}
     \hfill
     \begin{subfigure}[b]{0.45\textwidth}
         \centering
         \includegraphics[width=\textwidth]{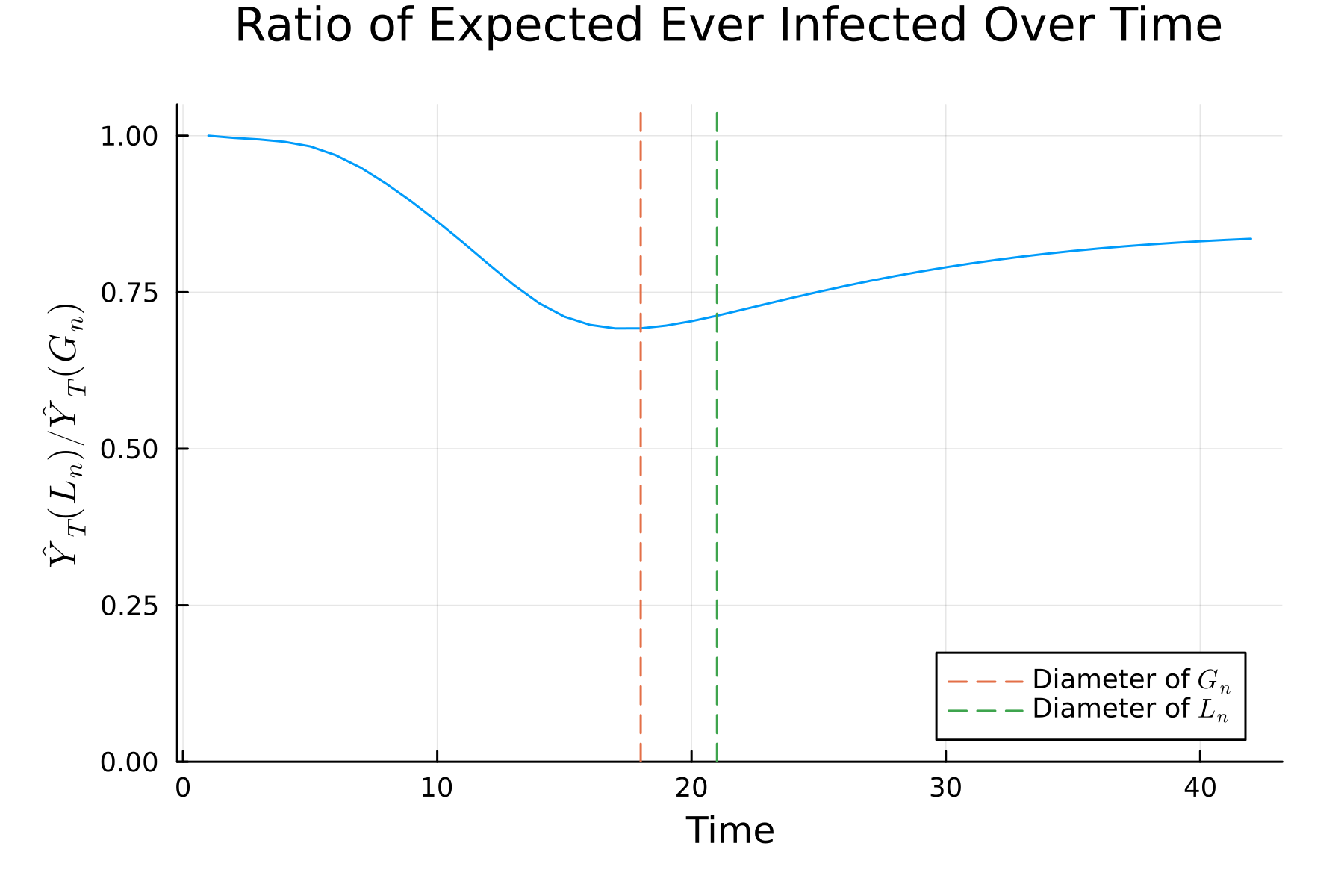}
         \caption{}
     \end{subfigure}
    \caption{Results using the COVID-19 travel data, with $G_n$ using $E_n$ generated i.i.d. with $\beta_n = \frac{1}{10n}$. Panel (A) shows the ratio $\hat Y_T(L_n)/\hat Y_T(G_n)$, while Panel (B) shows the Jaccard index $\calJ$. 
    Averages are taken over 2,500 Monte Carlo simulations.}
    \label{fig:sw-low-beta}
\end{figure}

Results are shown in Figure \ref{fig:sw-low-beta}. We take averages over 2,500 simulations. The top left panel shows the simulation of Theorem \ref{thm:sensitive-dep}. As in the main text, we choose the local neighborhood containing all $j_0$ conservatively: we chose the set to be all nodes within distance 2 of $i_0$. The distance from $i_0$ to $j_0$ is therefore 2, and the neighborhood that contains all possible $j_0$ contains 0.80 percent of the graph. Halfway to the diameter of $L_n$, the value of the average Jaccard index is 0.47 under $L_n$ and 0.44 under $G_n$, indicating largely distinct epidemics. The top right panel shows the simulation of Theorem \ref{thm:main-polynomial}. Note that in this case, the minimum ratio of $\hat Y_T(L_n)/\hat Y_T(G_n)$ is achieved at $T = 17$ and takes the value 0.69. This value is much larger than the values from the main text with either the pruned or i.i.d. errors, and comparable to the values with the same level of $\beta_n$ and graph dimension $q=4$ in the Monte Carlo simulations. 


\setcounter{figure}{0}  
\setcounter{table}{0}  
\section{NYC Micro-Cluster Application: Additional Results}\label{sec:nyc-appendix}

This appendix collects supplementary tables and figures for the NYC micro-cluster analysis in Section~\ref{sec:nyc-application}.

\begin{figure}[ht]
\centering
\includegraphics[width=\textwidth]{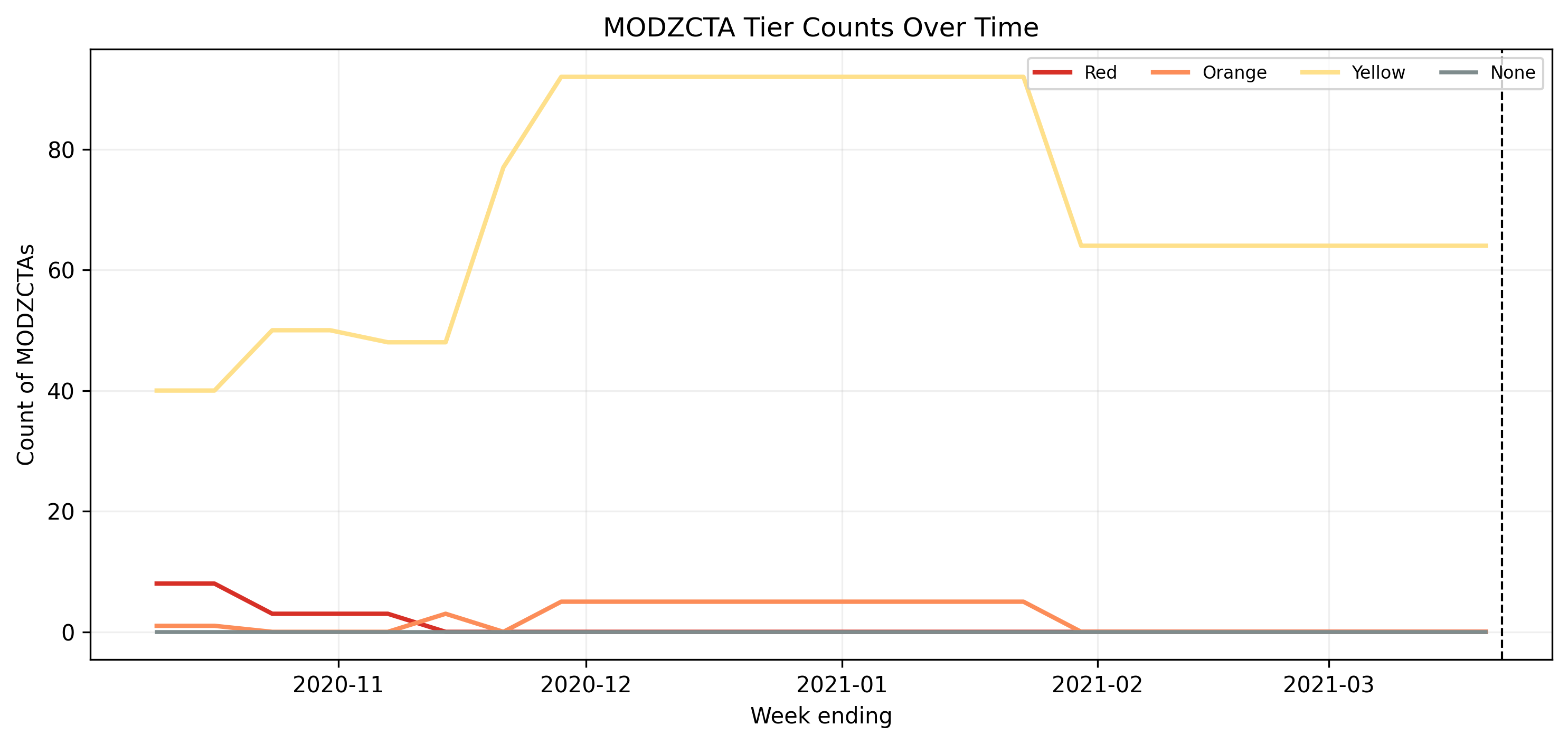}
\caption{Number of MODZCTAs assigned to each zone tier over the micro-cluster period.}
\label{fig:nyc-policy-timeline}
\end{figure}

\paragraph*{Policy timeline and summary statistics.} Figure~\ref{fig:nyc-policy-timeline} shows the count of MODZCTAs assigned to each tier over the program period. Table~\ref{tab:summary_stats} reports summary statistics for the panel.

\begin{table}[htbp]
\centering
\caption{Summary Statistics}
\label{tab:summary_stats}
\small
\begin{tabular}{lrrrrrrr}
\toprule
Variable & Mean & Std & Min & P25 & Median & P75 & Max \\
\midrule
  Case rate (per 100k) & 277.57 & 169.86 & 10.45 & 147.85 & 268.52 & 380.82 & 1827.50 \\
  Mobility risk & 0.16 & 0.14 & 0.00 & 0.00 & 0.15 & 0.29 & 0.74 \\
  Mobility risk (static) & 0.16 & 0.14 & 0.00 & 0.00 & 0.15 & 0.30 & 0.73 \\
  Neighbor risk & 0.15 & 0.16 & 0.00 & 0.00 & 0.13 & 0.33 & 0.80 \\
  Hidden threat & 0.01 & 0.05 & -0.25 & -0.02 & 0.00 & 0.03 & 0.28 \\
  Policy intensity & 0.15 & 0.18 & 0.00 & 0.00 & 0.00 & 0.33 & 1.00 \\
  $\Delta$ mobility risk & 0.01 & 0.07 & -0.55 & -0.01 & 0.00 & 0.01 & 0.71 \\
  Mobility case-rate exposure & 275.87 & 153.45 & 21.60 & 157.94 & 281.66 & 369.26 & 987.06 \\
  Neighbor case-rate exposure & 272.96 & 162.01 & 0.00 & 149.51 & 275.89 & 370.35 & 1827.50 \\
\bottomrule
\end{tabular}
\begin{tablenotes}\footnotesize
\item \textit{Notes:} Unit of observation is MODZCTA $\times$ week during the NYC micro-cluster program (Oct 2020--Mar 2021).
\end{tablenotes}
\end{table}

\begin{figure}[ht]
\centering
\includegraphics[width=0.7\textwidth]{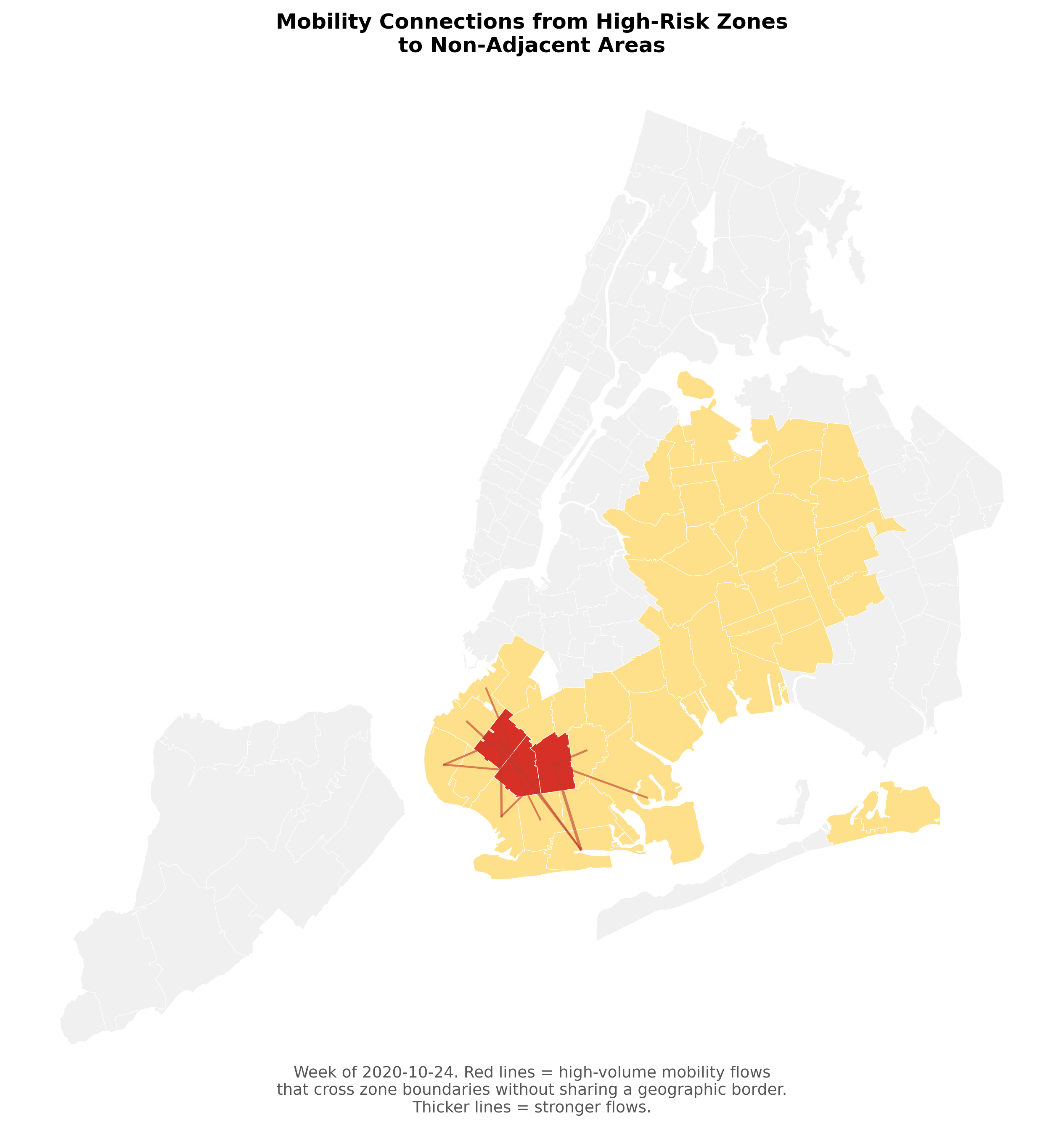}
\caption{Non-adjacent high-mobility connections from Red/Orange zone neighborhoods (week of 2020-10-24). These connections are invisible to the geographic adjacency network used by the policy.}
\label{fig:nyc-bridges}
\end{figure}

\begin{figure}[ht]
\centering
\includegraphics[width=0.8\textwidth]{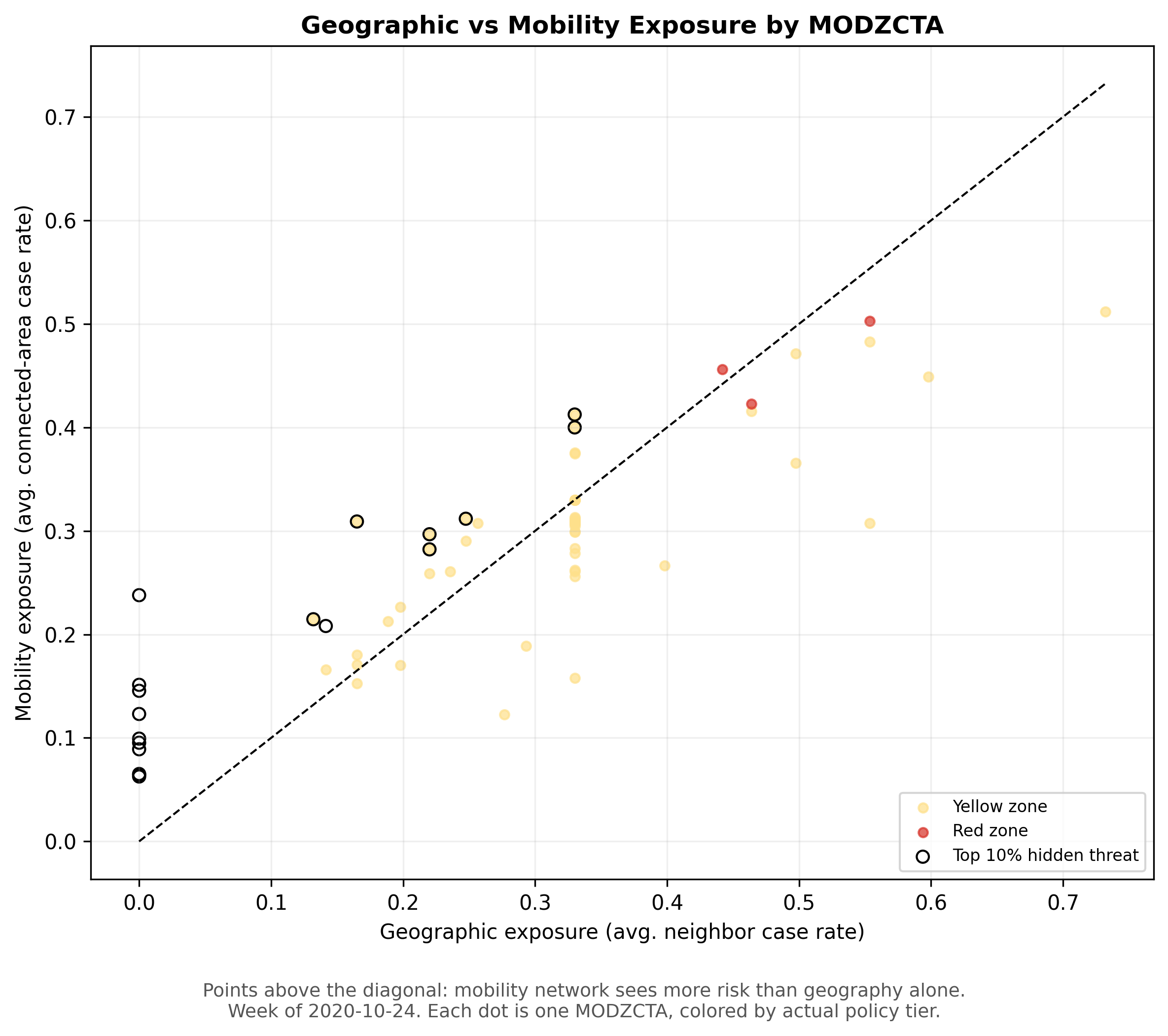}
\caption{Geographic vs.\ mobility exposure for a representative week, colored by zone tier.}
\label{fig:nyc-scatter}
\end{figure}

\paragraph*{Network mismatch and exposure descriptives.} Figure~\ref{fig:nyc-bridges} maps the high-volume mobility connections from Red/Orange zone neighborhoods that do not share a geographic border with the destination---these are the links invisible to the policy network. Figure~\ref{fig:nyc-scatter} plots geographic vs.\ mobility exposure for a representative week.  Table~\ref{tab:edge_mismatch} characterizes the edge mismatch between the two networks.

\begin{table}[htbp]
\centering
\caption{Edge Mismatch: Mobility vs.\ Geographic Adjacency}
\label{tab:edge_mismatch}
\begin{tabular}{lr}
\toprule
Metric & Value \\
\midrule
  High-mobility directed edges & 89 \\
  \quad of which non-adjacent & 7 \\
  Share non-adjacent & 0.079 \\
  P95 weight threshold & 0.254 \\
\bottomrule
\end{tabular}
\begin{tablenotes}\footnotesize
\item \textit{Notes:} High-mobility edges are those with row-normalized weight above the 95th percentile of the static (Oct 19, 2020) mobility matrix.
\end{tablenotes}
\end{table}

\paragraph*{Geo-superset horse-race.} Table~\ref{tab:superset_comparison} reports the full results for the $L$ vs.\ $G_{\text{super}}$ comparison referenced in Section~\ref{sec:nyc-application}.

\begin{table}[htbp]
\centering
\caption{Predictive Comparison: Geographic vs.\ Geo-Superset Mobility Exposure}
\label{tab:superset_comparison}
\small
\begin{tabular}{r cc ccc}
\toprule
 & \multicolumn{2}{c}{Spec 4: $G_{\text{super}}$ only} & \multicolumn{3}{c}{Spec 5: $L$ + $G_{\text{super}}$}  \\
\cmidrule(lr){2-3} \cmidrule(lr){4-6}
$h$ & $\hat\beta_{G_s}$ & $R^2$ & $\hat\beta_L$ & $\hat\beta_{G_s}$ & $R^2$ \\
\midrule
  1 & $21.579$ (2.821) & 0.186 & $0.242$ (3.524) & $21.372$ (4.169) & 0.188 \\
  2 & $21.995$ (3.658) & 0.259 & $-2.127$ (5.131) & $23.814$ (6.236) & 0.263 \\
  3 & $19.543$ (4.431) & 0.350 & $-3.394$ (4.930) & $22.445$ (6.720) & 0.352 \\
  4 & $15.226$ (4.466) & 0.459 & $-6.337$ (6.374) & $20.645$ (7.374) & 0.460 \\
\bottomrule
\end{tabular}
\begin{tablenotes}\footnotesize
\item \textit{Notes:} Dependent variable is $\Delta_{h,i,w}$. $G_{\text{super}}$: geo-superset mobility ($L \subset G_{\text{super}}$). Standardized coefficients; SEs (clustered by MODZCTA) in parentheses. All specifications include MODZCTA and week FE plus zone dummies and own case rate.
\end{tablenotes}
\end{table}

\paragraph*{Zone dummy coefficients.} Table~\ref{tab:nyc-zone-dummies} reports the zone dummy coefficients from Spec~3 of Table~\ref{tab:predictive_comparison}. Red and Orange zones show substantial negative effects at 2--4 week horizons, confirming the restrictions reduced case growth. Yellow zones show no economically significant effect.

\begin{table}[ht]
    \centering
    \caption{Zone dummy coefficients from Spec~3 ($L$ + $G_{\text{match}}$ horse-race).}
    \label{tab:nyc-zone-dummies}
    \footnotesize
    \begin{tabular}{lcccc}
        \toprule
        & 1-week & 2-week & 3-week & 4-week \\
        \midrule
        Red & $-$3.8 & $-$17.7 & $-$25.9$^{*}$ & $-$31.0$^{*}$ \\
         & \small(11.8) & \small(12.1) & \small(11.0) & \small(13.3) \\[2pt]
        Orange & $-$7.3 & $-$39.9$^{**}$ & $-$41.3$^{**}$ & $-$30.7$^{*}$ \\
         & \small(14.8) & \small(14.9) & \small(14.6) & \small(13.8) \\[2pt]
        Yellow & $-$2.1 & $-$1.2 & $-$0.1 & 1.9 \\
         & \small(2.9) & \small(4.3) & \small(5.0) & \small(5.9) \\
        \bottomrule
    \end{tabular}
    \begin{flushleft}
        \small\textit{Notes:} From Spec~3 of Table~\ref{tab:predictive_comparison}. Baseline: no zone. SEs clustered at MODZCTA level. $^{*}p<0.05$, $^{**}p<0.01$, $^{***}p<0.001$.
    \end{flushleft}
\end{table}





\begin{figure}[ht]
\centering
\includegraphics[width=0.7\textwidth]{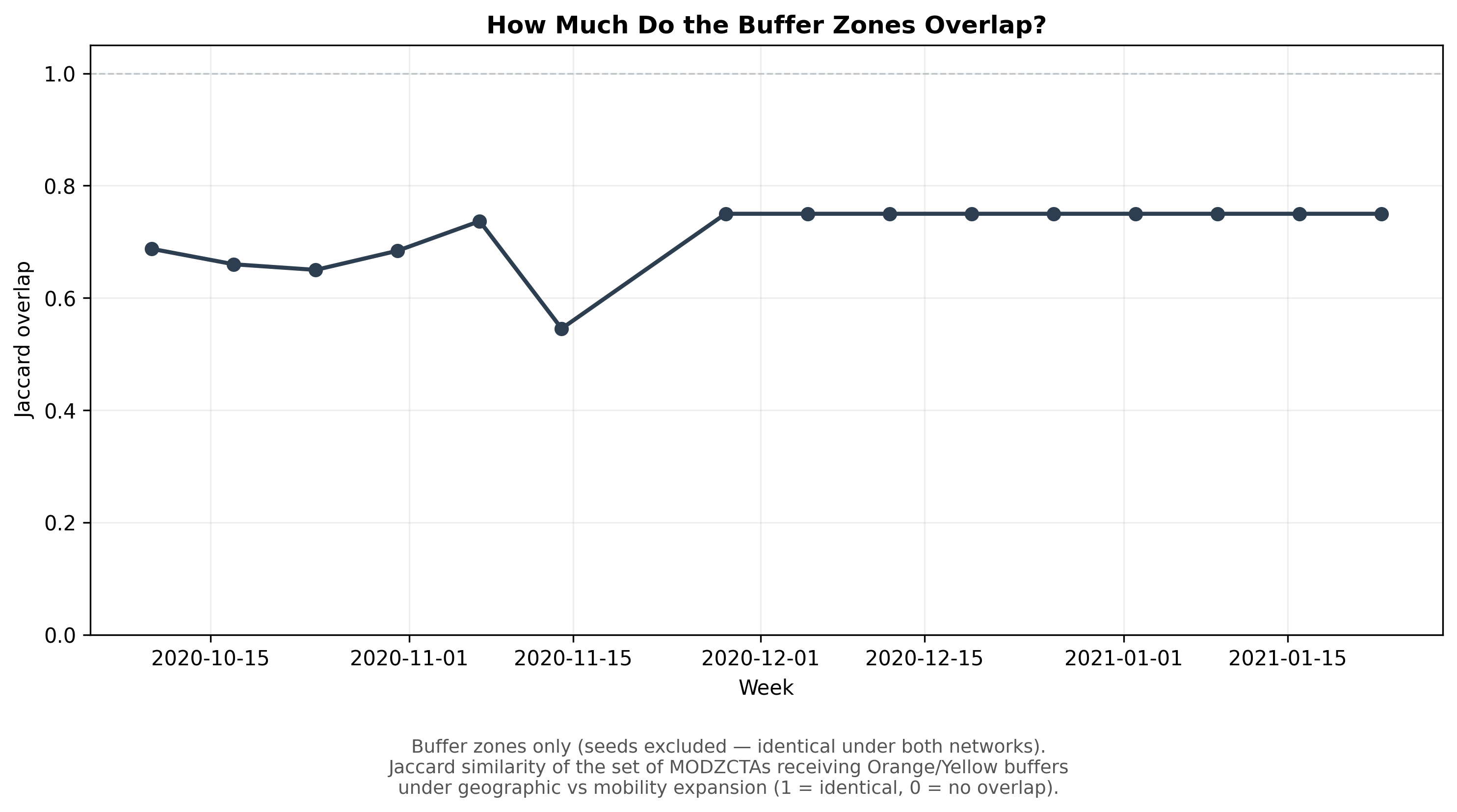}
\caption{Buffer Jaccard similarity over program weeks ($L$ vs.\ $G_{\text{match}}$, seeds excluded). Average overlap is 0.71, indicating ${\sim}30\%$ divergence in which neighborhoods receive containment resources.}
\label{fig:nyc-counterfactual-overlap}
\end{figure}

\paragraph*{Counterfactual zoning details.} Table~\ref{tab:counterfactual_superset} reports the week-by-week buffer coverage under all three networks.  Figure~\ref{fig:nyc-counterfactual-overlap} shows the Jaccard overlap over time.

\begin{table}[htbp]
\centering
\caption{Counterfactual Zoning: Buffer Coverage by Network Variant}
\label{tab:counterfactual_superset}
\small
\begin{tabular}{l r rrr rr}
\toprule
 & & \multicolumn{3}{c}{Buffer nodes} & \multicolumn{2}{c}{Flips vs.\ geo} \\
\cmidrule(lr){3-5} \cmidrule(lr){6-7}
Week & Zoned & $L$ & $G_{\text{match}}$ & $G_{\text{super}}$ & $G_{\text{match}}$ & $G_{\text{super}}$ \\
\midrule
  2020-10-10 & 54 & 46 & 33 & 46 & 19 & 5 \\
  2020-10-17 & 54 & 46 & 33 & 46 & 20 & 6 \\
  2020-10-24 & 22 & 19 & 13 & 19 & 10 & 5 \\
  2020-10-31 & 22 & 19 & 13 & 19 & 10 & 4 \\
  2020-11-07 & 22 & 19 & 14 & 19 & 8 & 6 \\
  2020-11-14 & 12 & 12 & 9 & 12 & 3 & 0 \\
  2020-11-28 & 9 & 9 & 8 & 9 & 1 & 0 \\
  2020-12-05 & 9 & 9 & 8 & 9 & 1 & 0 \\
  2020-12-12 & 9 & 9 & 8 & 9 & 1 & 0 \\
  2020-12-19 & 9 & 9 & 8 & 9 & 1 & 0 \\
  2020-12-26 & 9 & 9 & 8 & 9 & 1 & 0 \\
  2021-01-02 & 9 & 9 & 8 & 9 & 1 & 0 \\
  2021-01-09 & 9 & 9 & 8 & 9 & 1 & 0 \\
  2021-01-16 & 9 & 9 & 8 & 9 & 1 & 0 \\
  2021-01-23 & 9 & 9 & 8 & 9 & 1 & 0 \\
\midrule
  Total & 267 & 242 & 187 & 242 & 79 & 26 \\
\bottomrule
\end{tabular}
\begin{tablenotes}\footnotesize
\item \textit{Notes:} Buffer nodes are MODZCTAs assigned Orange or Yellow tier under each network. $G_{\text{match}}$: degree-matched mobility (top-$d_i$ by weight). $G_{\text{super}}$: geo-superset mobility (per-node threshold, $G \supseteq L$). Flips count MODZCTAs with a different tier assignment than under $L$.
\end{tablenotes}
\end{table}

\begin{figure}[ht]
\centering
\includegraphics[width=0.7\textwidth]{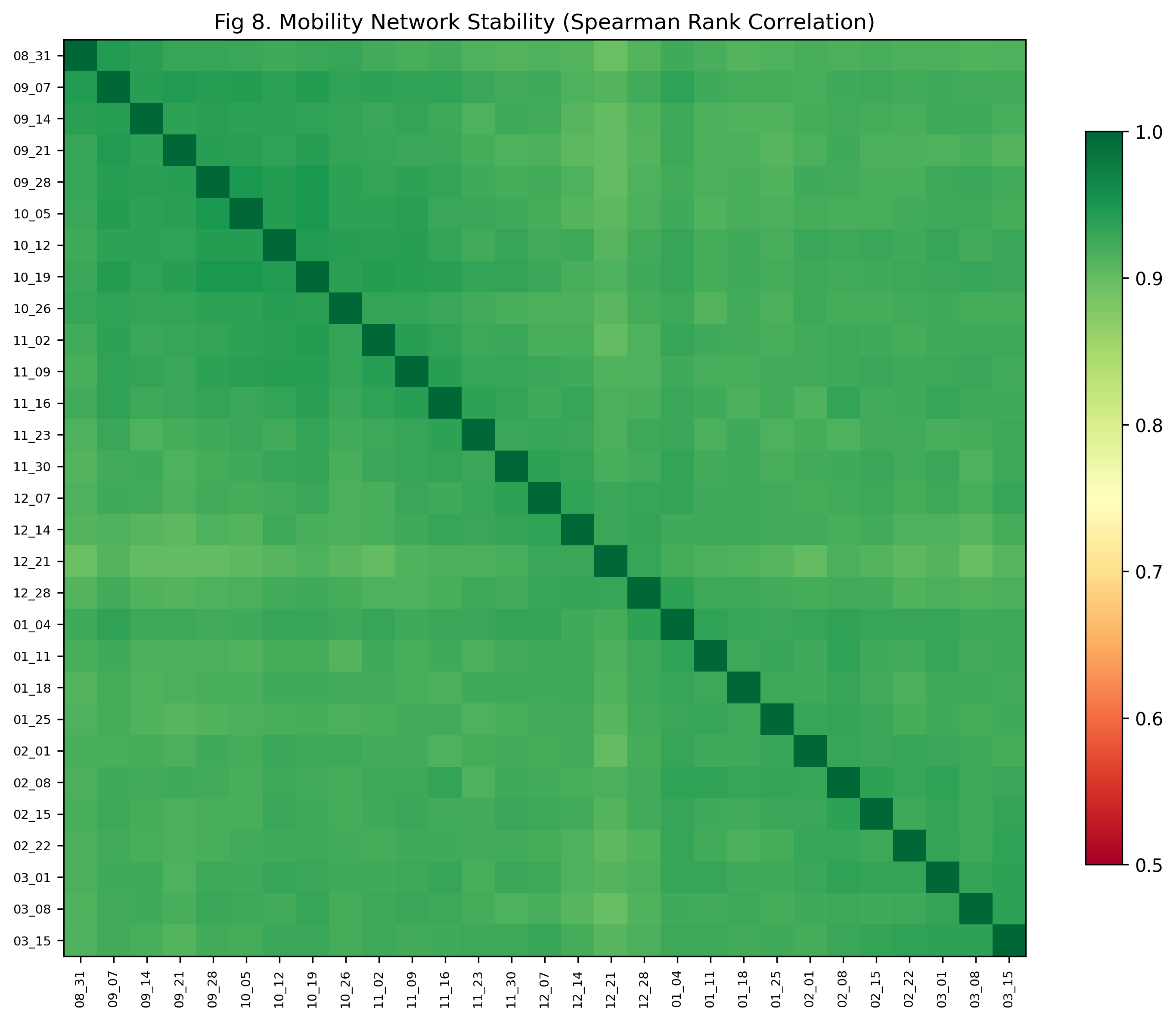}
\caption{Week-to-week Spearman rank correlation between mobility flow matrices. Average $\rho \approx 0.93$.}
\label{fig:nyc-mobility-stability}
\end{figure}

\paragraph*{Mobility stability.} Figure~\ref{fig:nyc-mobility-stability} shows the pairwise Spearman correlation between mobility flow matrices across all week pairs, averaging 0.93. The stability of the mobility network rules out the concern that week-to-week noise in flow data drives the predictive advantage.

\end{document}